\documentclass[useAMS,usenatbib]{mn2e}

\usepackage{graphicx}
\usepackage{epstopdf}
\usepackage{amsmath}
\usepackage{relsize}
\usepackage{graphics}
\usepackage{multirow}
\usepackage{booktabs}
\usepackage{supertabular}
\usepackage{lscape}
\usepackage{longtable}
\usepackage[figuresright]{rotating}
\usepackage{multicol}
\usepackage{natbib}
\usepackage{rotating}
\usepackage{epsfig}
\usepackage{amsfonts}
\usepackage{threeparttable}
\usepackage[english]{babel}
\usepackage{textcomp}
\usepackage{natbib}
\usepackage{subfigure}
\usepackage{tabularx}

\newcolumntype{h}{>{\hsize=1.5\hsize}X}

\usepackage{pifont}

\usepackage{amssymb} 
\usepackage{url}
\usepackage{float}

\newcolumntype{R}[1]{>{\raggedleft\arraybackslash}p{#1}}
\newcolumntype{C}[1]{>{\centering\arraybackslash}p{#1}}
\newcolumntype{L}[1]{>{\raggedright\arraybackslash}p{#1}}

\title[Newly confirmed M-dwarfs in BPMG and ABDMG]{Spectroscopic confirmation of M-dwarf candidate members of the Beta Pictoris and AB Doradus Moving Groups}
\author[A. S. Binks \& R. D. Jeffries]{A. S. Binks\thanks{E-mail: a.s.binks@keele.ac.uk} \& R. D. Jeffries\\
Astrophysics Group, School of Chemistry and Physics, Keele University, Keele, Staffordshire ST5 5BG}
\begin{document}

\date{Accepted. Received; in original form}

\pagerange{\pageref{firstpage}--\pageref{lastpage}} \pubyear{2015}

\maketitle

\label{firstpage}

\begin{abstract}
Optical spectroscopic observations are reported for 24 and 23, nearby, proper-motion-selected M-dwarf candidate members of the Beta Pictoris and AB Doradus moving groups (BPMG and ABDMG). Using kinematic criteria, the presence of both H$\alpha$ emission and high X-ray-to-bolometric luminosity, and position in absolute colour-magnitude diagrams, 10 and 6 of these candidates are confirmed as likely members of the BPMG and ABDMG respectively. Equivalent widths or upper limits for the Li~{\sc i}~6708\AA\ line are reported and the lithium depletion boundary (LDB) age of the BPMG is revisited. Whilst non-magnetic evolutionary models still yield an estimated age of $21 \pm 4$\,Myr, models that incorporate magnetic inhibition of convection imply an older age of $24 \pm 4$ Myr. A similar systematic increase would be inferred if the stars were 25 per cent covered by dark magnetic starspots. Since young, convective M-dwarfs are magnetically active and do have starspots, we suggest that the original LDB age estimate is a lower limit. The LDB age of the ABDMG is still poorly constrained -- non-magnetic evolutionary models suggest an age in the range 35--150\,Myr, which could be significantly tightened by new measurements for existing candidate members.
\end{abstract}


\section{Introduction}\label{S_Intro}

Since the 1990s, at least 10 kinematically-coherent, but spatially-dispersed young (10--100\,Myr) groups of stars have been discovered in the Solar neighbourhood (within 100\,pc). These `moving groups' (herein MGs, see, for example, \citealt{2004a_Zuckerman}; \citealt{2008a_Torres}; \citealt{2013a_Malo}) are important because their members are nearby, offering excellent opportunities to spatially resolve young, and hence relatively luminous, low-mass companions. Assuming their members are coeval, their ages can be estimated using techniques similar to those deployed for age-dating clusters and can then be used to test stellar evolutionary models.

Two of the most important groups are known as the Beta Pictoris MG (BPMG, \citealt{1999a_Barrado_y_Navascues}; \citealt{2001a_Zuckerman}) and the AB Doradus MG (ABDMG, \citealt{2004a_Zuckerman}; \citealt{2013a_Barenfeld}). The BPMG, including $\gtrsim\,40$ members with measured parallaxes, is one of the closest MGs (most members are between 10 and 70\,pc) and has an age of $21-26$\,Myr (\citealt{2014a_Binks}; \citealt{2014a_Malo}; \citealt{2014a_Mamajek}). Studies of BPMG members have led to the discovery of numerous sub-stellar companions and directly imaged circumstellar material. These include: the disc around $\beta$ Pic and its $<20\,M_{\rm Jup}$ companion (\citealt{1984a_Smith}; Lagrange 2010; 2011; \citealt{2014a_Bonnefoy}), the free-floating late-L dwarf PSO~J318.5-22 (\citealt{2013a_Liu}), the sub-stellar companion to PZ~Tel (\citealt{2010a_Biller}; \citealt{2012a_Jenkins}; \citealt{2012a_Mugrauer}) and the imaged disc around the M-dwarf AU Mic (\citealt{2004a_Kalas}; \citealt{2006a_Augereau} and \citealt{2014a_MacGregor}). Recently, \cite{2015a_Macintosh} directly imaged a $2-12\,M_{\rm Jup}$ companion to the F0 star 51~Eri in the BPMG. 

\nocite{2010a_Lagrange}
\nocite{2011a_Lagrange}

There are $\gtrsim\,50$ ABDMG members with measured parallaxes; surveys in the past 5 years have extended the number of likely candidates to $\sim\,100$ (\citealt{2012a_Schlieder}; \citealt{2013a_Malo}; Gagn\'e et al. 2014; 2015a). The group was reported to include the the $4-7\,M_{\rm Jup}$ free-floating planet CFBDSIR 2149-0403 (\citealt{2012a_Delorme}), the $\sim$ T5 brown dwarf SDSS~1110+0116 (Gagn\'e et al. 2015b) and the eponymous member, AB Dor, is a quadruple system including the very-low mass AB Dor C ($0.090 \pm 0.005\,M_{\odot}$, \citealt{2005a_Close}). Reported ages for the group have ranged from 50 to 150\,Myr, however, an analysis of a sample of K-dwarf members by \cite{2013a_Barenfeld} found that the main-sequence turn on constrains the age to $\geq 110\,$Myr, and recent work has suggested coevality with the Pleiades at 125\,Myr (\citealt{2005a_Luhman}; \citealt{2007a_Ortega}; \citealt{2014a_McCarthy}; \citealt{2015a_Bell}). 

\nocite{2015b_Gagne}

Based on their populations of solar-type and high-mass stars, then for a standard IMF we would expect large numbers of low-mass stars to be members of these MGs. In recent years there has been a focus on finding these objects, mainly through proper-motion selection. These low-mass members are valuable because they offer the best opportunity to identify and investigate even lower mass brown dwarf and planetary companions with optimal contrast and spatial resolution. The known age of MGs means that these objects become benchmarks against which to test the uncertain physics of very low-mass stellar, substellar and planetary evolution and atmospheres (\citealt{2013a_Biller}; \citealt{2013a_Dent}; \citealt{2014a_Brandt}; \citealt{2015a_Bowler}). Confirmed low-mass members can also be used to refine the ages of MGs using the lithium depletion boundary (LDB) technique - the lowest luminosity at which Li remains present in the photospheres of these fully convective objects leads to an age that is precise and may be less model-dependent than rival techniques (\citealt{1997a_Bildsten}; \citealt{2001a_Jeffries}; \citealt{2004a_Burke}; \citealt{2015a_Tognelli}). \cite{2014a_Binks} used M-dwarf members of the BPMG to calculate an LDB age of $21 \pm 4\,$Myr, which is larger than earlier reported ages based on isochrones in the Hertzsprung-Russell diagram (\citealt{2001a_Zuckerman}) and kinematic traceback (\citealt{2002a_Song}). \cite{2014b_Malo} used a similar sample, but obtained an age of $26 \pm 3$\,Myr from evolutionary models featuring the influence of magnetic fields (see section~\ref{S_LDB_BPMG}). The ABDMG is sparsely populated with known M-dwarf members and no LDB has been reported for it as yet.

In this paper we present spectroscopy of proper-motion selected M-dwarf candidates of the BPMG (24 candidates) and ABDMG (23 candidates) and test their membership status based on kinematic and age-dependent criteria. Eight of the BPMG candidates were presented in \cite{2014a_Binks}, but the details of the observations are presented here for the first time. In $\S$\ref{S_Candidate_Observations} we describe the initial target selection and in $\S$\ref{S_Results} we present all the spectroscopic observations and compare our measurements with previously published values. We assess membership status in $\S$\ref{S_Confirming} and in $\S$\ref{S_LDBs} we discuss the implications for the LDBs of BPMG and ABDMG in light of the updated M-dwarf samples in each group. Concluding remarks are provided in $\S$\ref{S_Conclusions}.

\section{Candidate selection and observations}\label{S_Candidate_Observations}
M-dwarf candidates of the BPMG and ABDMG were selected for observation from possible members listed in the proper-motion-based surveys of \cite{2012a_Shkolnik}, \cite{2012a_Schlieder} and \cite{2013a_Malo}. Spectra for ten BPMG candidates were obtained on 28-29 December 2012 using the 2.56-m Nordic Optical Telescope (NOT) and Fibre-fed Echelle Spectrograph (FIES, $R\,\sim\,46000$), calibrated with simultaneous ThAr lamp spectra. The wavelength range covered $\lambda\lambda\,3630-7260$\AA\ and spectra were flat-fielded, extracted, wavelength calibrated and blaze-corrected using {\sc FIEStool} (\citealt{2005a_Stempels}). Spectroscopy for a further 14 BPMG and 23 ABDMG candidates were obtained on a second observing run on 20-26 March 2013 using the 2.5-m Isaac Newton Telescope (INT) and Intermediate Dispersion Spectrograph. The H1800V grating and a 1.4 arcsec slit gave a 2-pixel resolution of 0.7\AA\ in the range $\lambda\lambda\,6540-7170$\AA. INT spectra were bracketed with Cu-Ne lamp exposures and extracted and wavelength calibrated using standard tasks from the IRAF package; we observed the spectro-photometric standard Hiltner~600 at twilight to obtain relative flux-calibrated spectra.

Heliocentric radial velocities (RVs) were determined by cross-correlation with the M-dwarf RV standards HD~190007, GJ~411 and GJ~526 (M0V, M2V and M4V, respectively) at the NOT and GJ~686, HD~119850, HD~265866, GJ~273 and GJ~699 at the INT (M1V, M1.5V, M3V, M3.5V and M4V, respectively), where the RVs of the standards are published in \cite{2012a_Chubak}. Typical RV precisions were 1.2 and $1.5\,{\rm km\,s}^{-1}$ at the NOT and INT respectively. The external accuracy of the measurements, judged against the standards, was about $0.3\,{\rm km\,s}^{-1}$.

Our INT spectra have relatively low resolving power ($\sim\,7000$), but we have attempted to estimate the $v\sin i$ of our targets, where no better measurement exists in the literature. We fitted a quadratic relationship between the measured width of the cross-correlation functions used to estimate the RVs and the published $v\sin i$ values for 9 of our targets that also had high resolution spectra in \cite{2014a_Malo}. The rms discrepancy from the fit was $3\,{\rm km\,s}^{-1}$. This relationship was then used to estimate a $v\sin i$ for all the other targets. The limited resolving power of our spectra meant we were unable to discern rotational broadening below $20\,{\rm km\,s}^{-1}$ and the calibration would be an extrapolation above $60\,{\rm km\,s}^{-1}$. In such cases we quote upper or lower limits to $v\sin i$ respectively. All $v\sin i$ measurements, whether obtained from the literature or from our calibration, are presented in Tables~\ref{T_BPMG} and \ref{T_ABDMG}.

The equivalent widths (EWs) of the H$\alpha$ and Li lines were determined by direct integration above/below a continuum. Uncertainty in continuum placement leads to an estimated uncertainty of $\sim 0.1$\AA\ in the H$\alpha$ EWs, whilst the Li EW errors were calculated using the formulation in \cite{1988a_Cayrel_de_Strobel}. Where no Li line could be located, we quote 2$\sigma$ upper limits (see Tables~\ref{T_BPMG} and~\ref{T_ABDMG}).

Spectral types were estimated to a precision of half a sub-class using narrow-band TiO5 spectral indices at wavelengths of 7042--7046\,\AA\ and 7126--7135\,\AA\ (\citealt{1997a_Gizis}).

\nocite{2014a_Binks}

\section{Results}\label{S_Results}

\begin{figure*}
    \begin{minipage}[b]{\textwidth}
    \begin{center}
      \begin{tabular}{cc}
\includegraphics[width=0.42\textwidth]{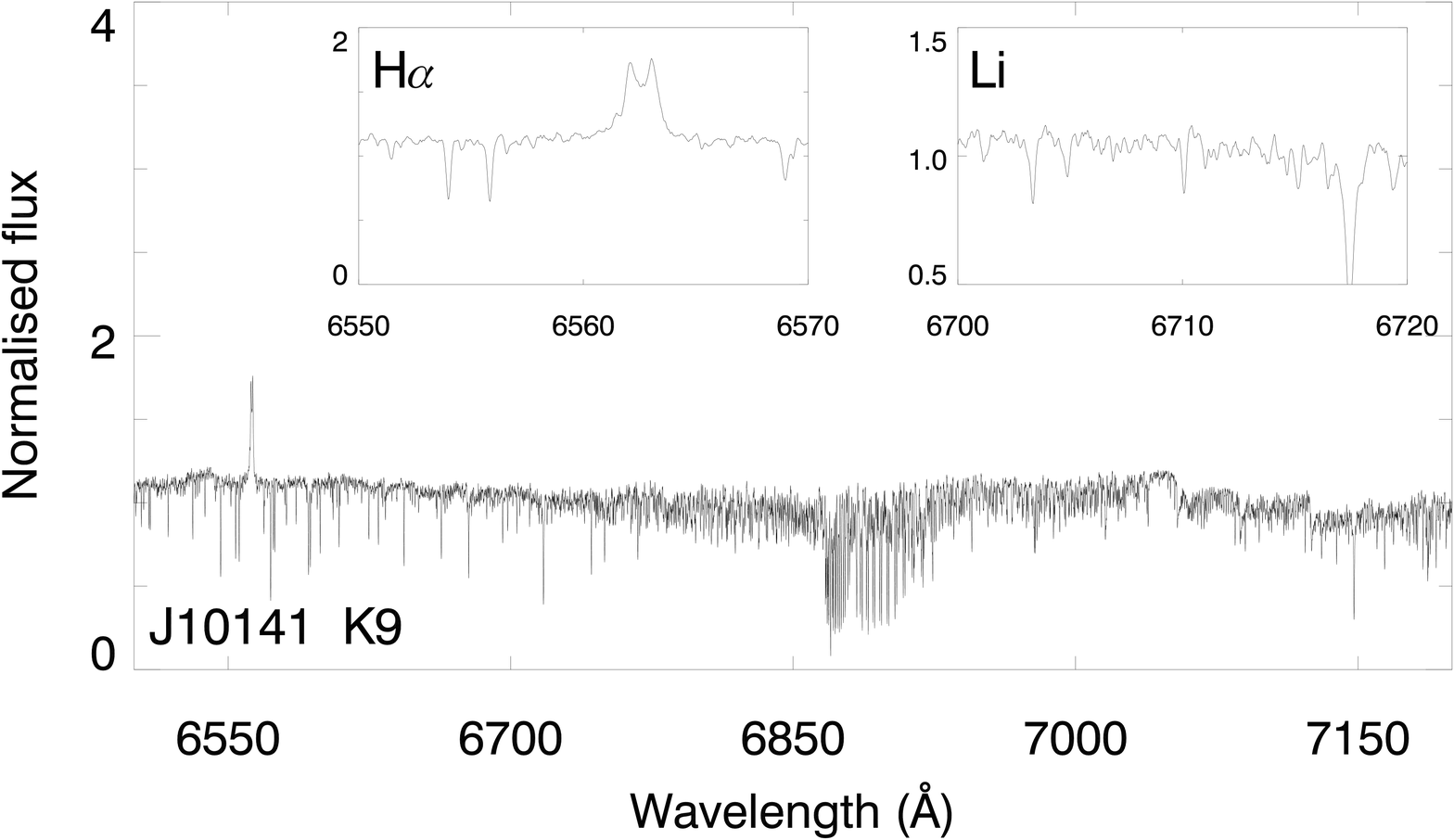} & \includegraphics[width=0.42\textwidth]{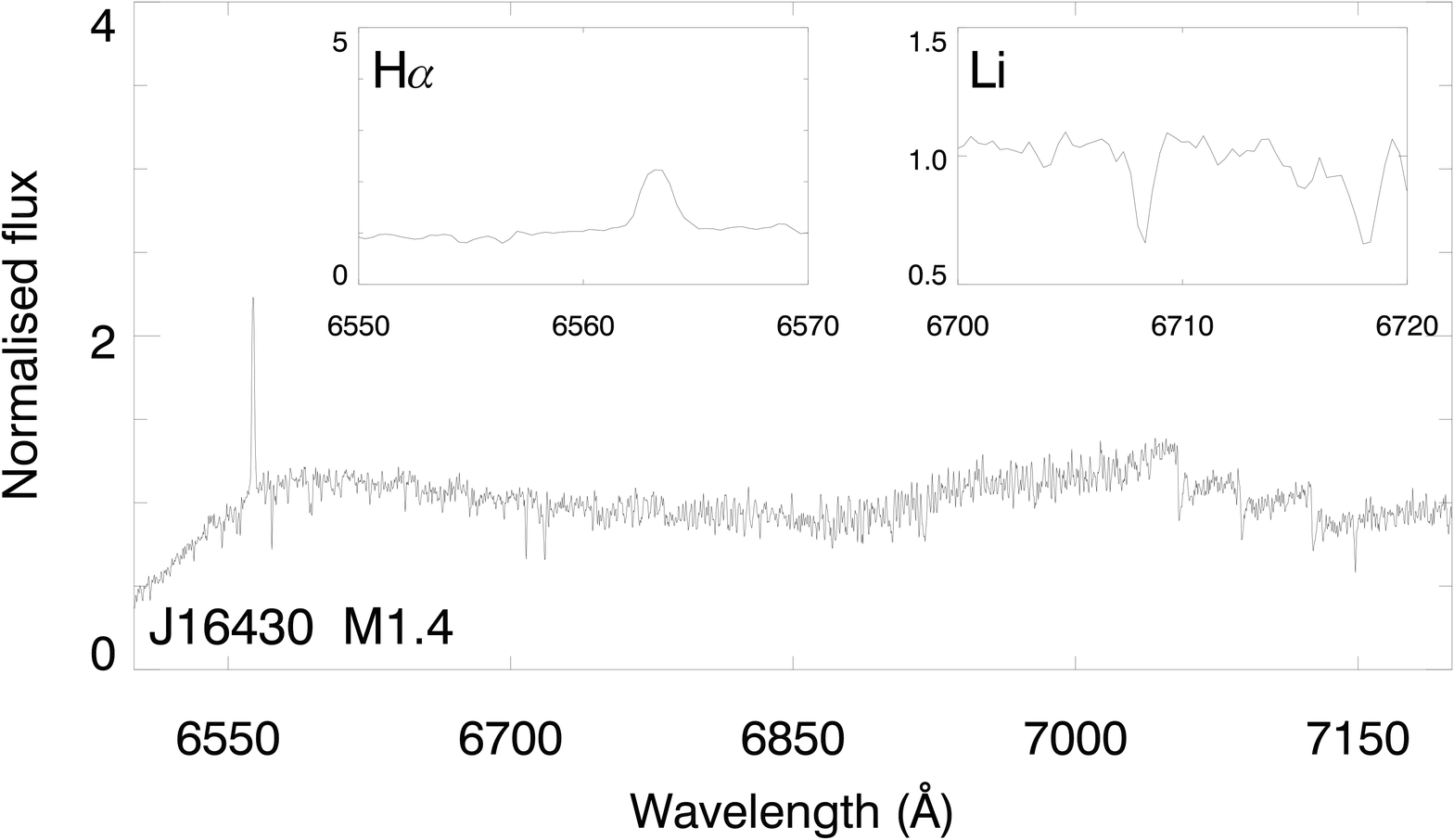} \\
\includegraphics[width=0.42\textwidth]{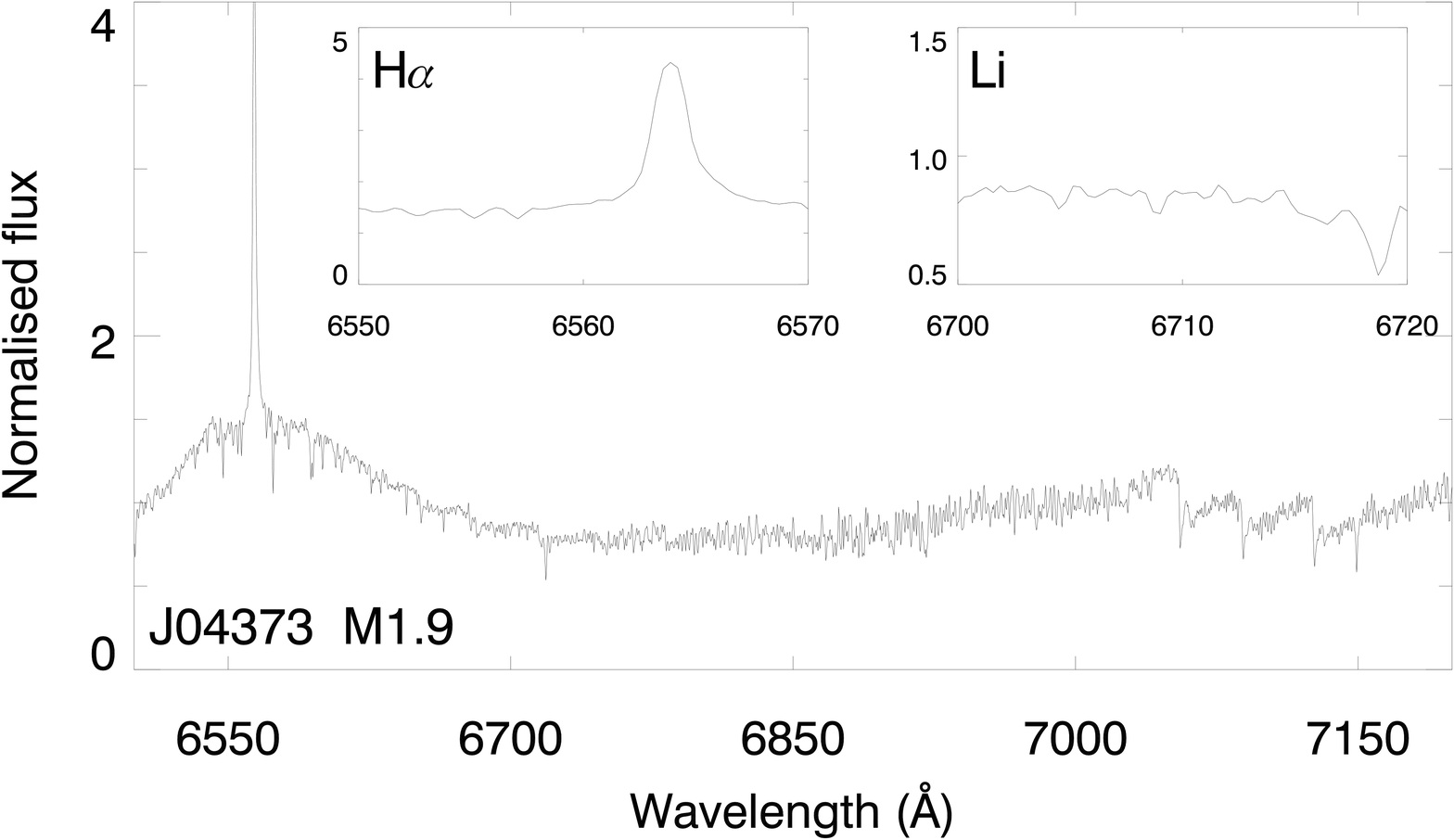} & \includegraphics[width=0.42\textwidth]{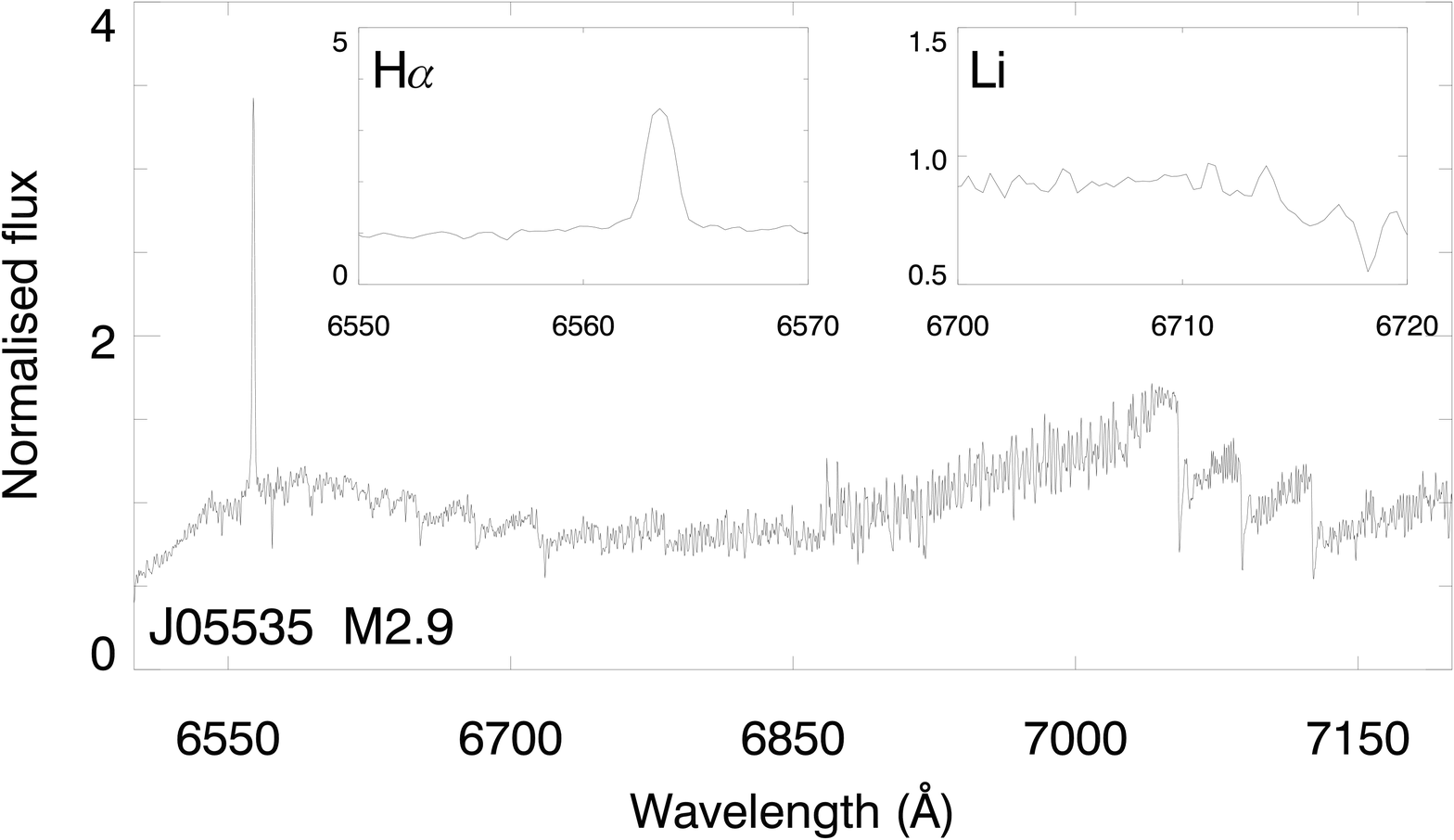} \\
\includegraphics[width=0.42\textwidth]{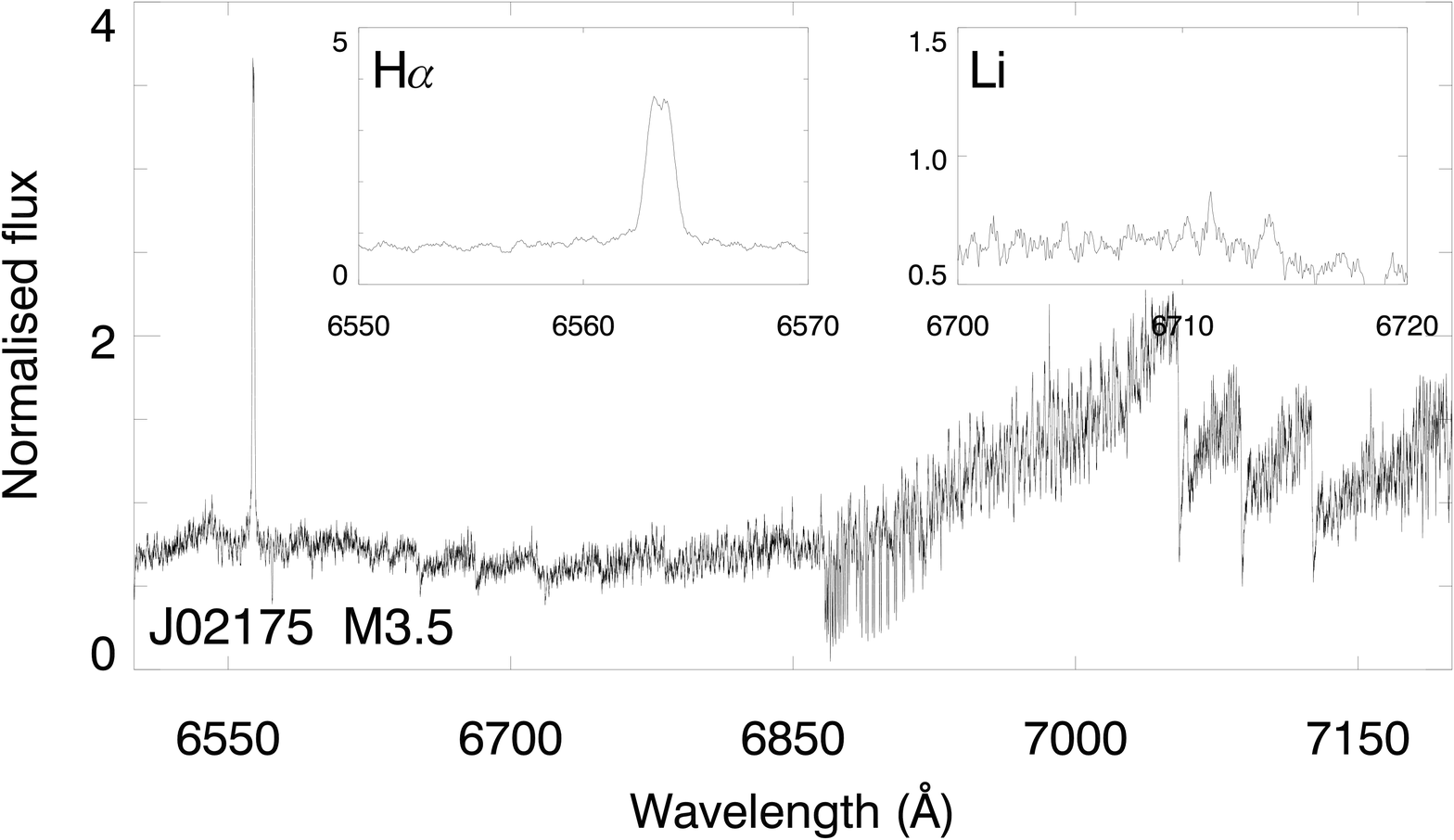} & \includegraphics[width=0.42\textwidth]{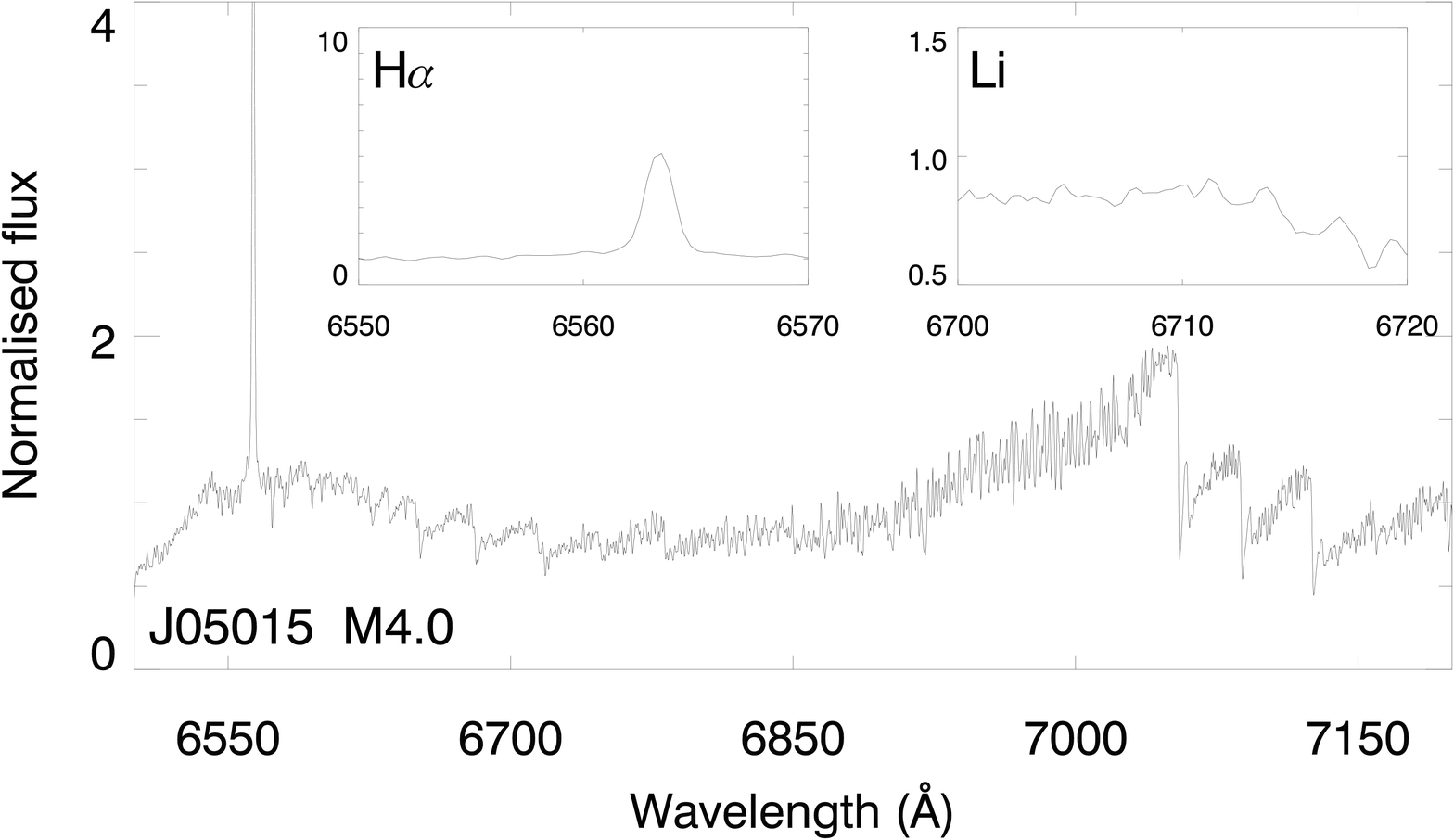} \\
\includegraphics[width=0.42\textwidth]{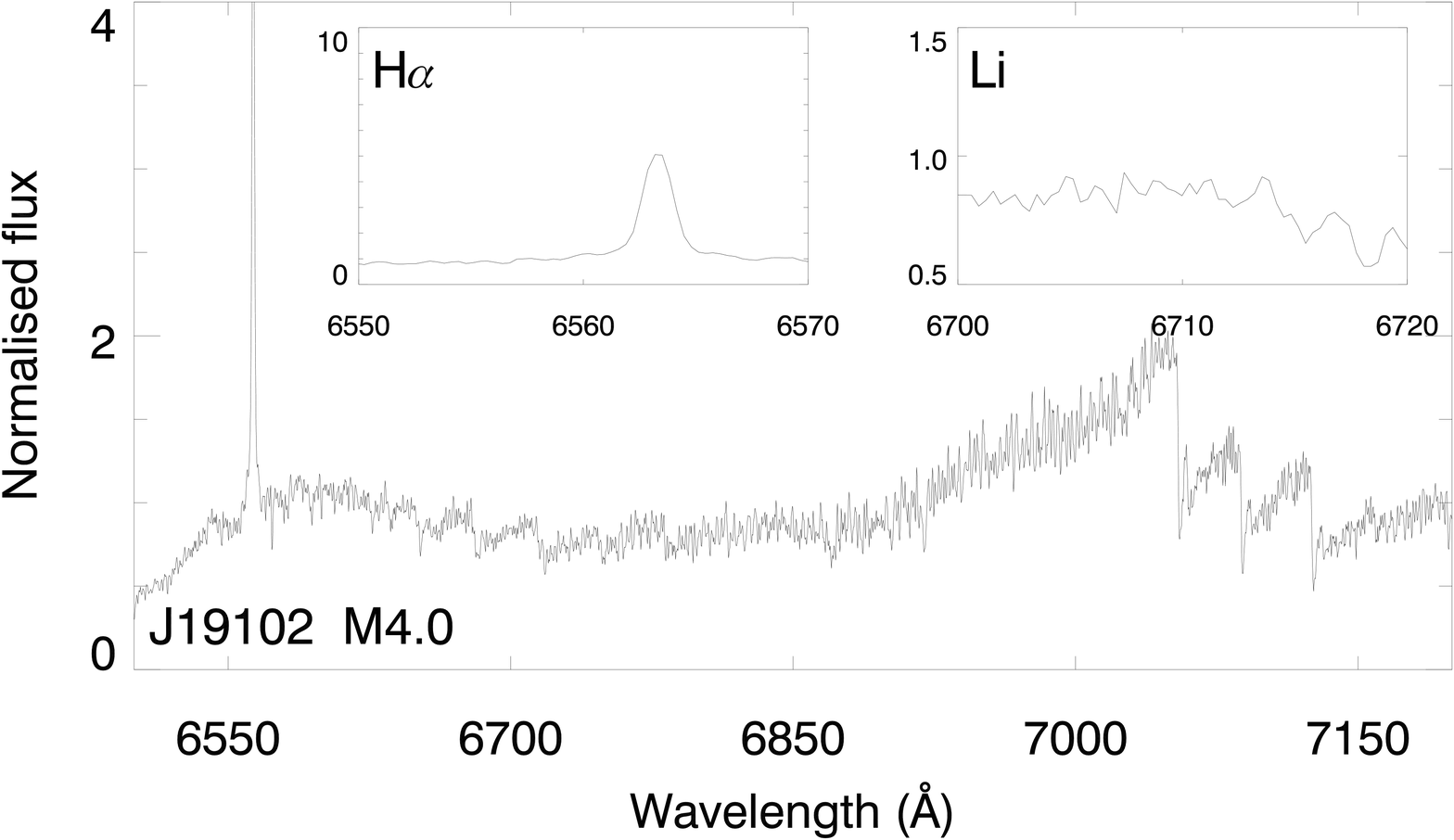} & \includegraphics[width=0.42\textwidth]{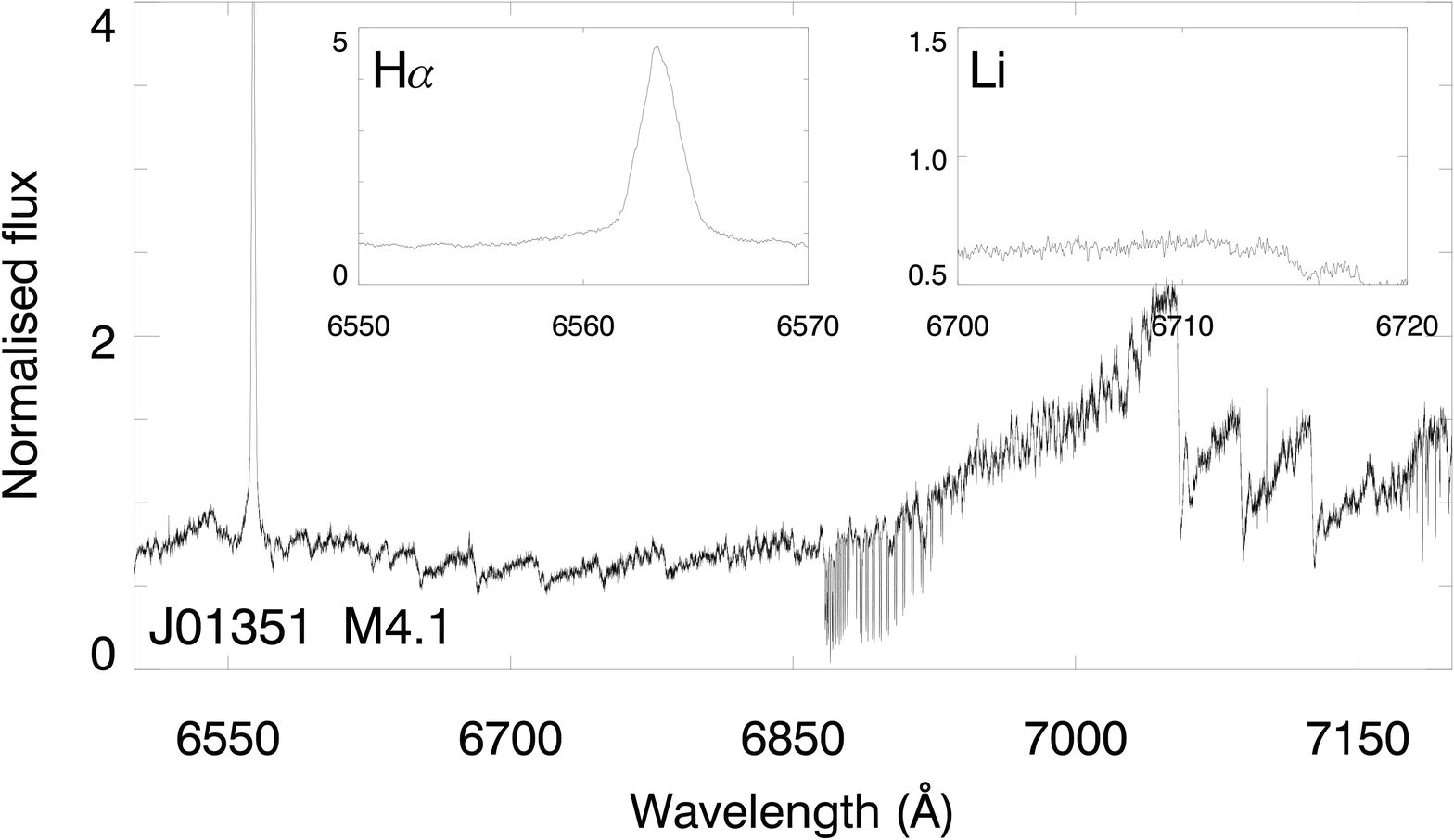} \\
\includegraphics[width=0.42\textwidth]{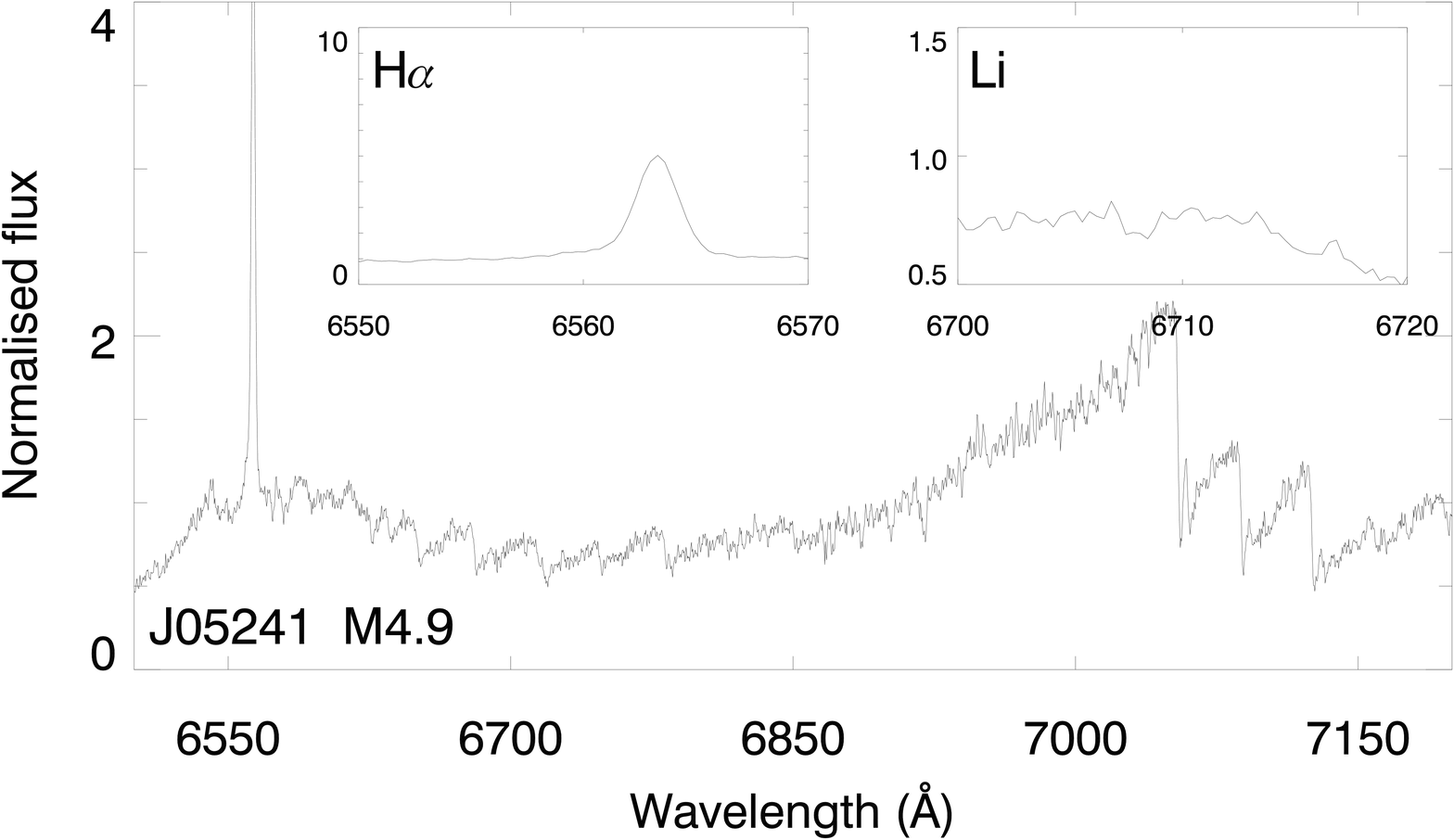} & \includegraphics[width=0.42\textwidth]{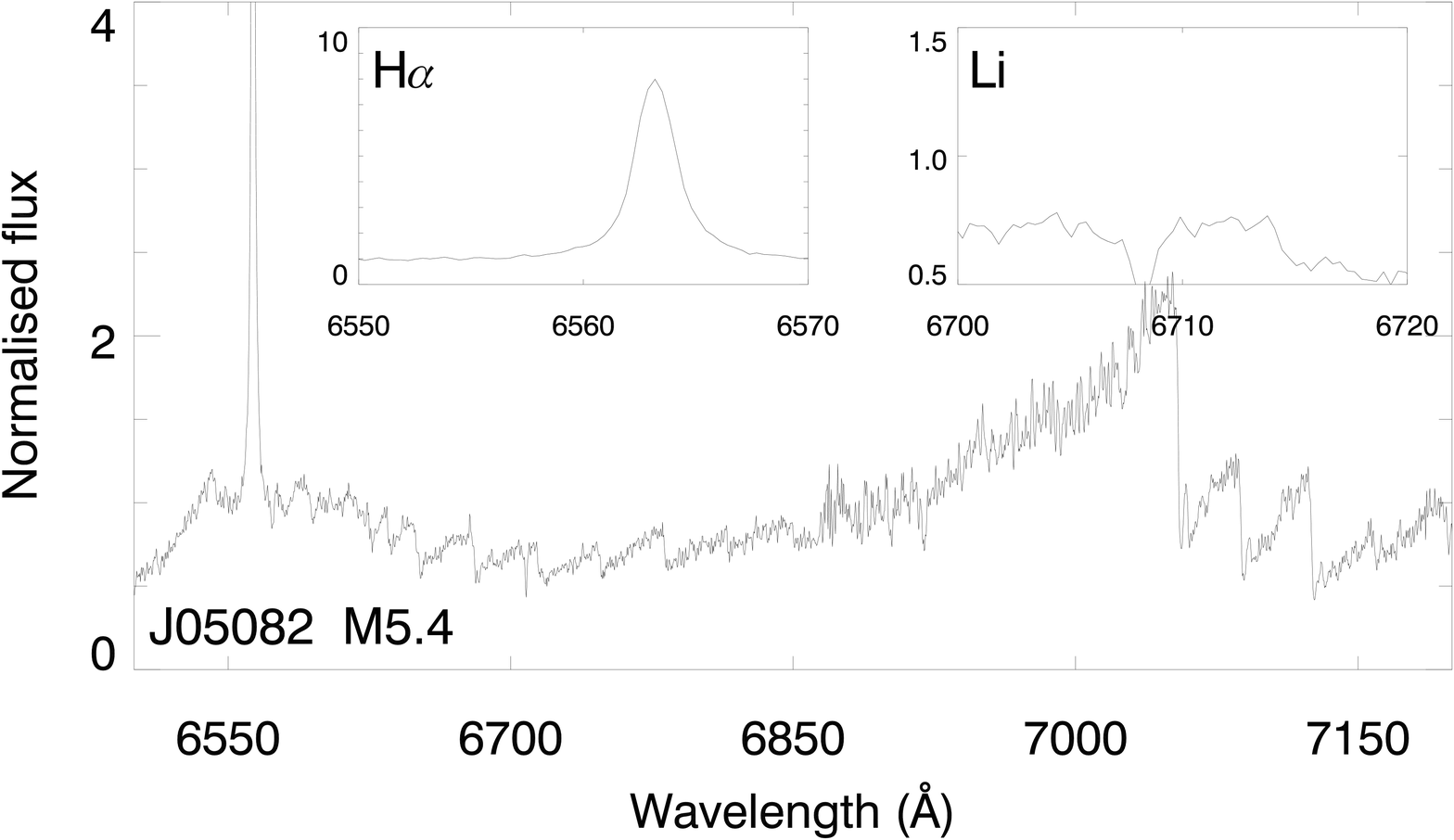} \\
      \end{tabular}
      \end{center}
\caption{The ten confirmed BPMG members which also qualify from the requirements of H$\alpha$ in emission and suitable placement on the CMD. Full 2MASS names are given in Table~\ref{T_BPMG}. The inserts in each plot are nomalised spectra in the regions of the H$\alpha$ and Li~{\sc i} 6708\AA\ line. All spectra (excluding objects `J0135' and `J0217', observed at the NOT, which have been blaze-corrected) have been subject to relative flux-calibration and telluric correction.}
\label{F_BPMG_Good}
    \end{minipage}
\end{figure*}

\begin{figure*}
    \begin{minipage}[b]{\textwidth}
    \begin{center}
      \begin{tabular}{cc}
\includegraphics[width=0.42\textwidth]{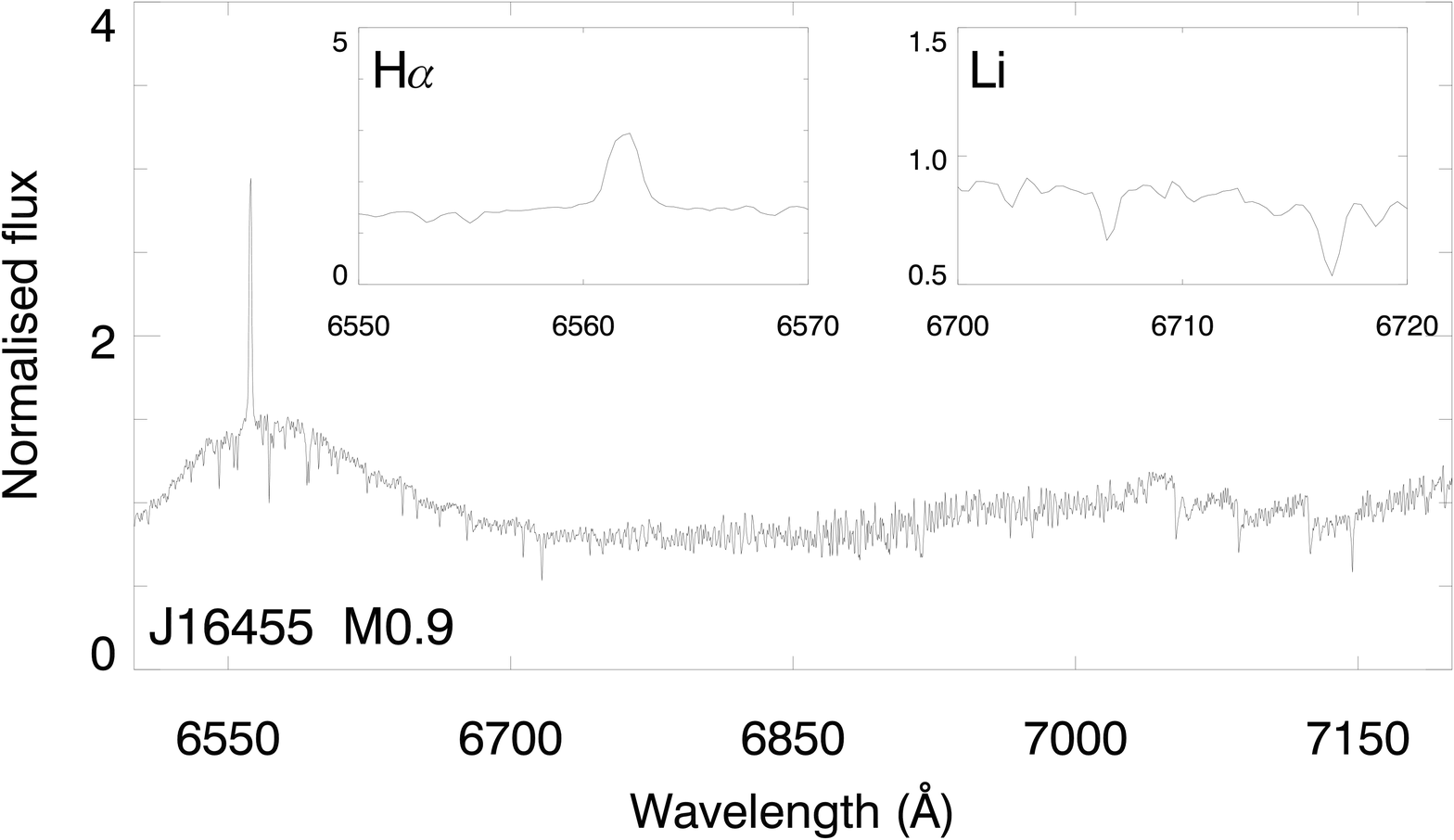} & \includegraphics[width=0.42\textwidth]{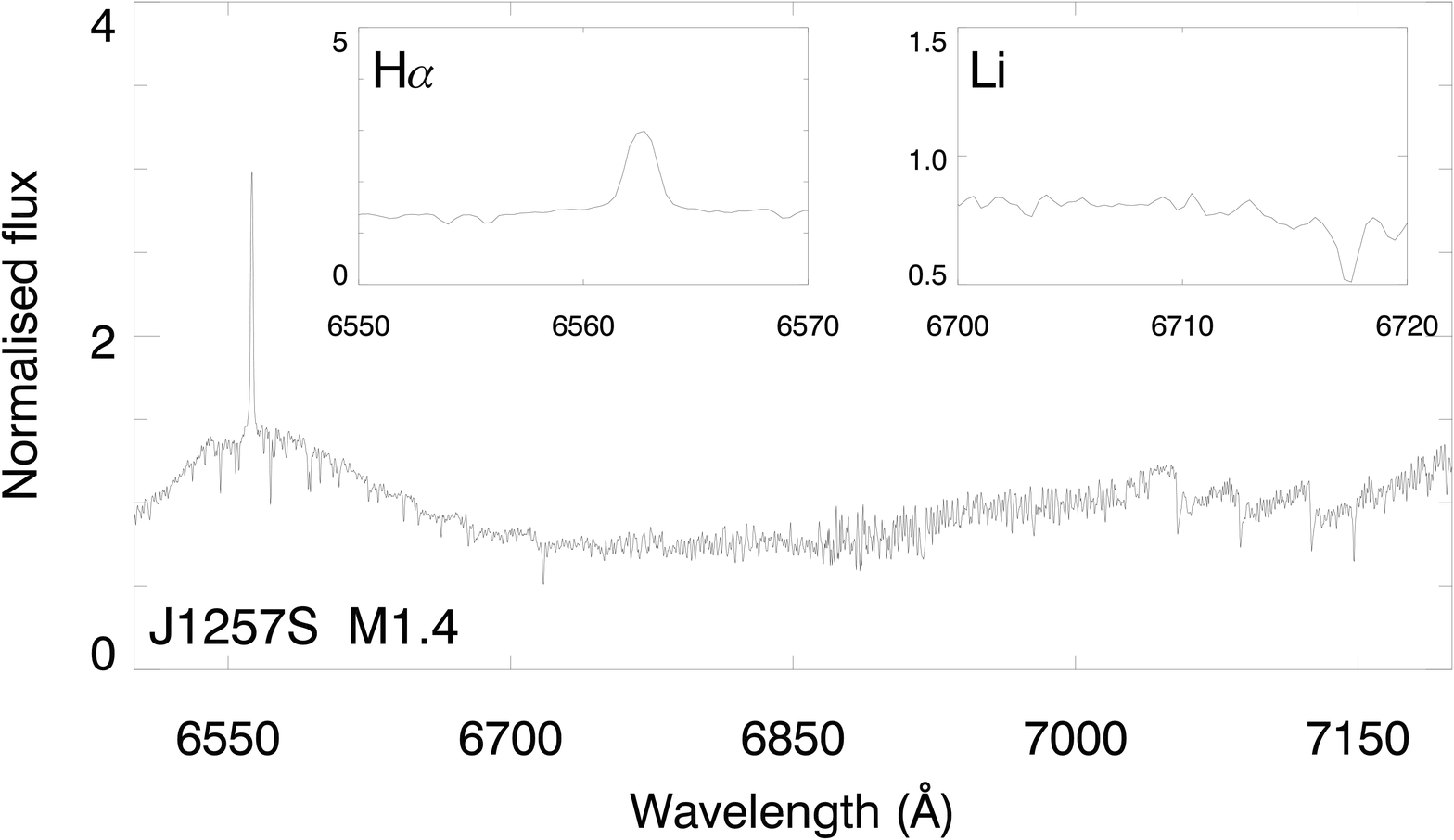} \\
\includegraphics[width=0.42\textwidth]{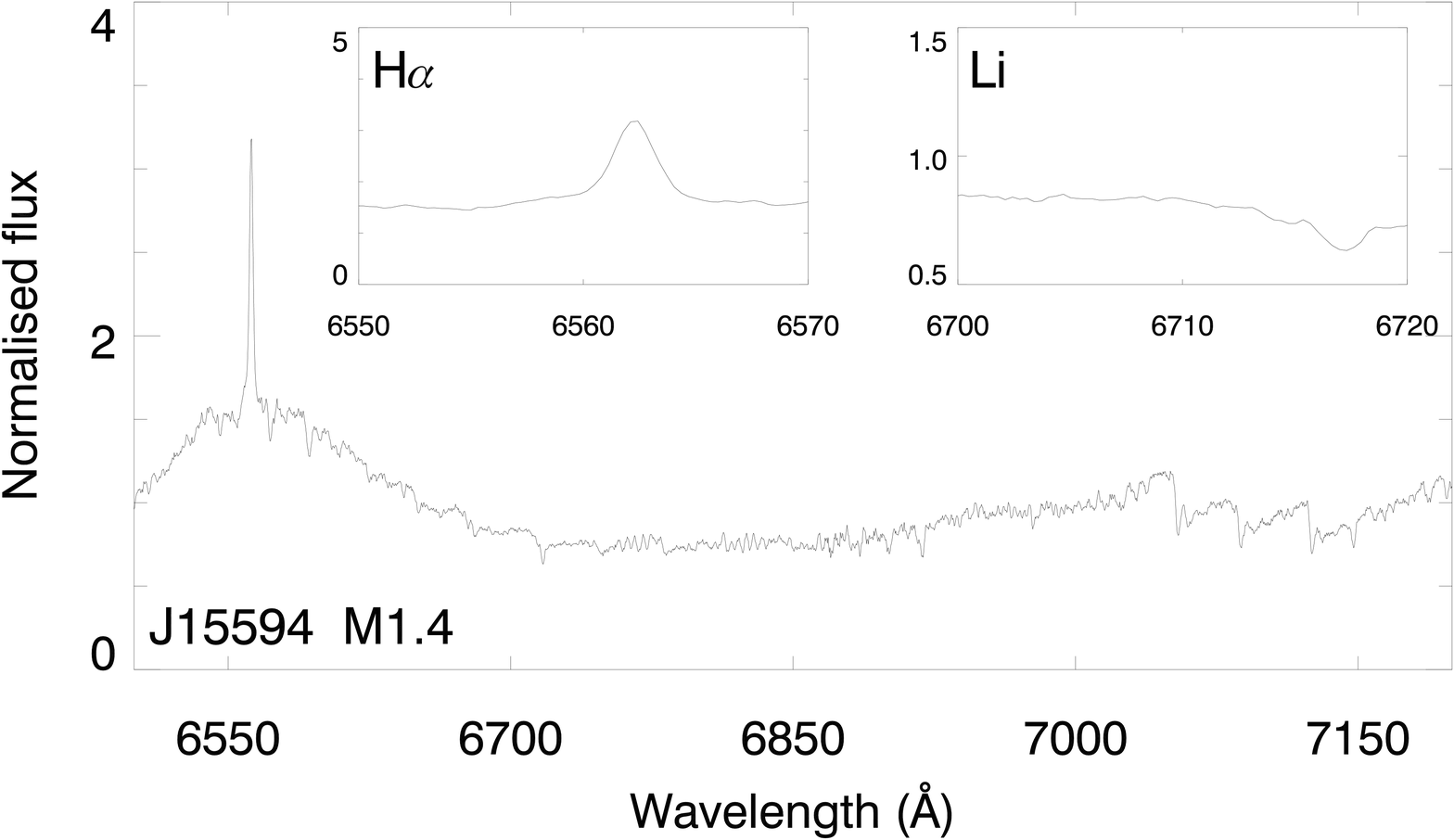} & \includegraphics[width=0.42\textwidth]{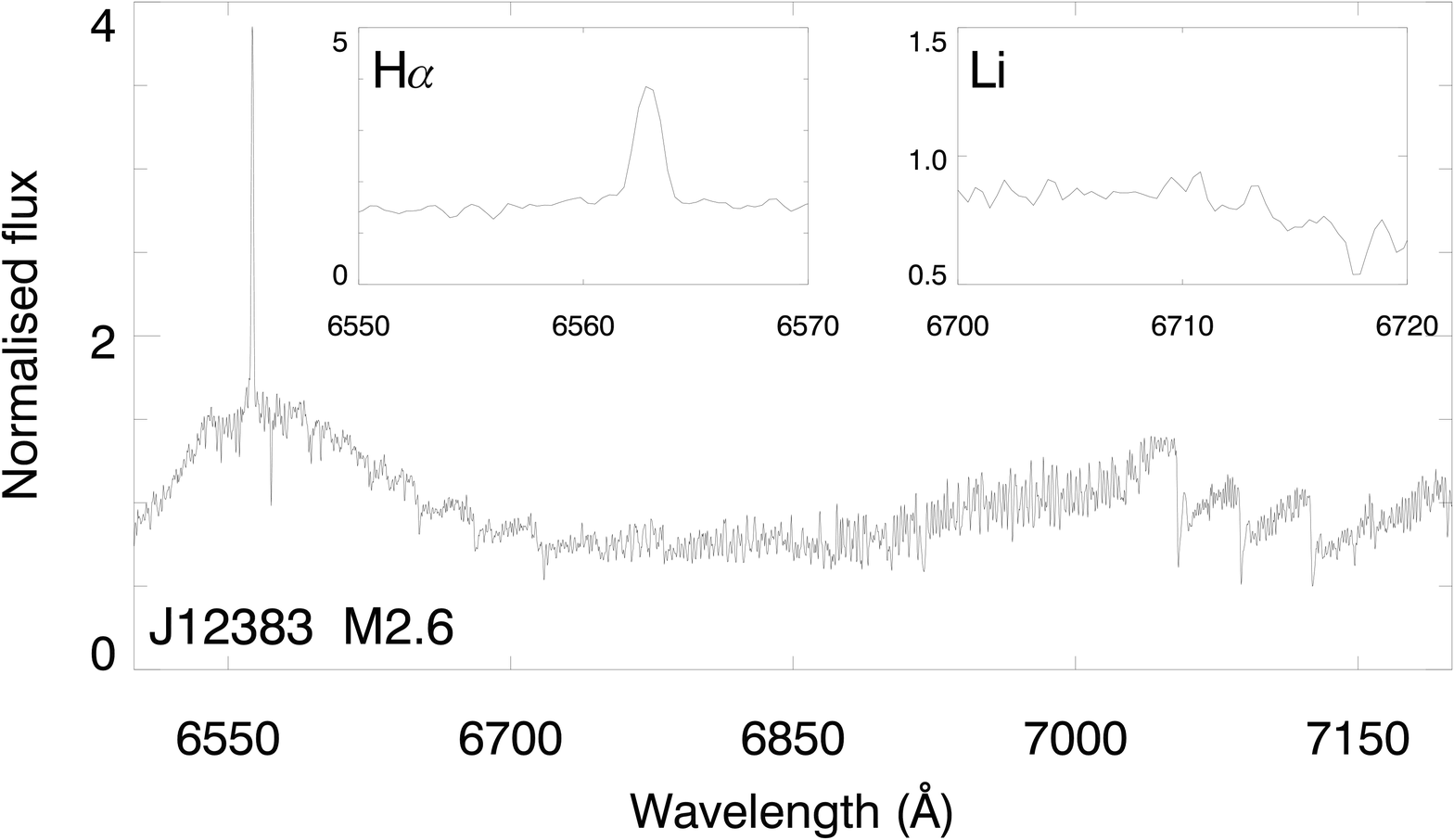} \\
\includegraphics[width=0.42\textwidth]{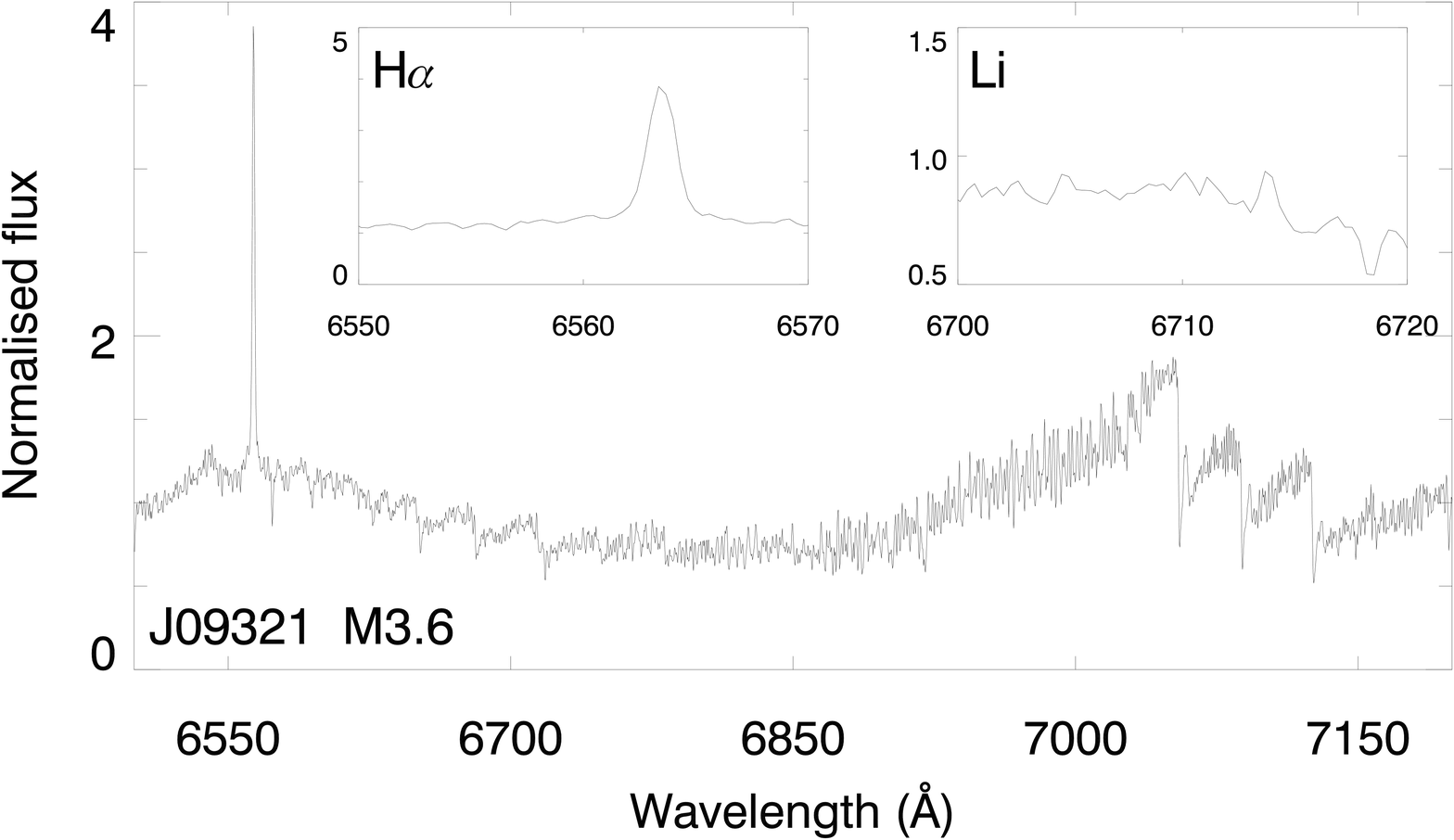} & \includegraphics[width=0.42\textwidth]{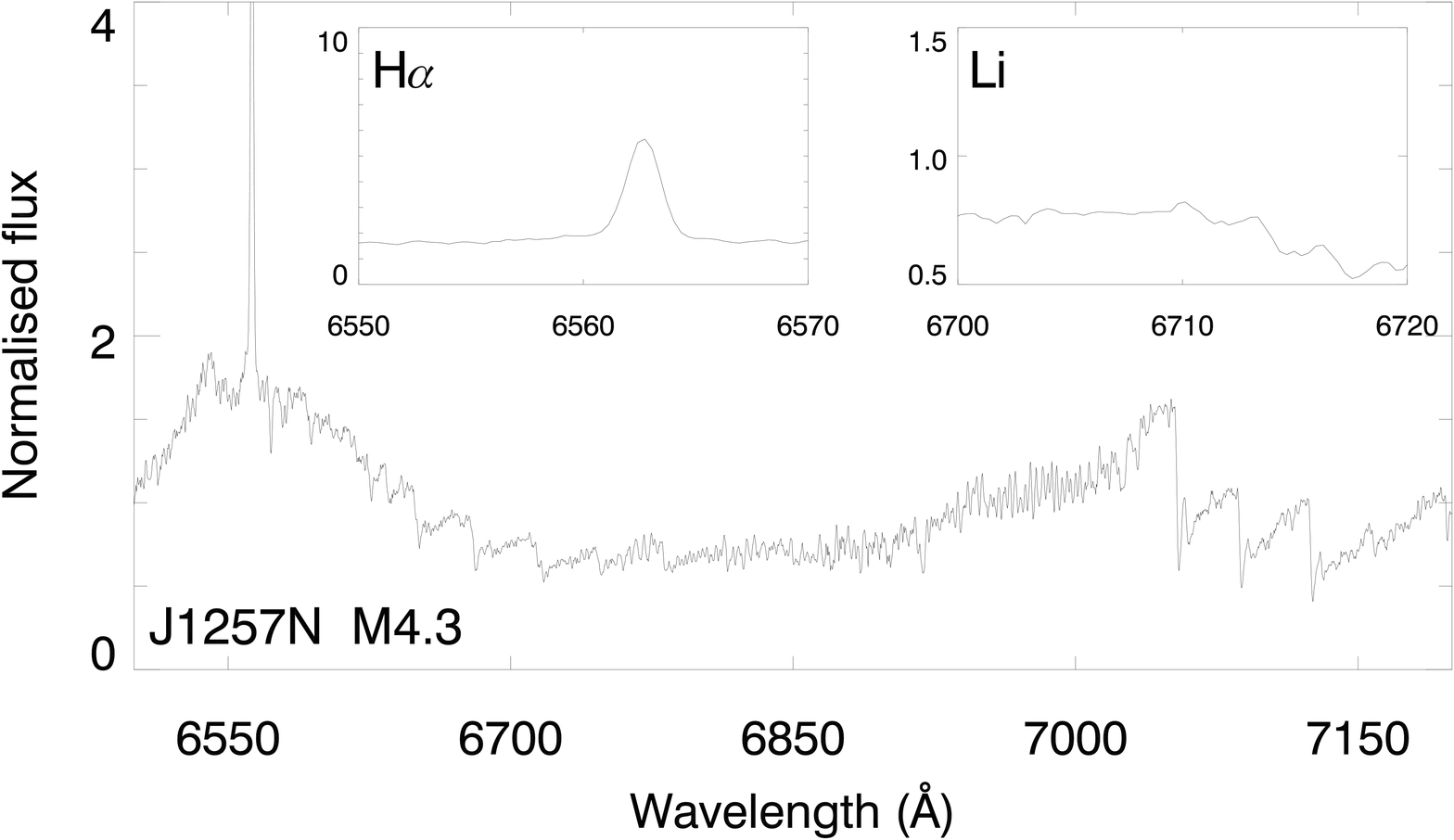} \\
      \end{tabular}
      \end{center}
\caption{The six M-dwarf RV-confirmed ABDMG members which also qualify from the requirements of H$\alpha$ in emission and suitable placement on the CMD. Full 2MASS names are given in Table~\ref{T_ABDMG}. The inserts in each plot are normalised spectra in the regions of the H$\alpha$ and Li~{\sc i} 6708\AA\ line. All spectra have been subject to relative flux-calibration and telluric correction.}
\label{F_ABDMG_Good}
    \end{minipage}
\end{figure*}

\begin{figure*}
    \begin{minipage}[b]{\textwidth}
    \begin{center}
      \begin{tabular}{cc}
        \includegraphics[width=0.42\textwidth]{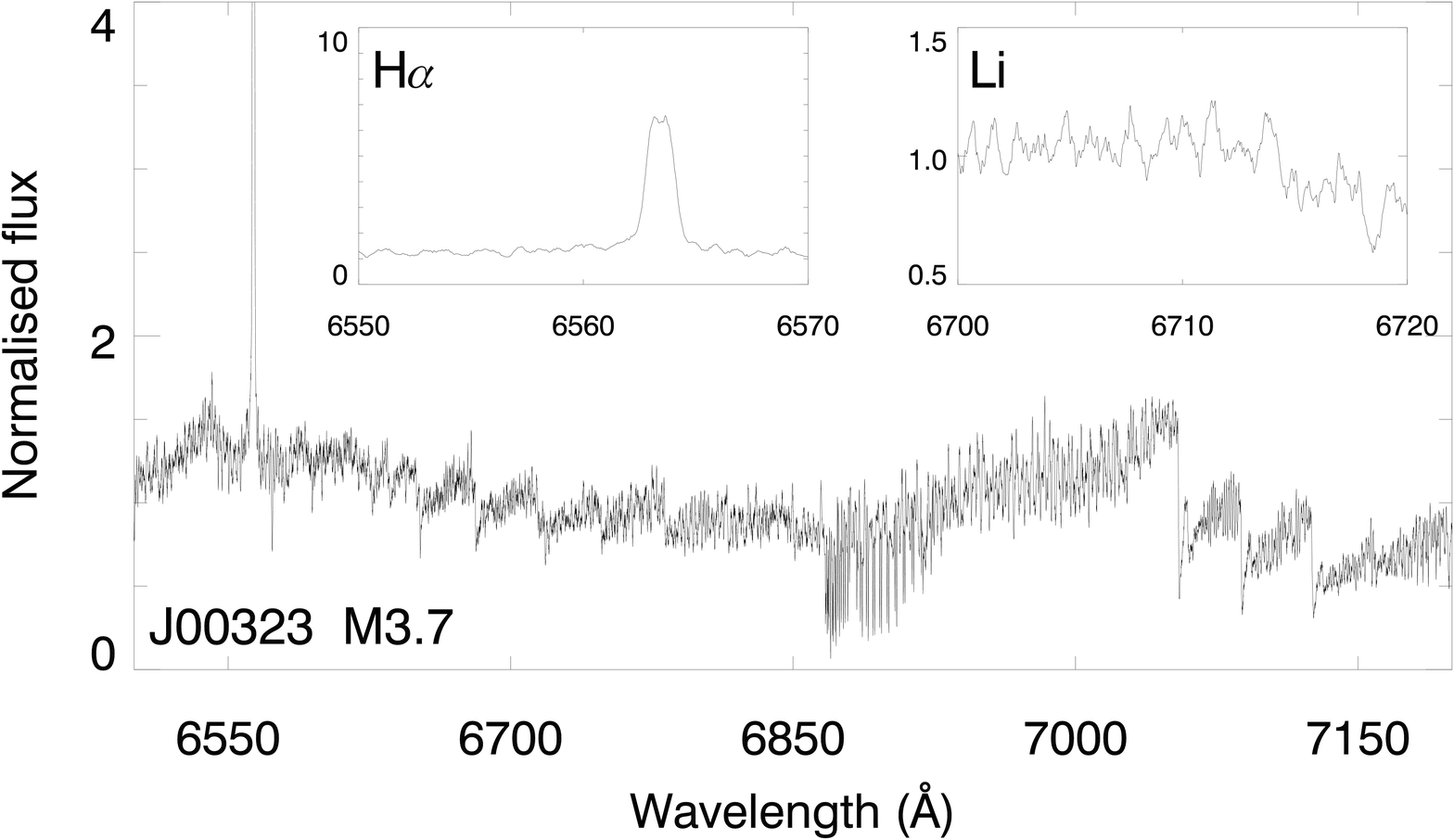} & \includegraphics[width=0.42\textwidth]{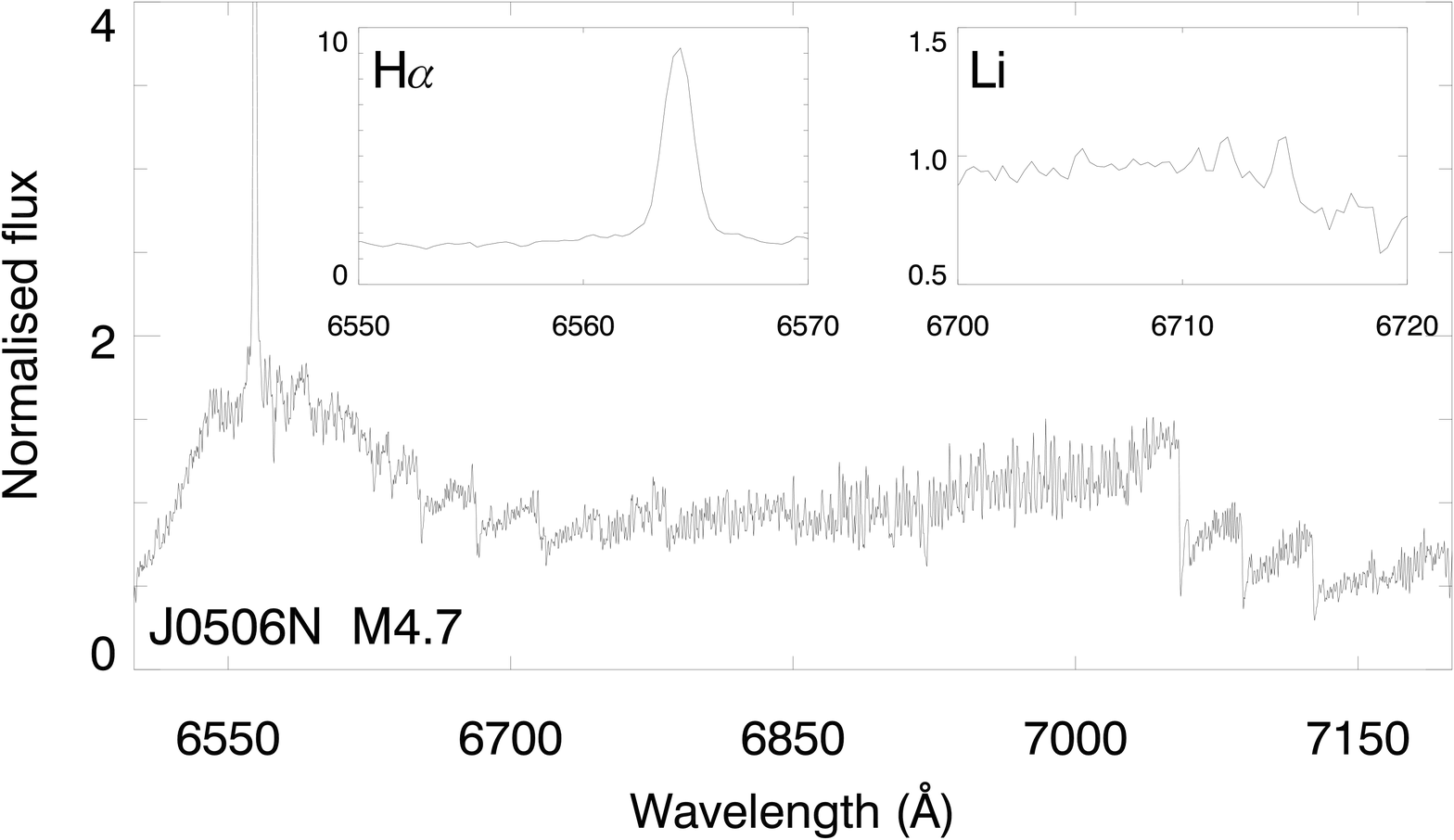} \\
        \includegraphics[width=0.42\textwidth]{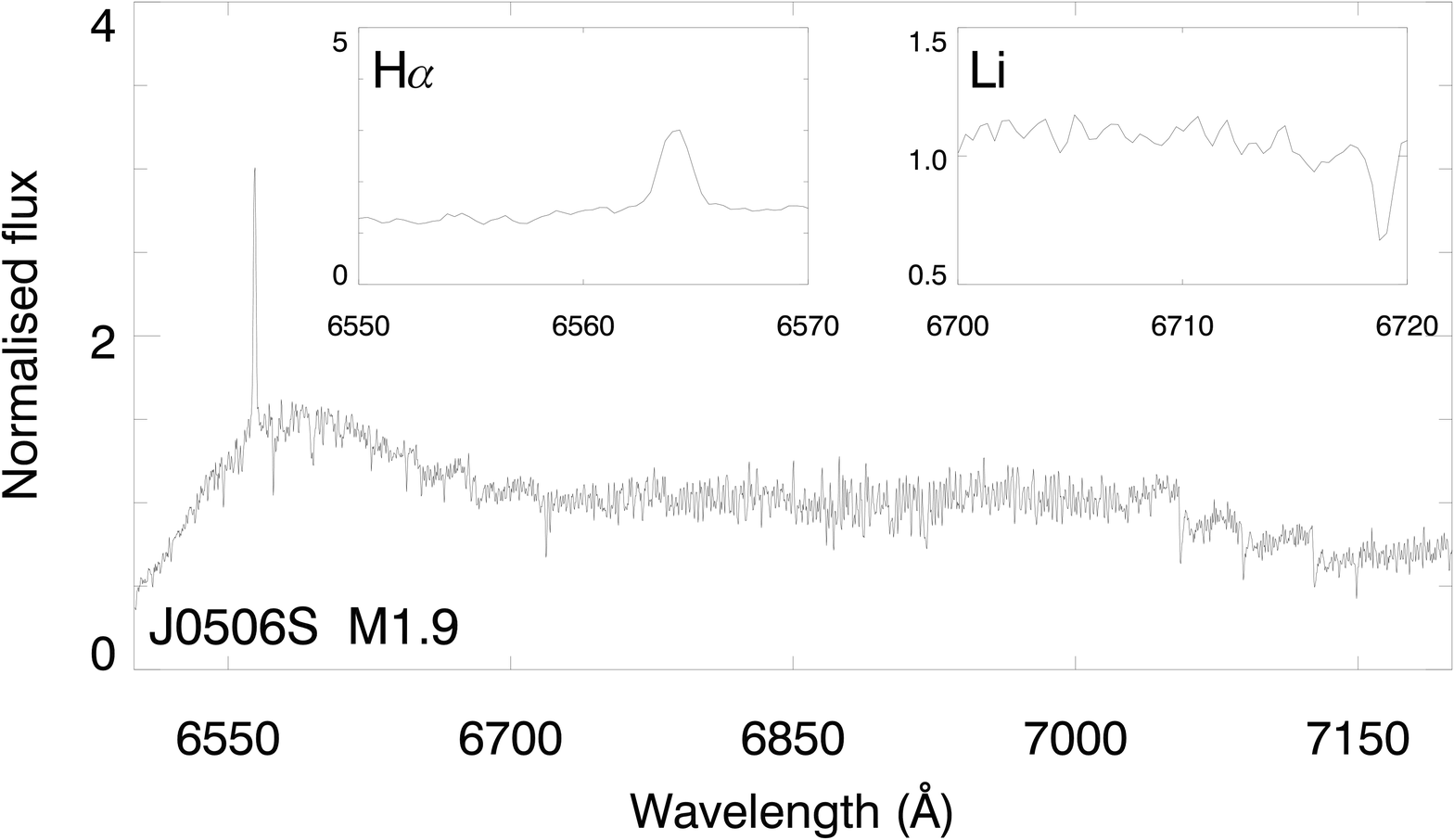} & \includegraphics[width=0.42\textwidth]{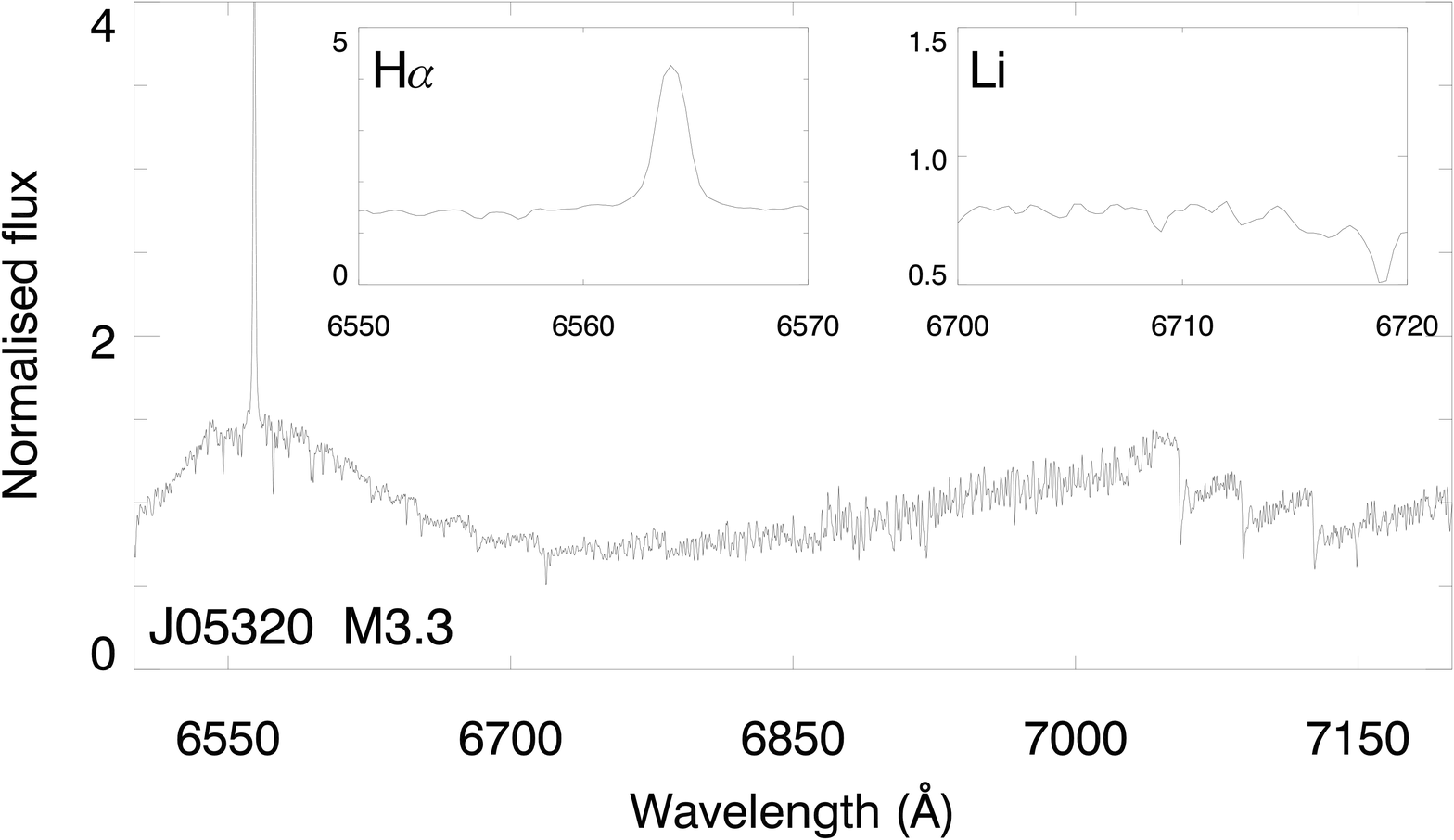} \\
	\includegraphics[width=0.42\textwidth]{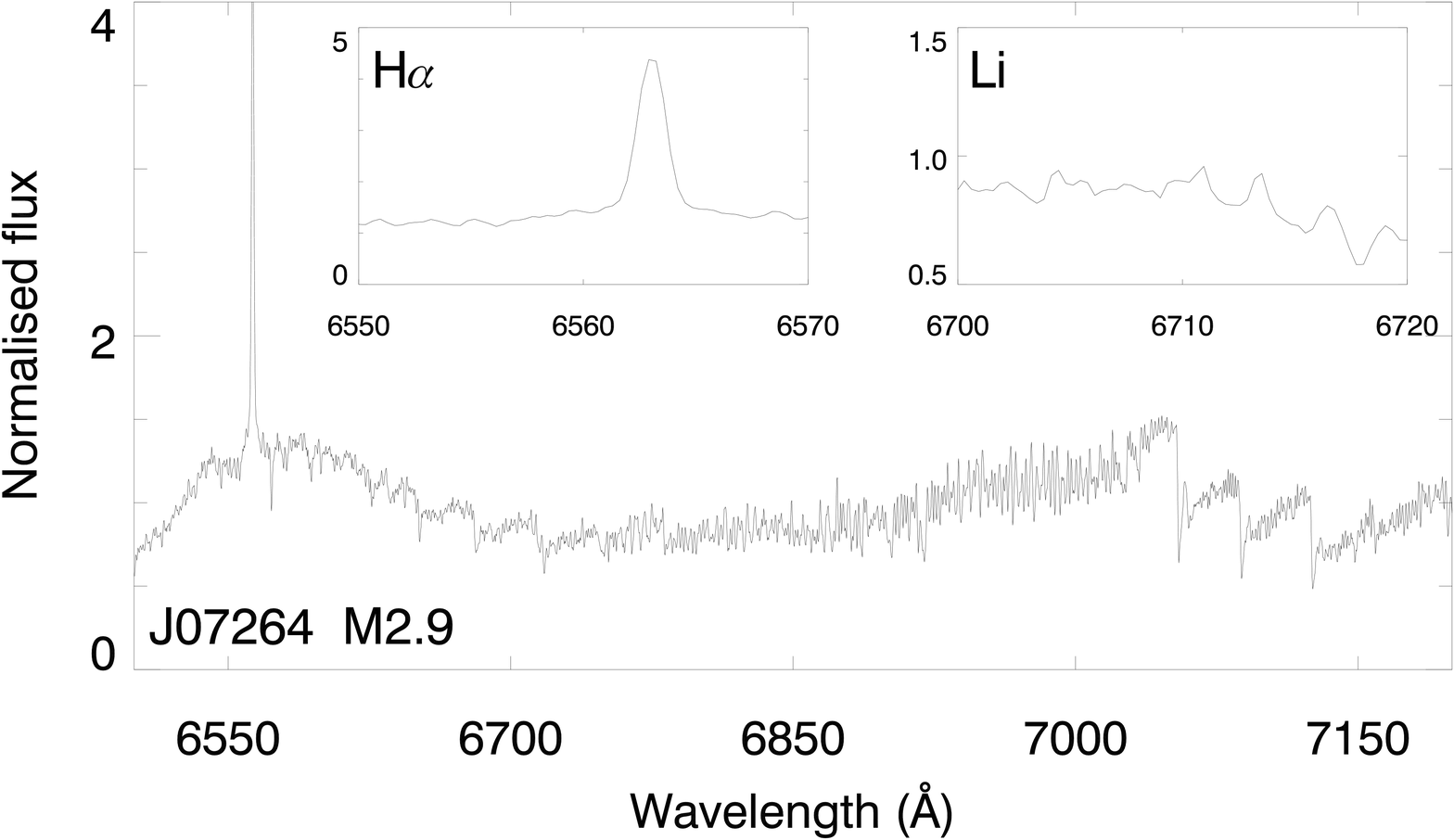} & \includegraphics[width=0.42\textwidth]{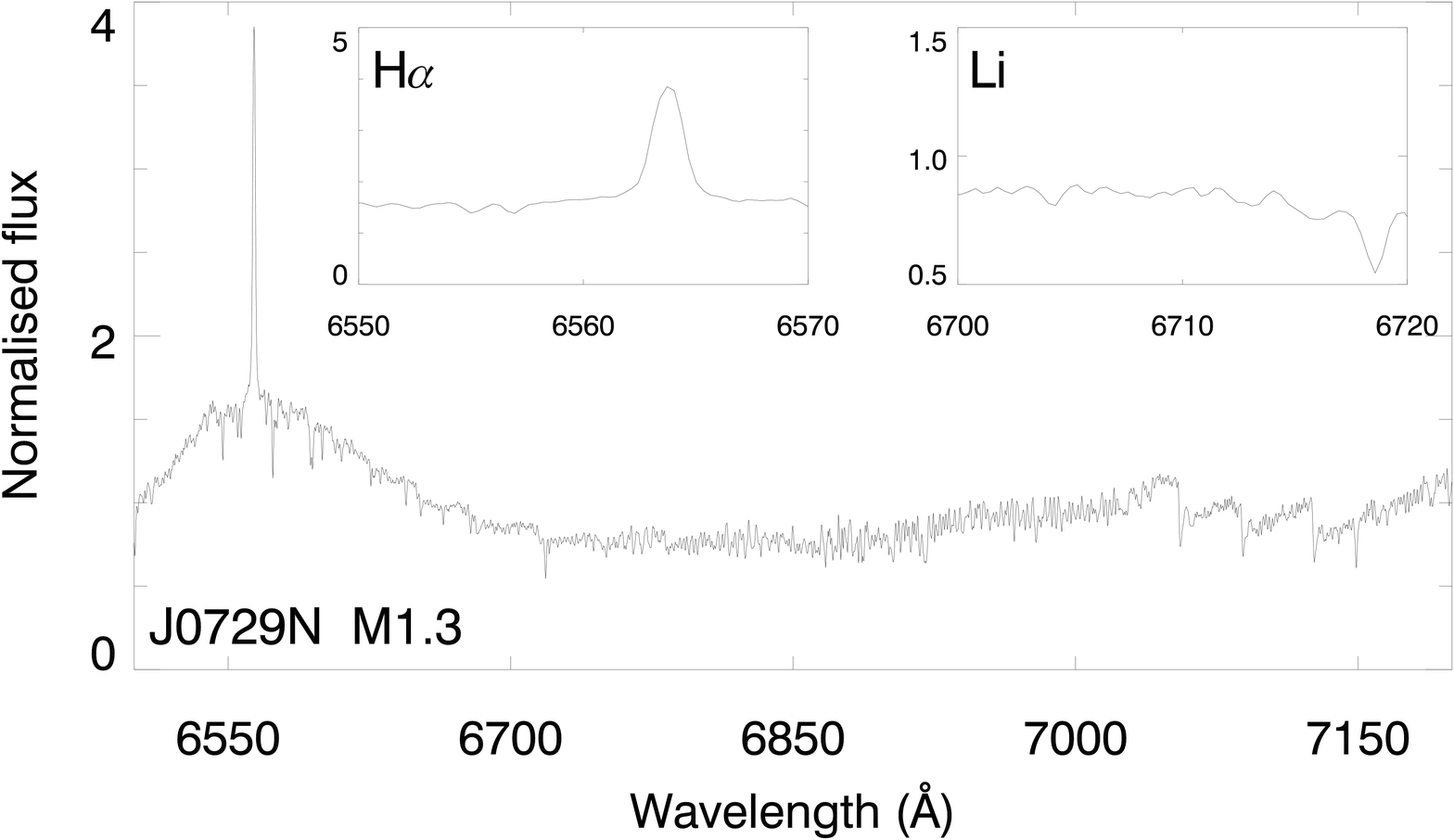} \\
        \includegraphics[width=0.42\textwidth]{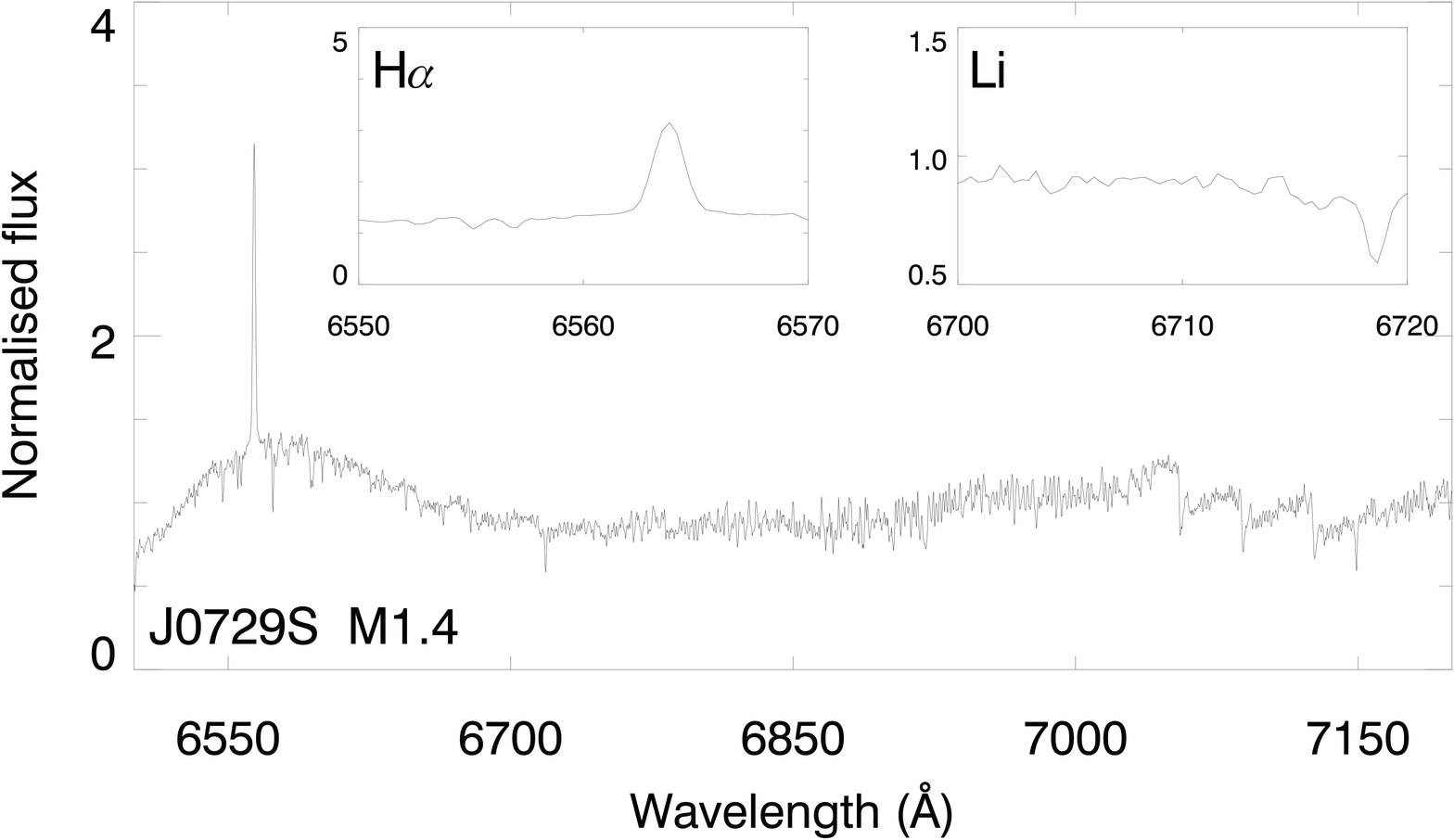} & \includegraphics[width=0.42\textwidth]{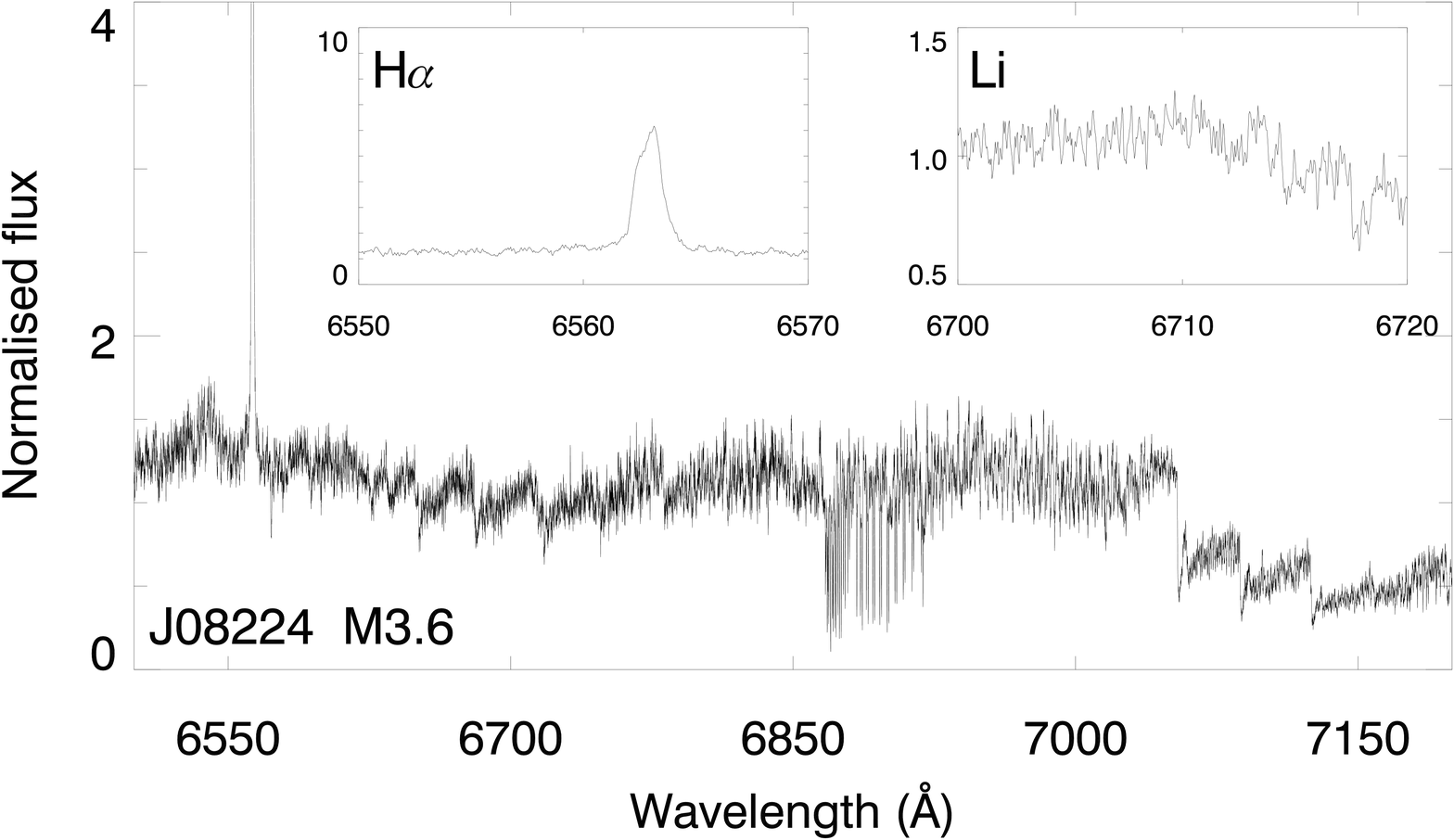} \\
	\includegraphics[width=0.42\textwidth]{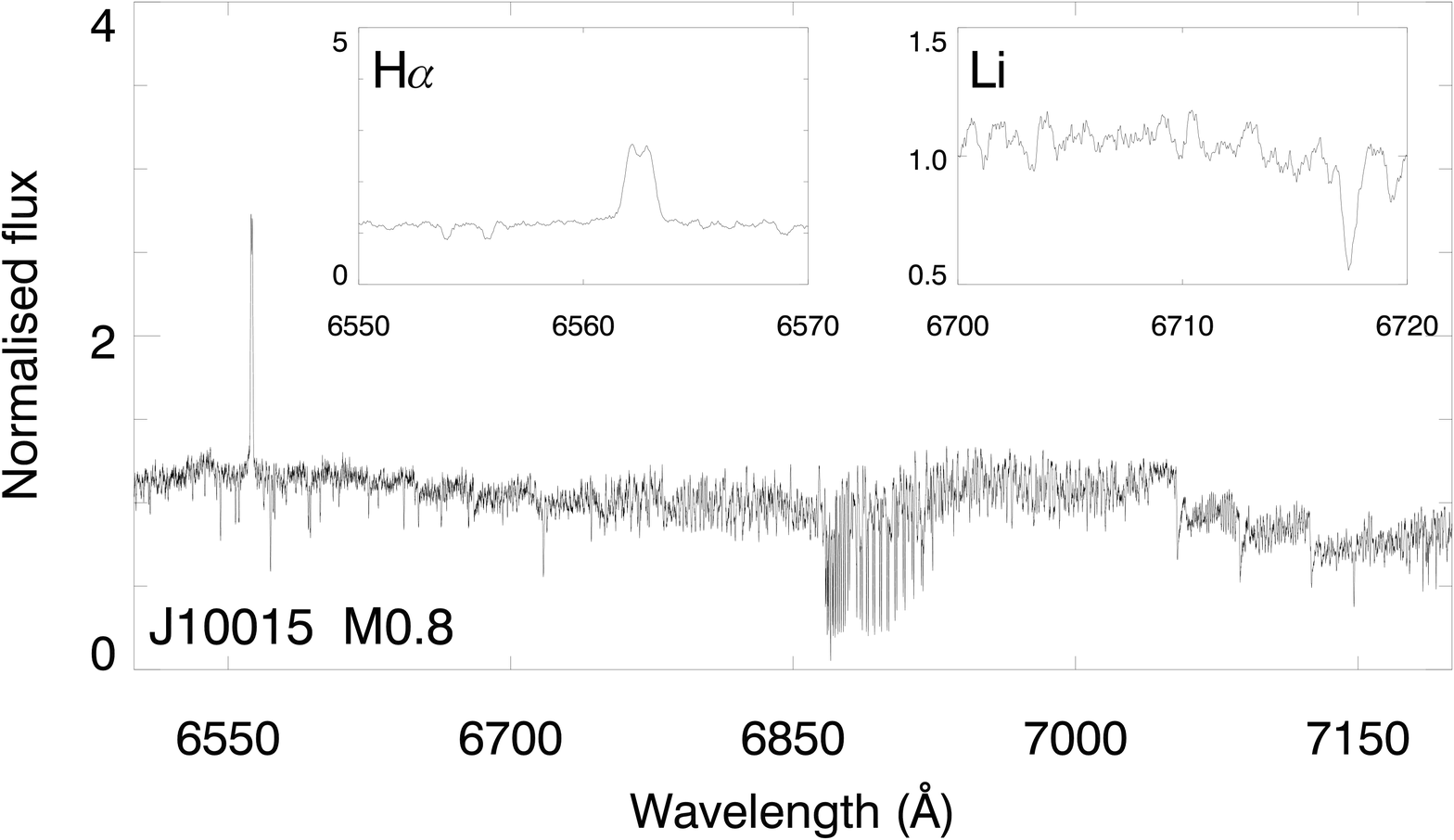} & \includegraphics[width=0.42\textwidth]{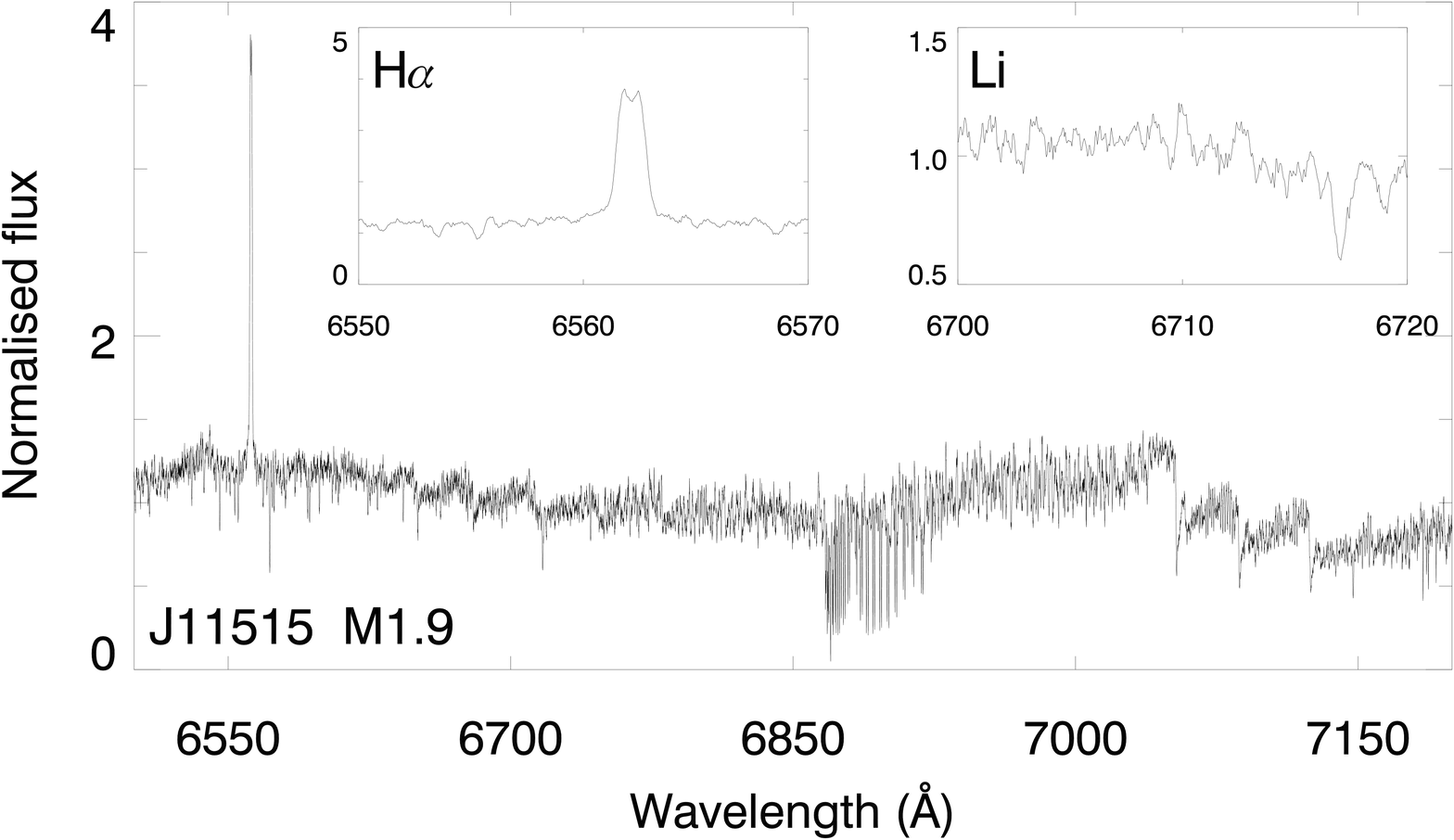} \\
      \end{tabular}
      \end{center}
      	      \caption{BPMG candidates that fail our membership criteria. All objects fail membership on the grounds of RV, except for J0501, which satisfies RV criterion, but has H$\alpha$ in absorption. All spectra (excluding `J0032', `J0501', `J0822', `J1001', `J1151', `J1211' and `J2351', observed at the NOT, which have been blaze-corrected) have been subject to relative flux-calibration and telluric correction.}
      \label{F_BPMG_Fail}
    \end{minipage}
\end{figure*}

Spectra for the confirmed BPMG and ABDMG M-dwarfs are presented in Figures~\ref{F_BPMG_Good} and~\ref{F_ABDMG_Good} respectively; the spectra of candidates that fail membership tests are presented in Figures~\ref{F_BPMG_Fail} and~\ref{F_ABDMG_Fail} (see $\S$\ref{S_Confirming} for a description of membership criteria).

\newpage

\begin{figure*}
    \begin{minipage}[b]{\textwidth}
    \begin{center}
      \begin{tabular}{cc}
        \includegraphics[width=0.42\textwidth]{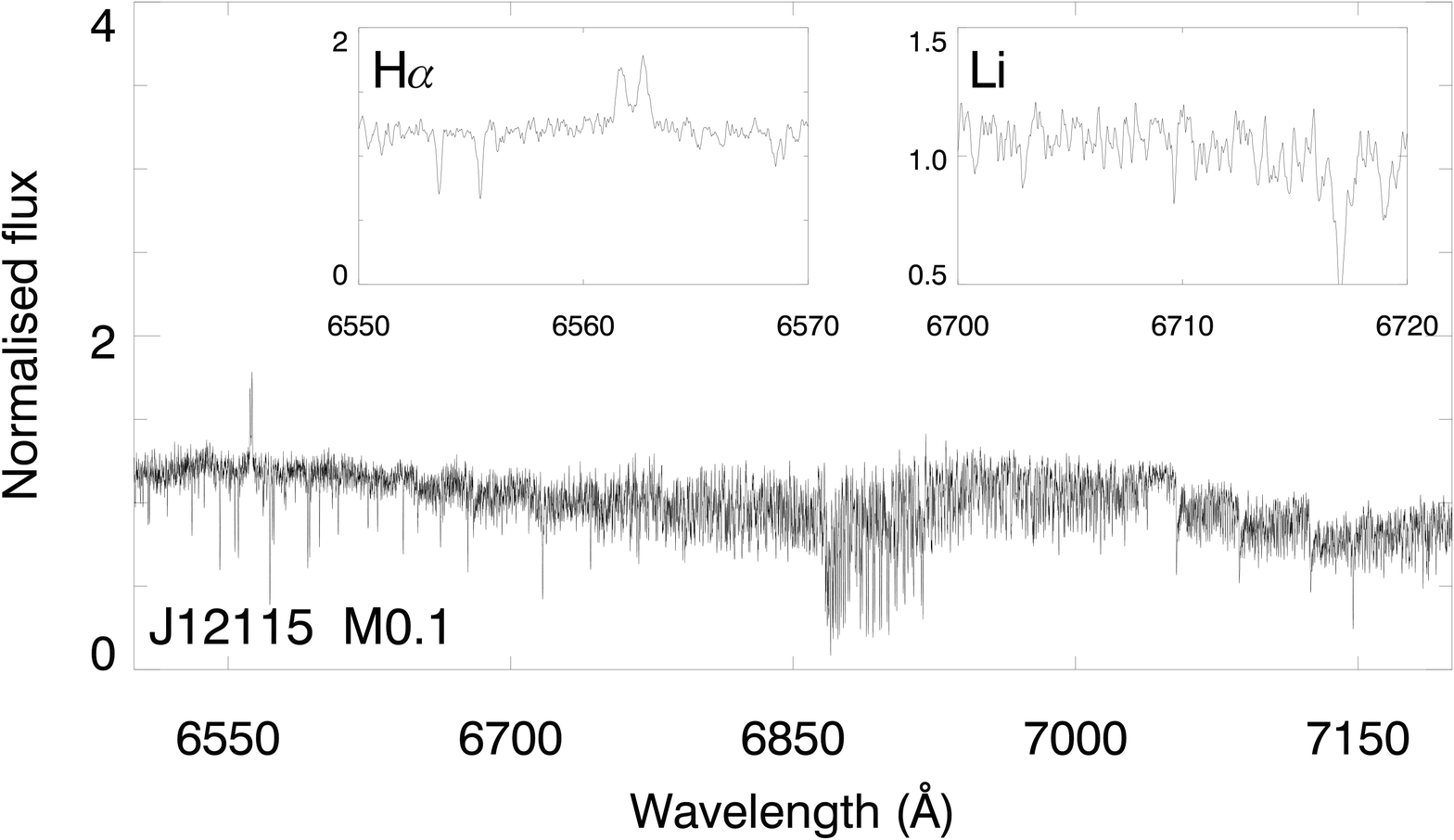} & \includegraphics[width=0.42\textwidth]{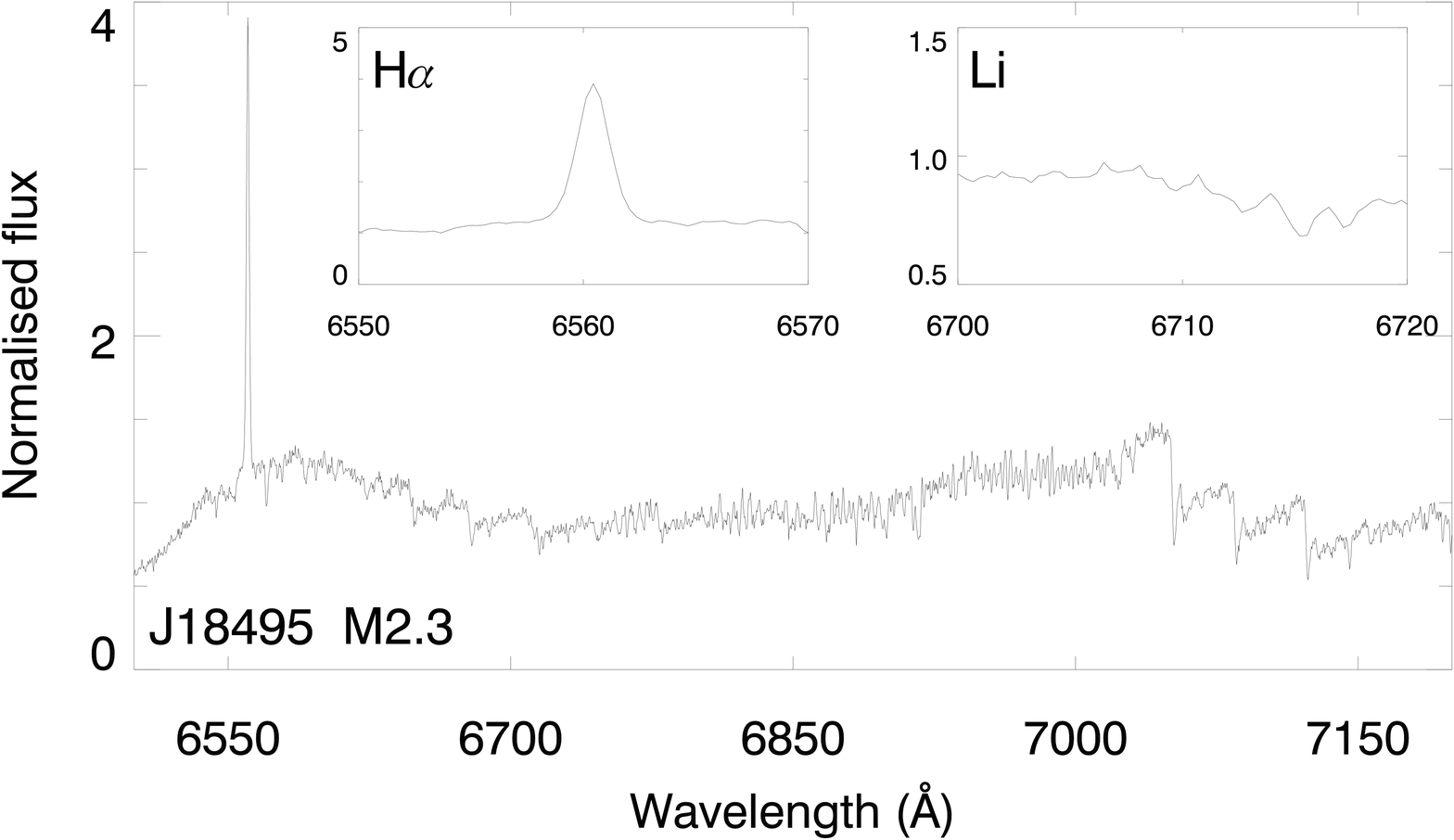} \\
        \includegraphics[width=0.42\textwidth]{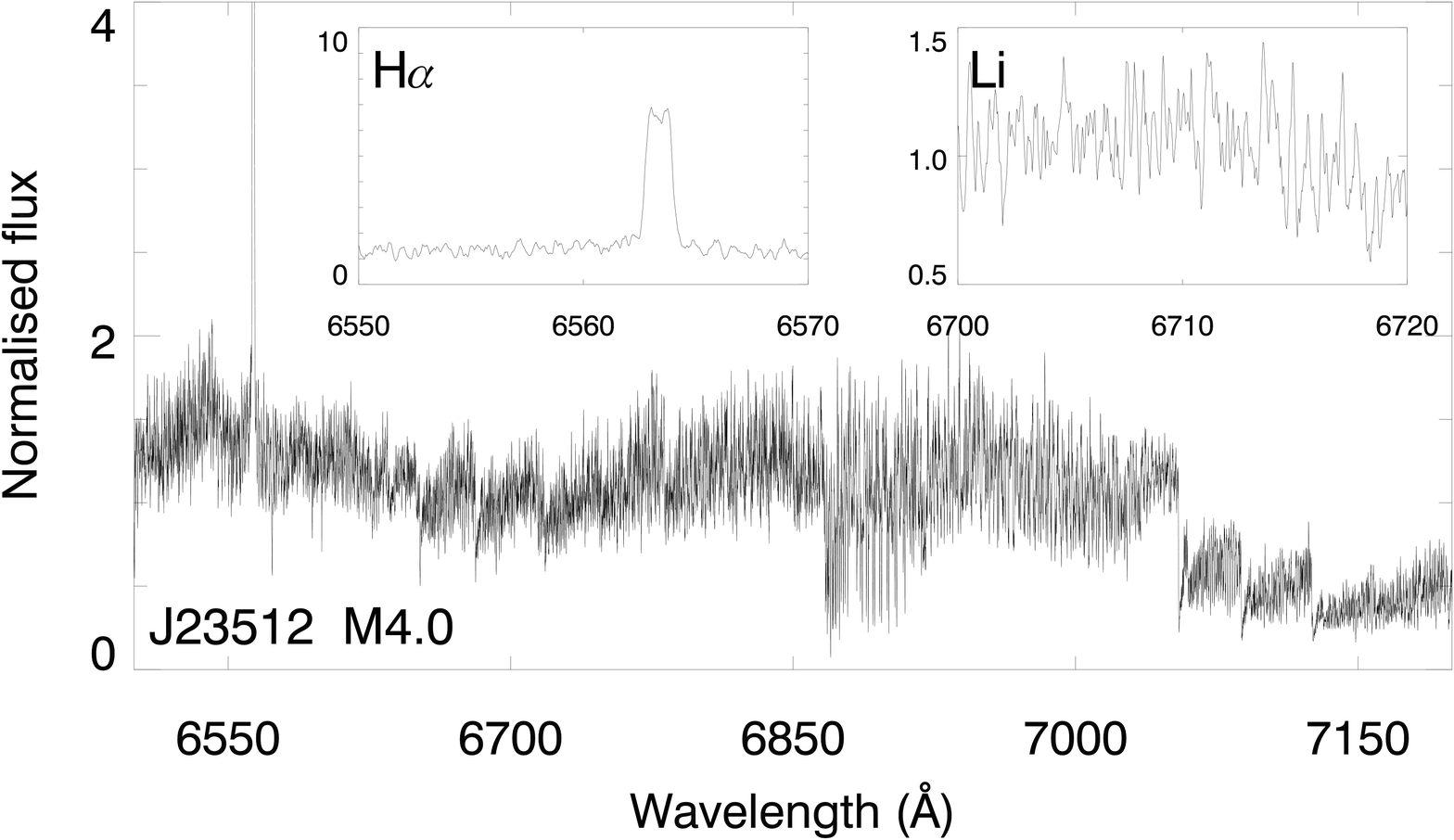} & \includegraphics[width=0.42\textwidth]{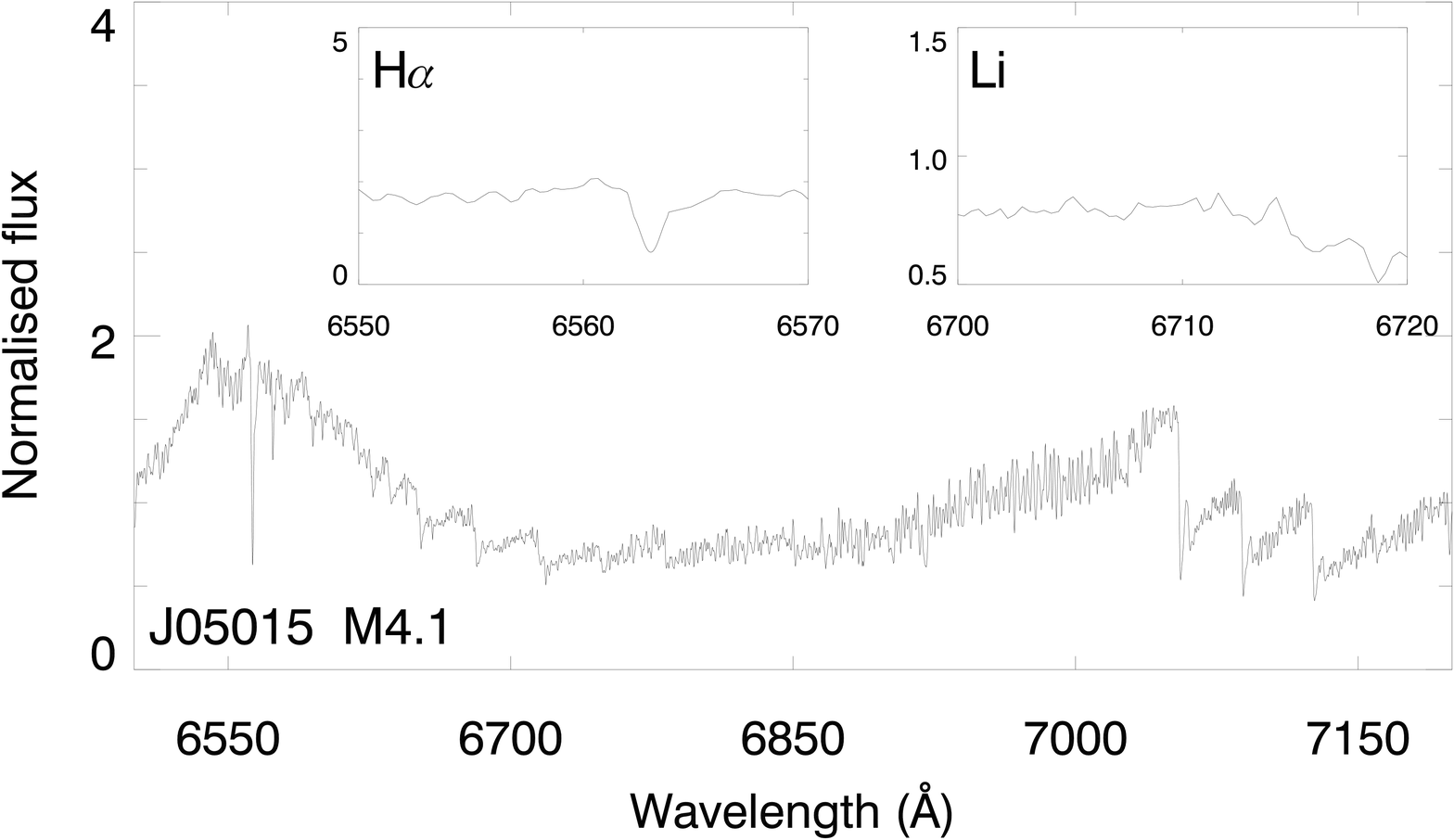} \\
                        \multicolumn{2}{c}{{\bf Figure 3.} \textit{continued.}} \\
      \end{tabular}
      \end{center}
    \end{minipage}
\end{figure*}

\begin{figure*}
    \begin{minipage}[b]{\textwidth}
    \begin{center}
      \begin{tabular}{cc}
\includegraphics[width=0.42\textwidth]{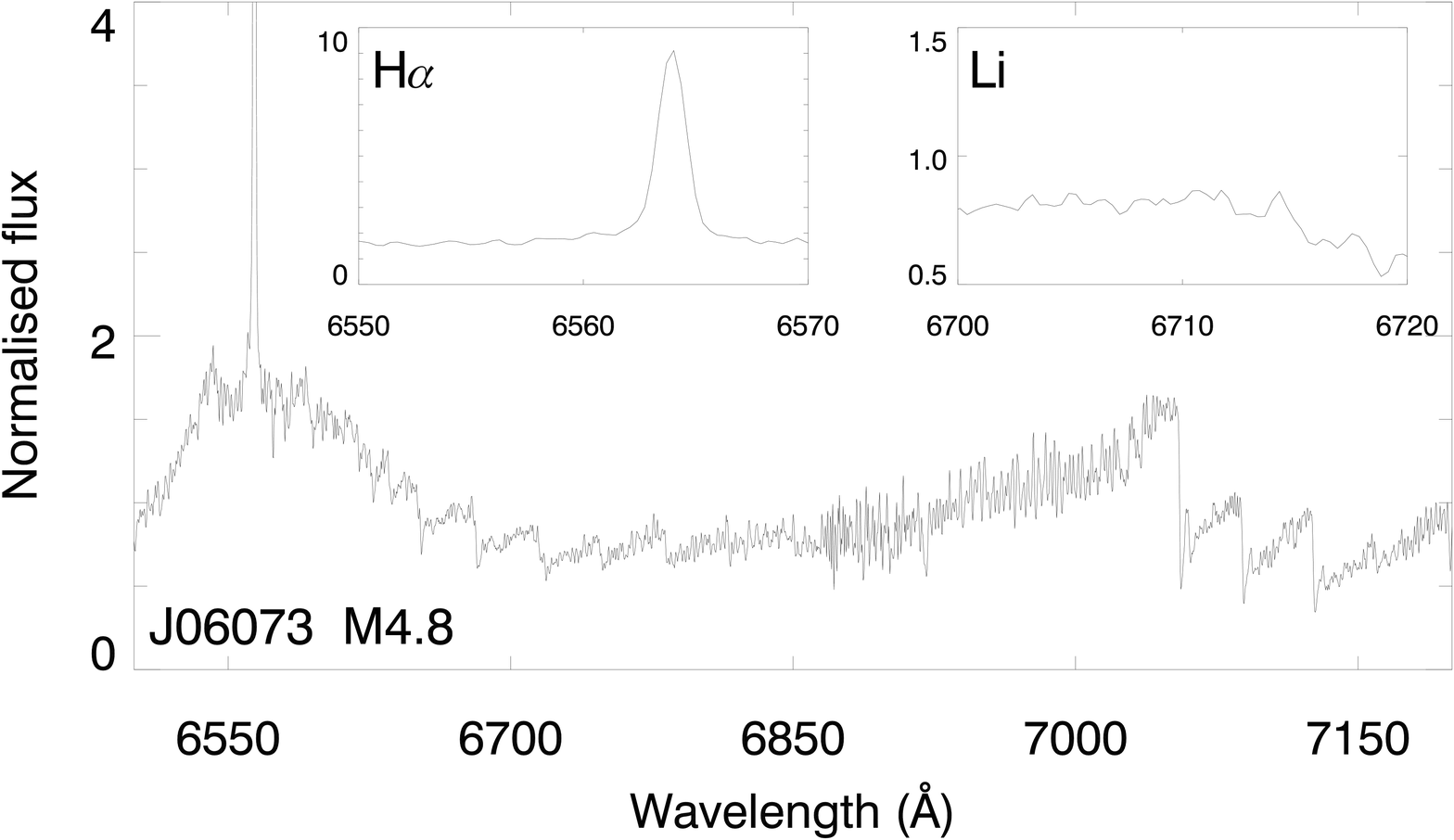} & \includegraphics[width=0.42\textwidth]{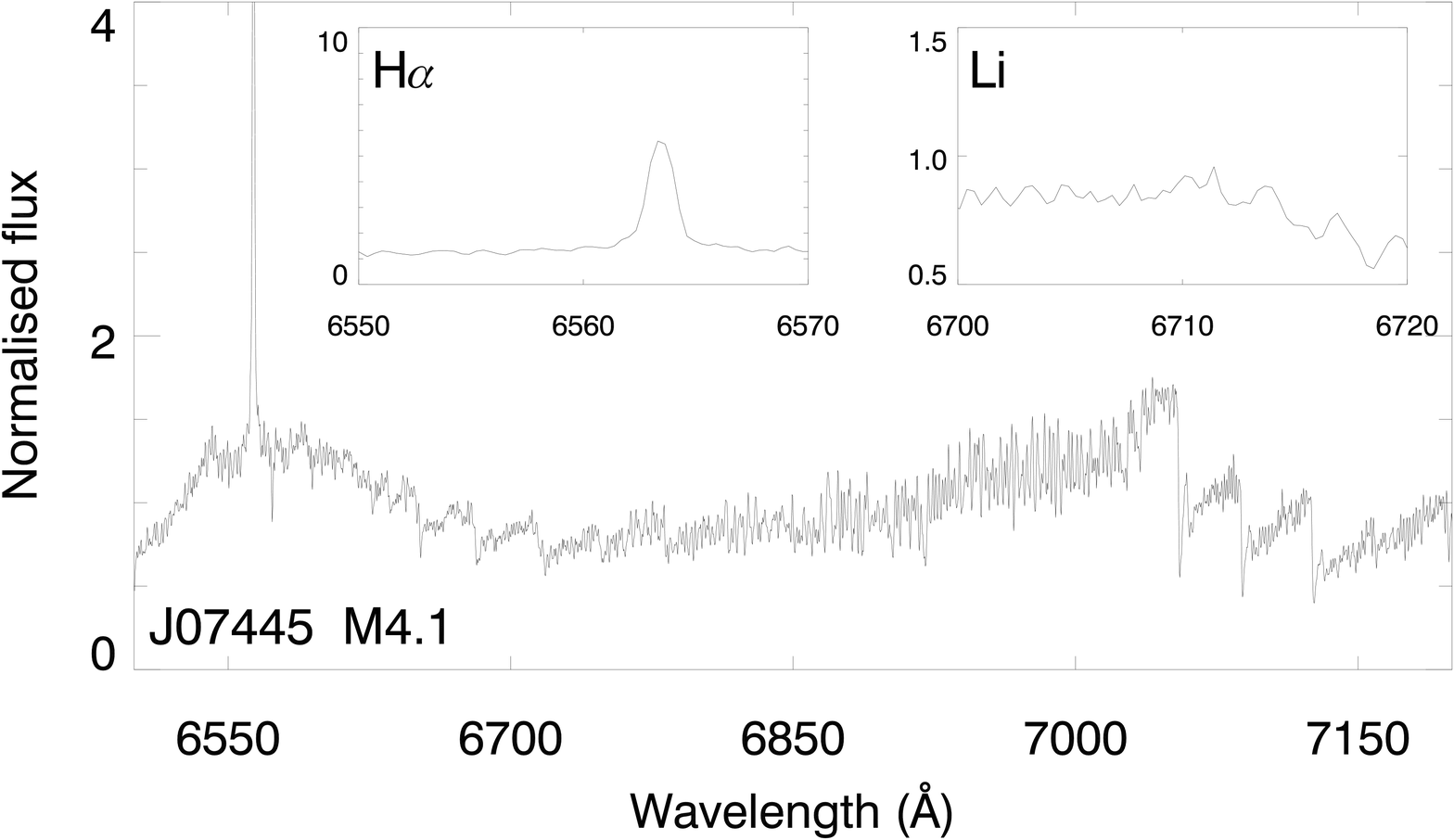} \\
 \includegraphics[width=0.42\textwidth]{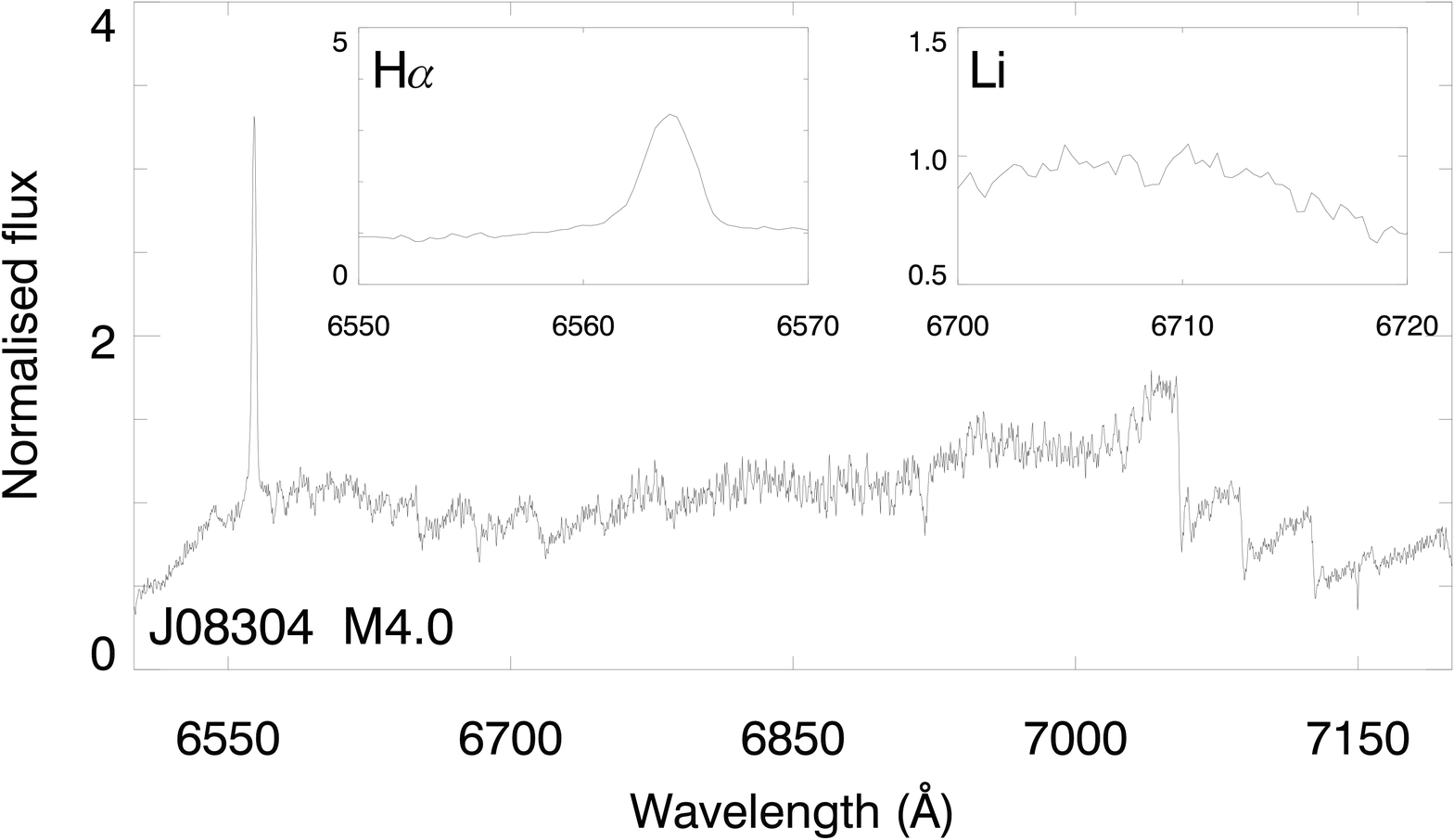} & \includegraphics[width=0.42\textwidth]{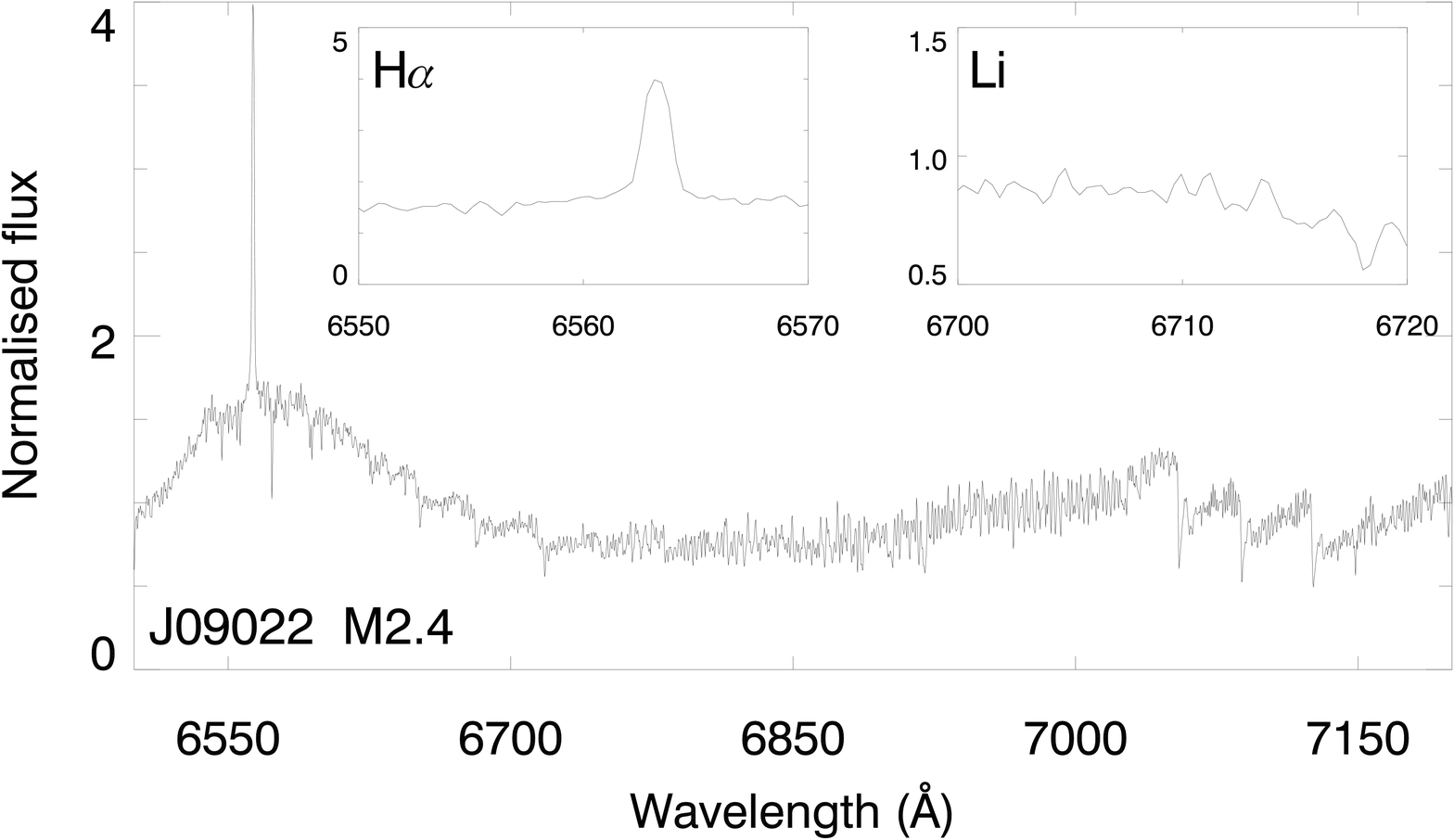} \\
  \includegraphics[width=0.42\textwidth]{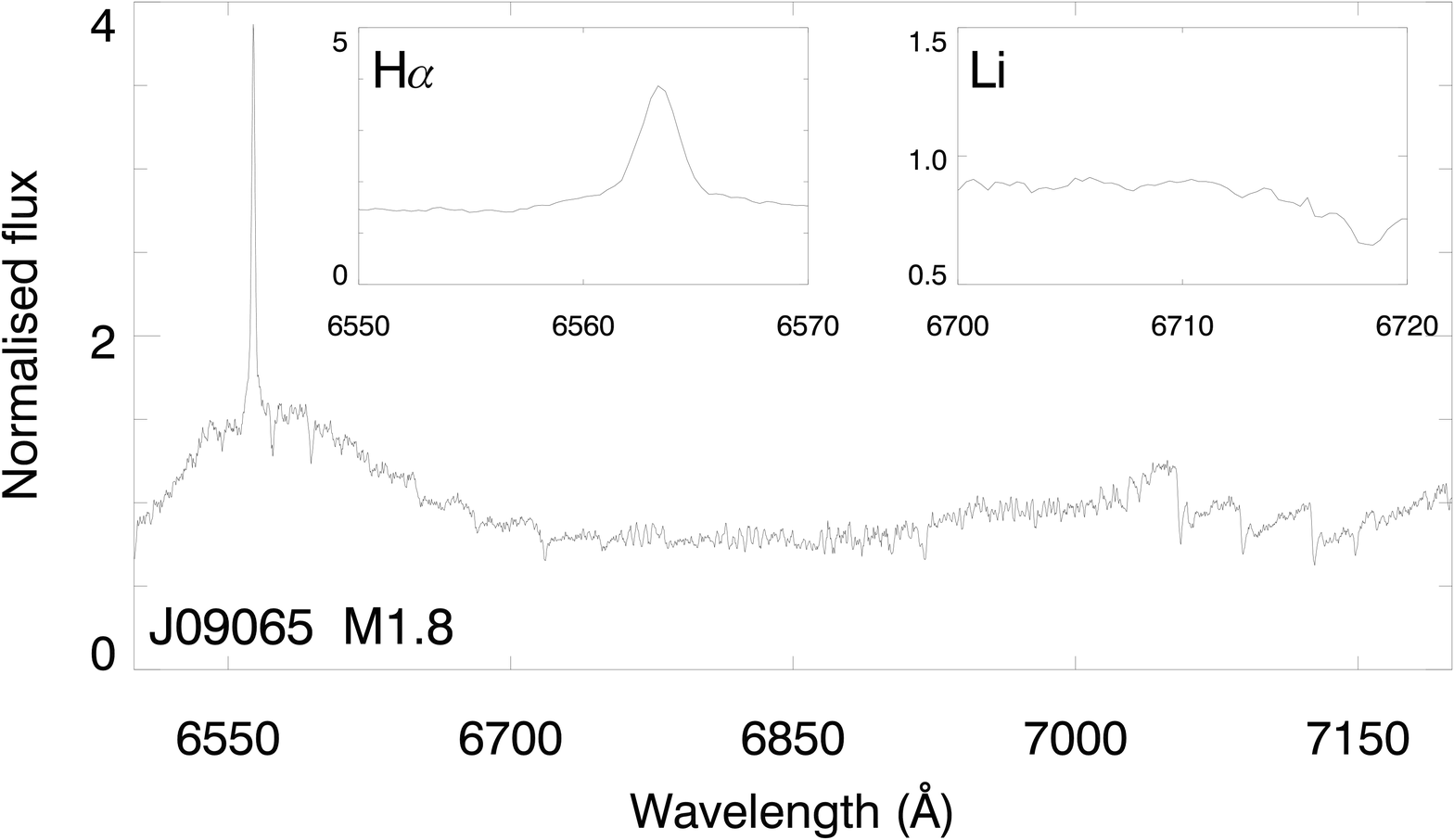} & \includegraphics[width=0.42\textwidth]{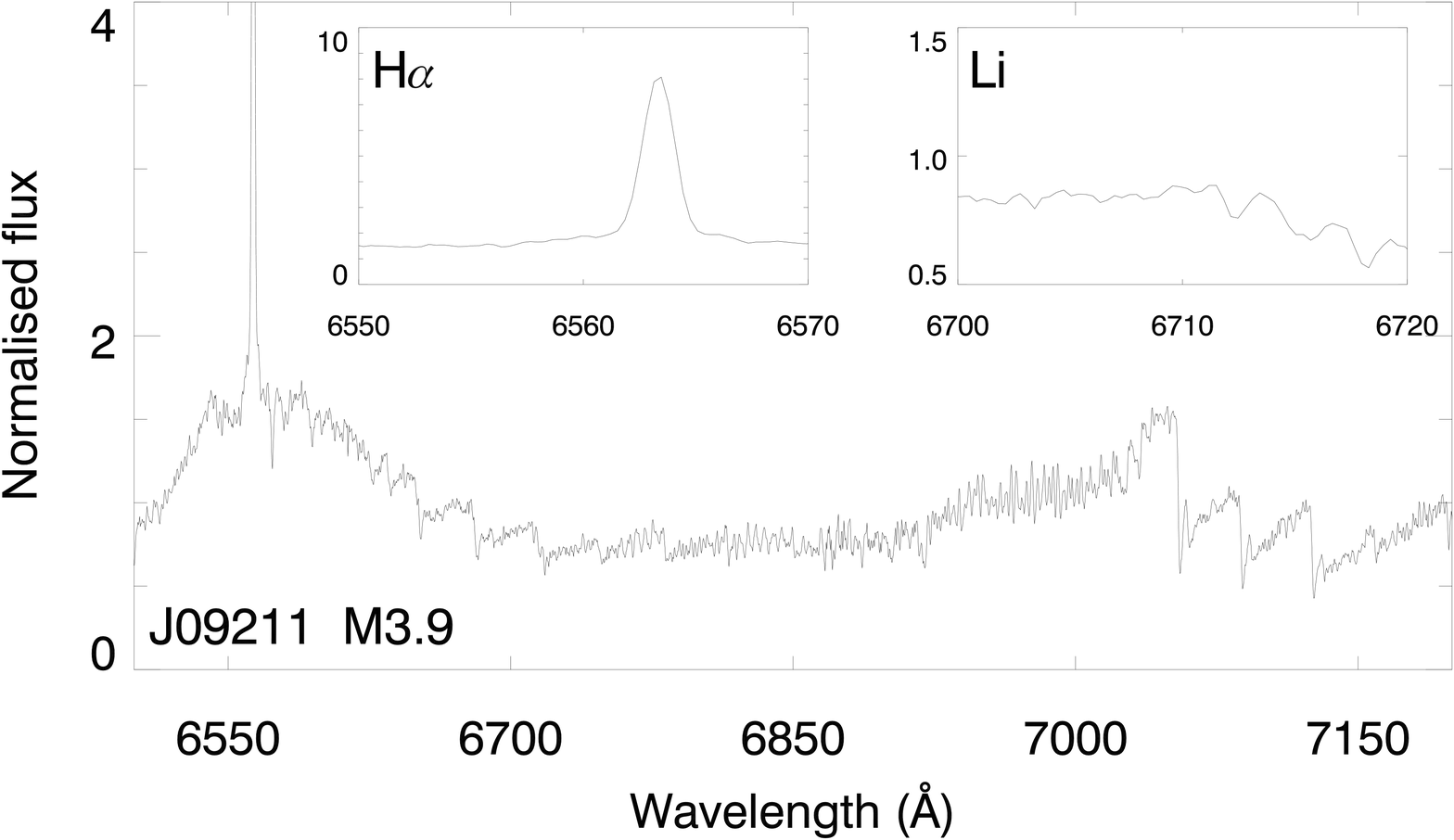} \\
\includegraphics[width=0.42\textwidth]{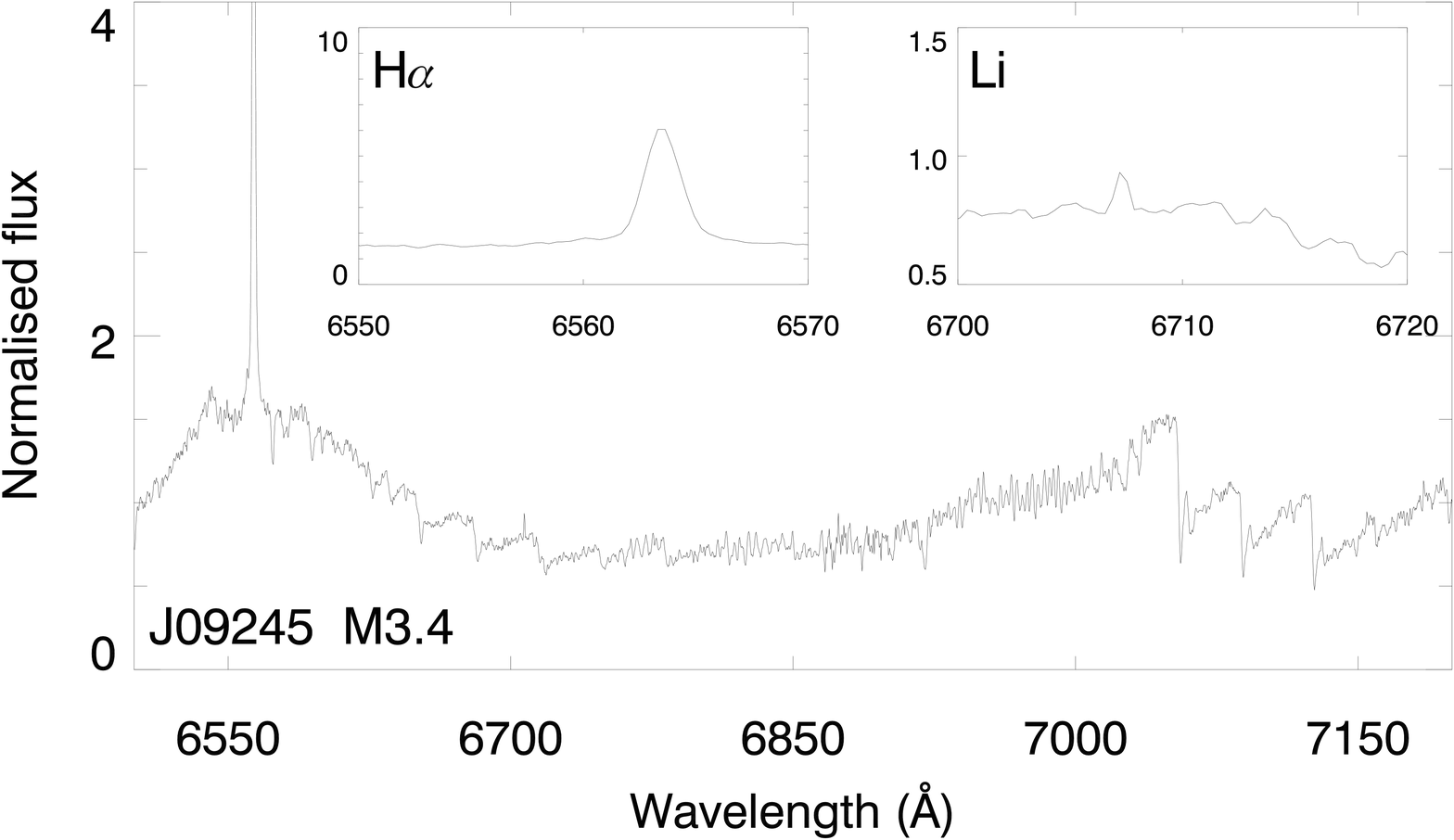} & \includegraphics[width=0.42\textwidth]{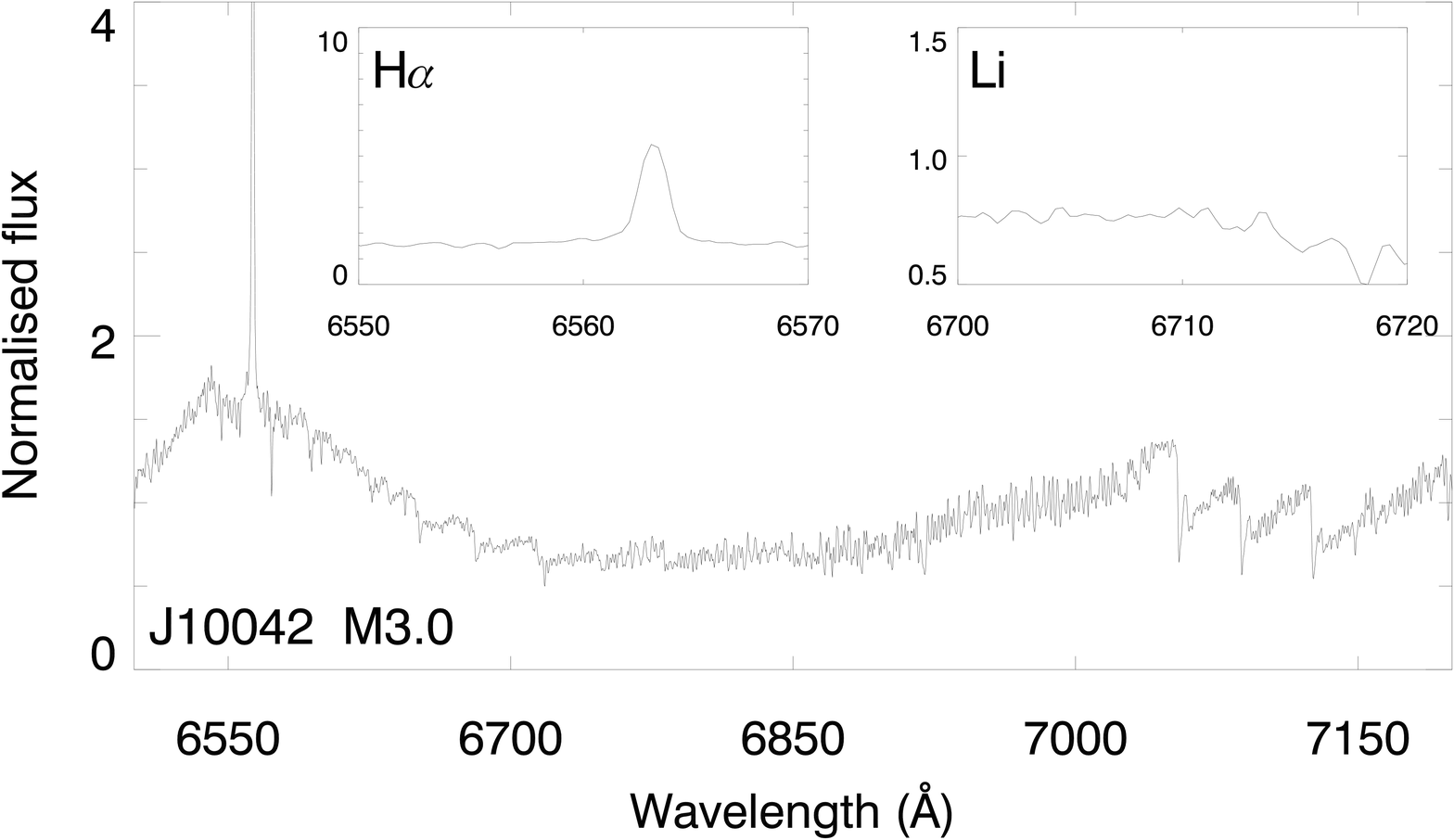} \\
 \includegraphics[width=0.42\textwidth]{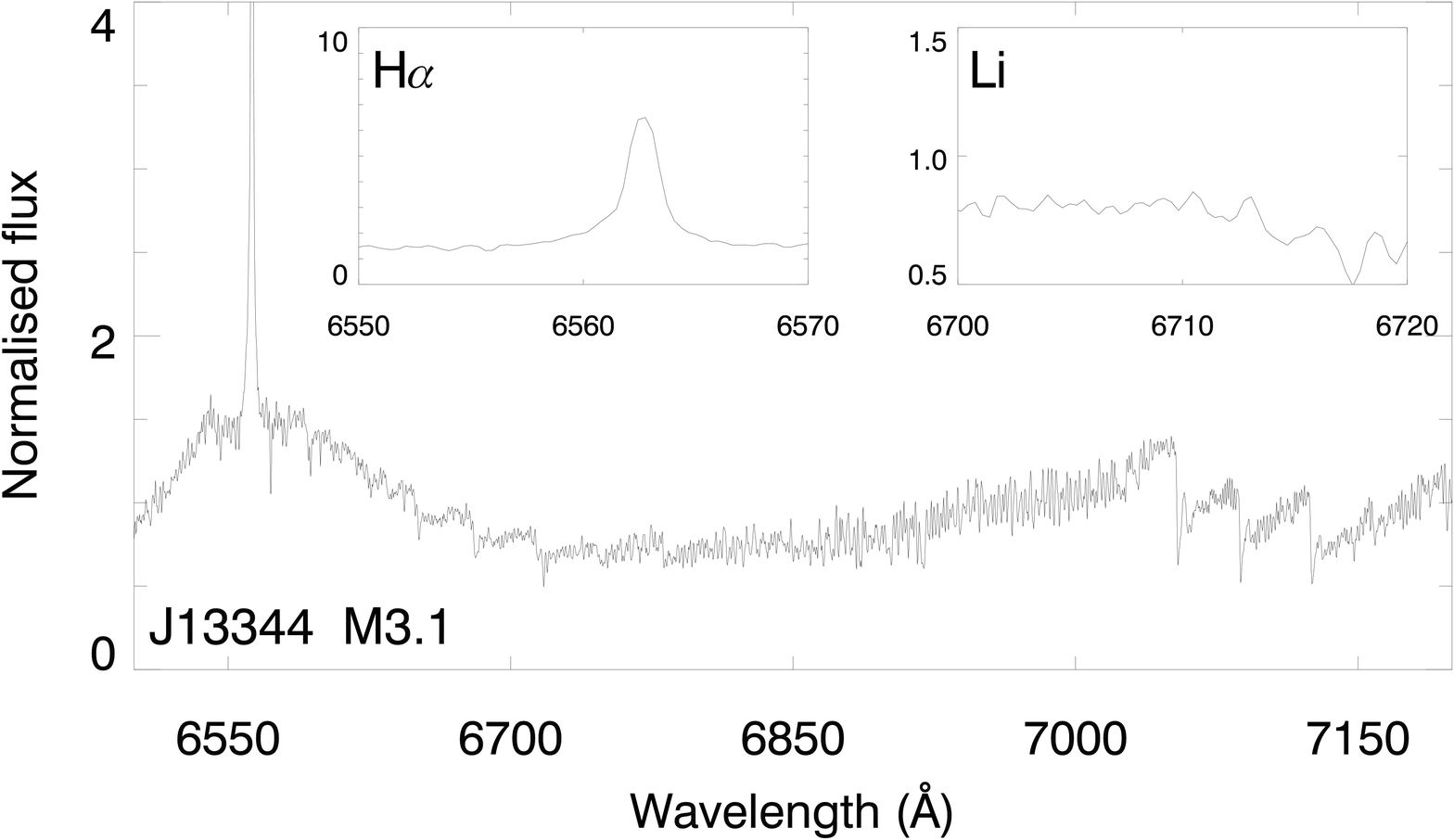} & \includegraphics[width=0.42\textwidth]{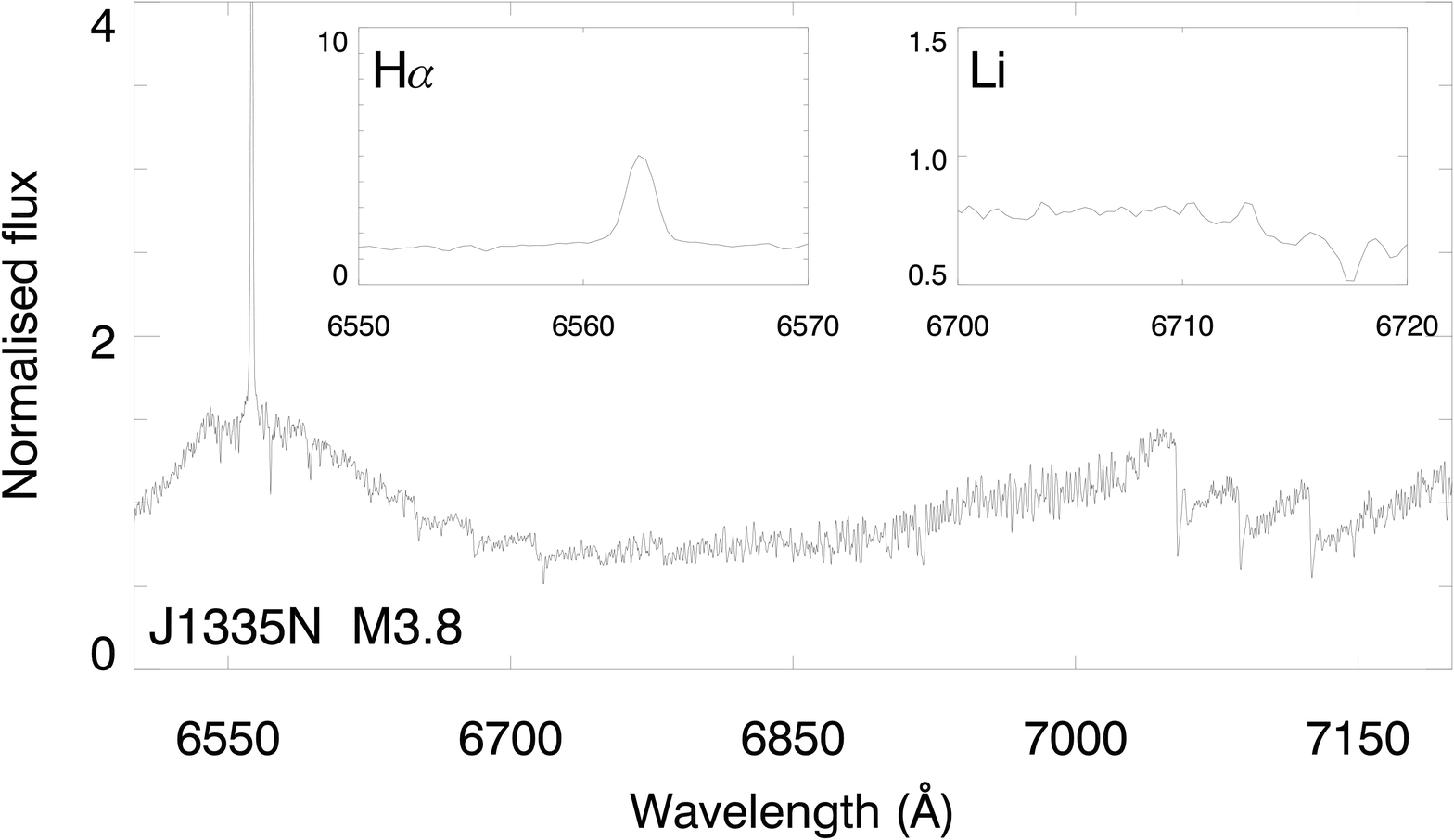} \\
       \end{tabular}
      \end{center}
      \caption{ABDMG candidates that fail membership. All objects fail membership on the grounds of RV, except for J1012, which satisfies RV criteria, but fails membership because it has H$\alpha$ in absorption. All spectra have been subject to relative flux-calibration and telluric correction.}
      \label{F_ABDMG_Fail}
    \end{minipage}
\end{figure*}

\newpage

\begin{figure*}
    \begin{minipage}[b]{\textwidth}
    \begin{center}
      \begin{tabular}{cc}
  \includegraphics[width=0.42\textwidth]{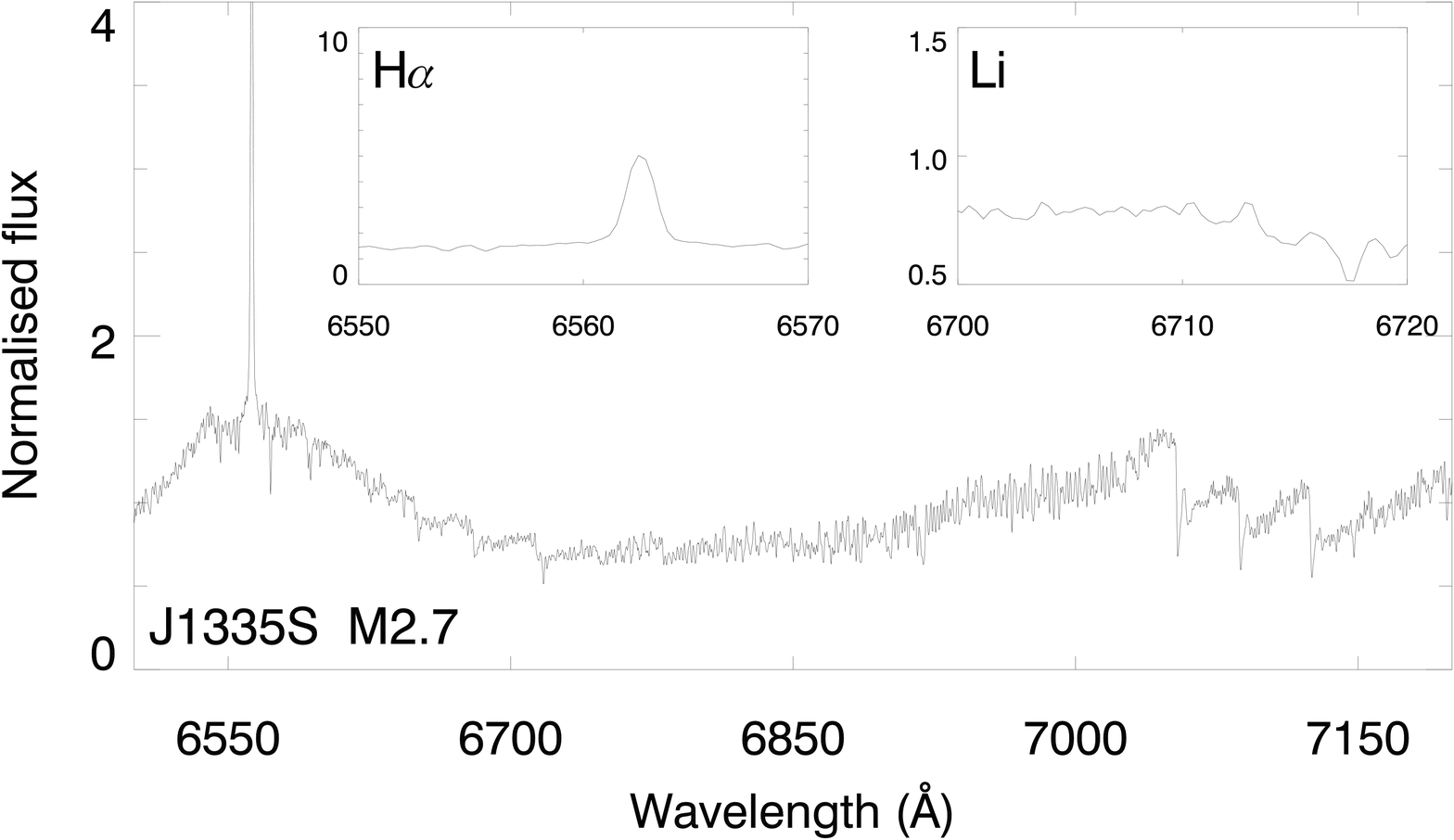} & \includegraphics[width=0.42\textwidth]{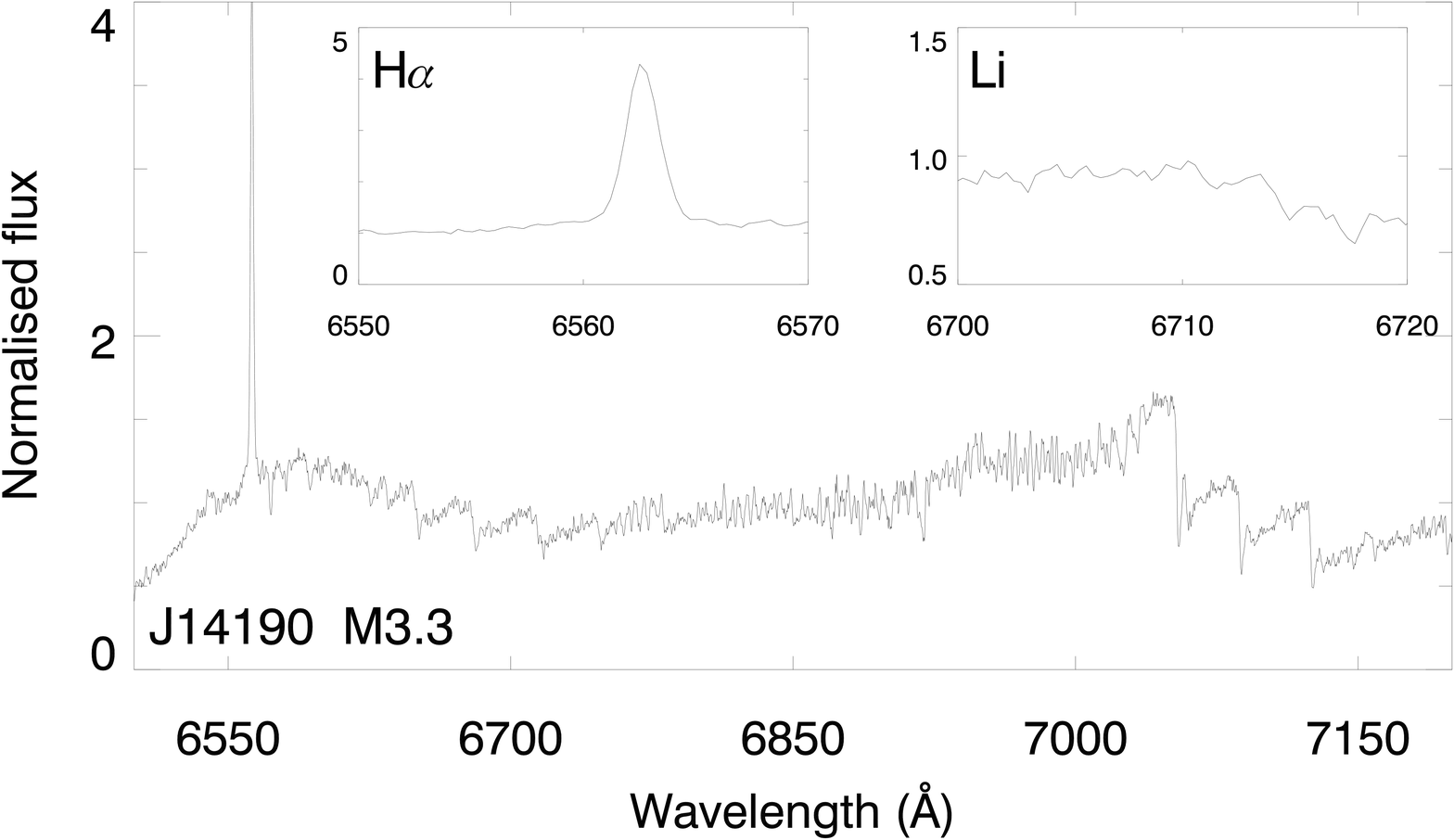} \\
\includegraphics[width=0.42\textwidth]{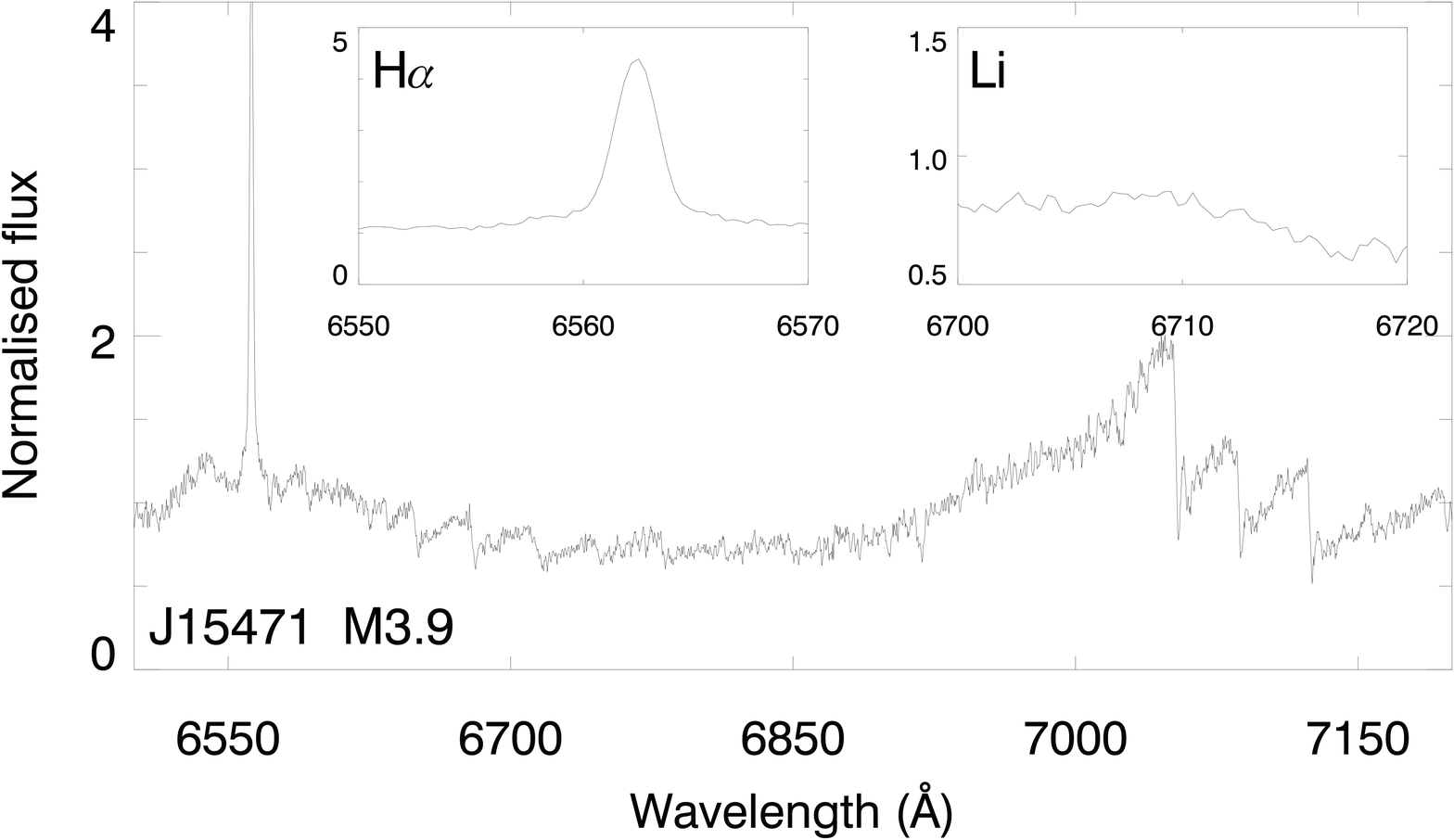} & \includegraphics[width=0.42\textwidth]{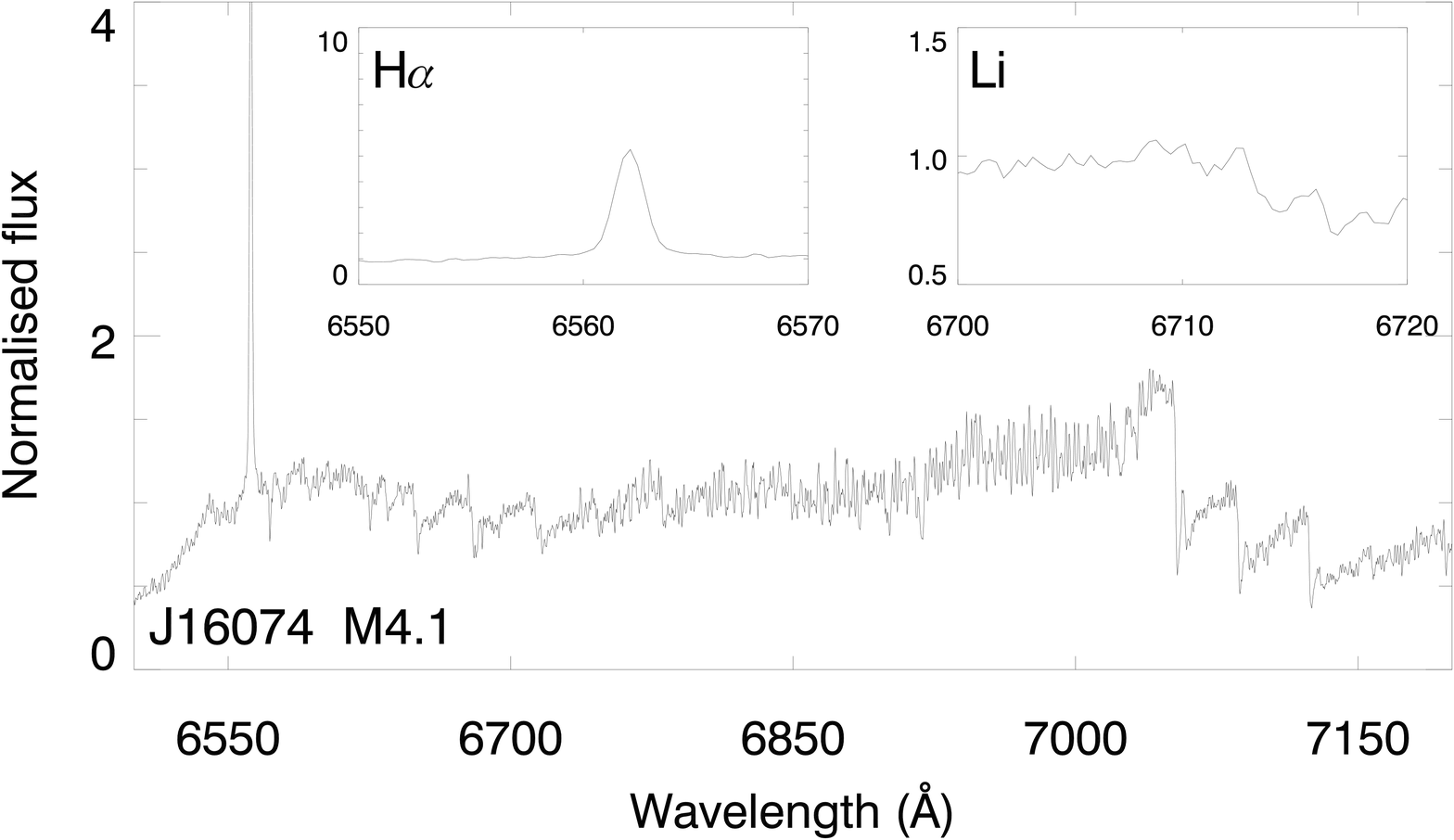} \\
 \includegraphics[width=0.42\textwidth]{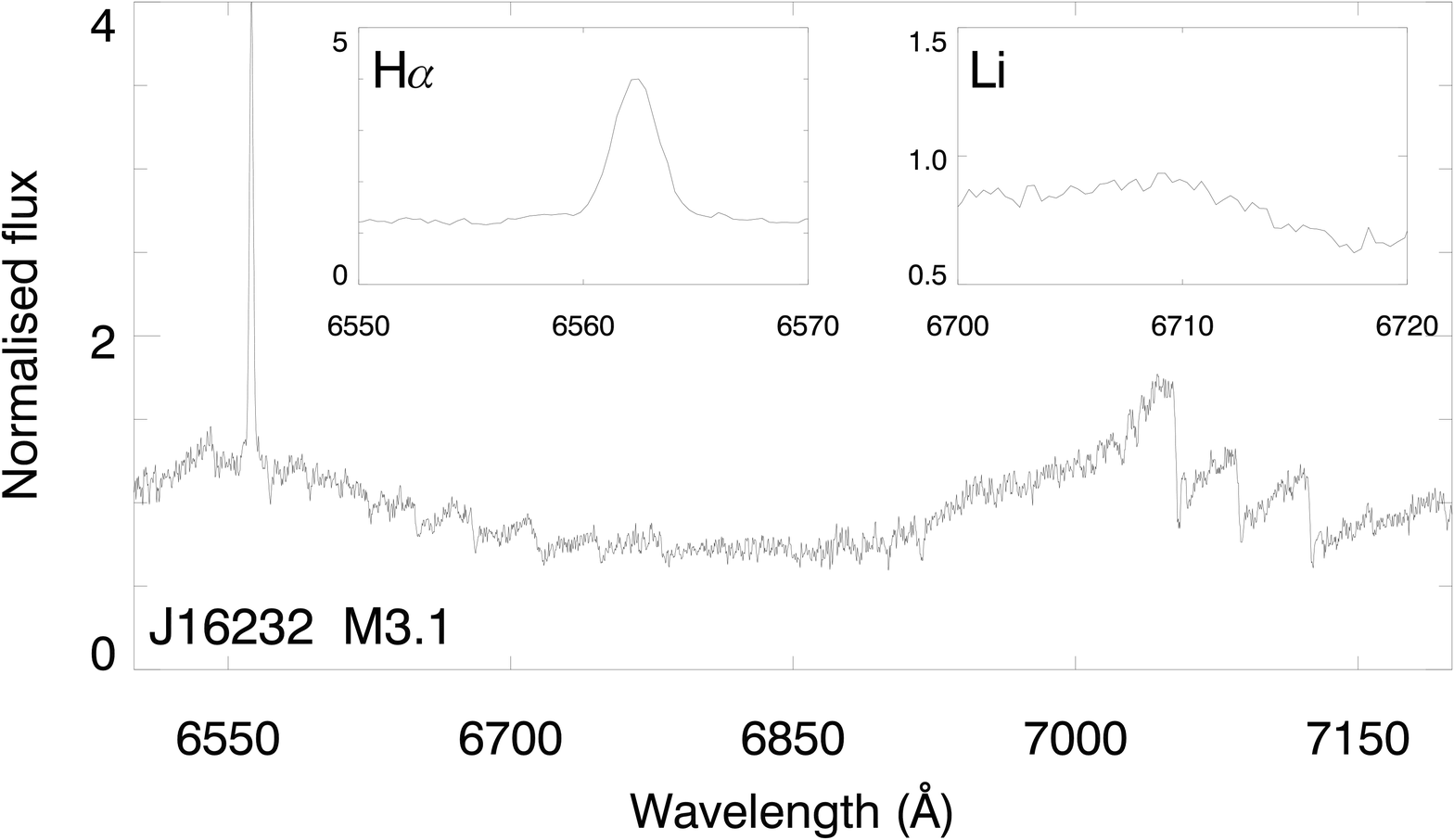} & \includegraphics[width=0.42\textwidth]{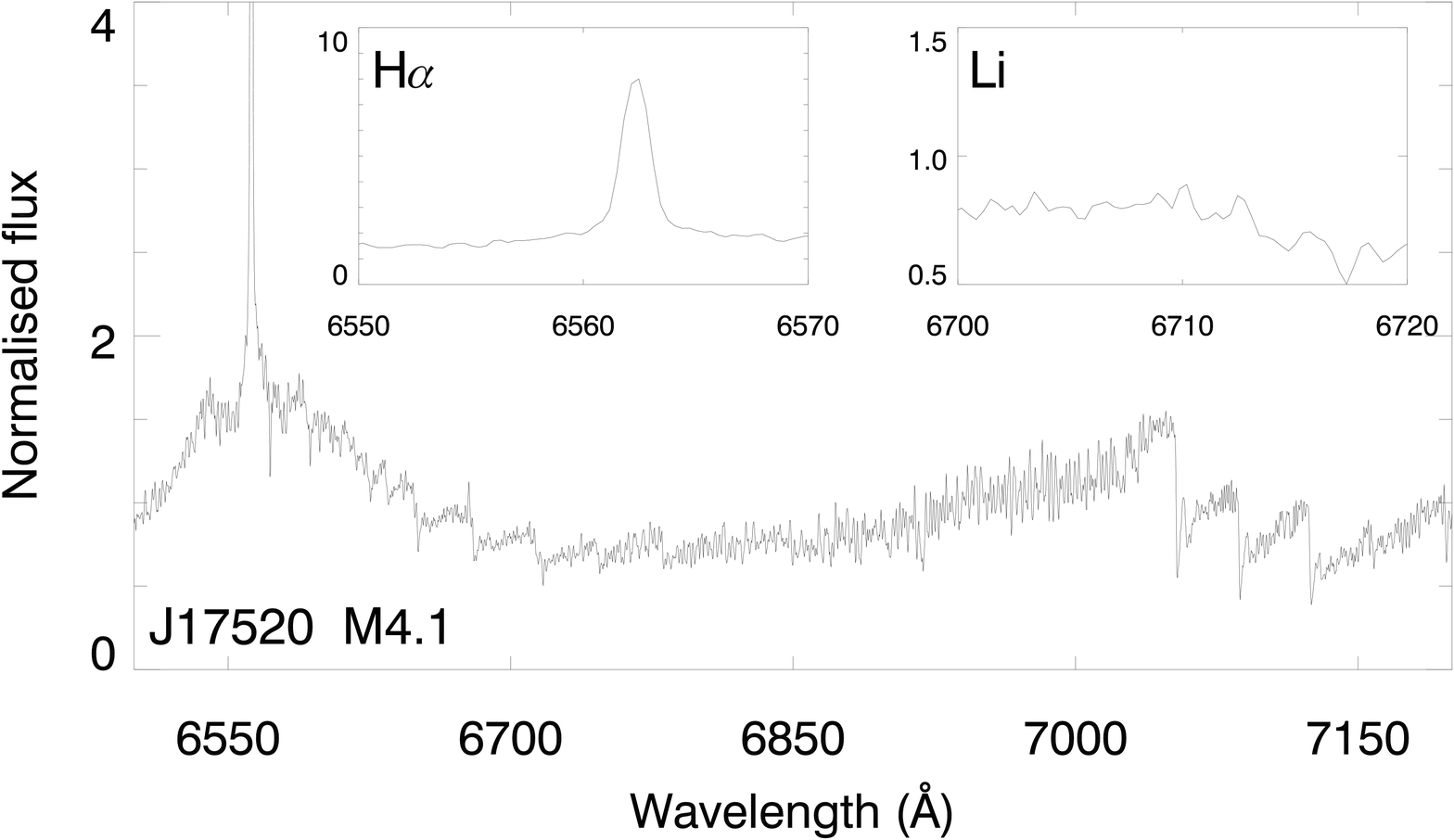} \\
      \end{tabular}
      \begin{tabular}{c}
  \includegraphics[width=0.42\textwidth]{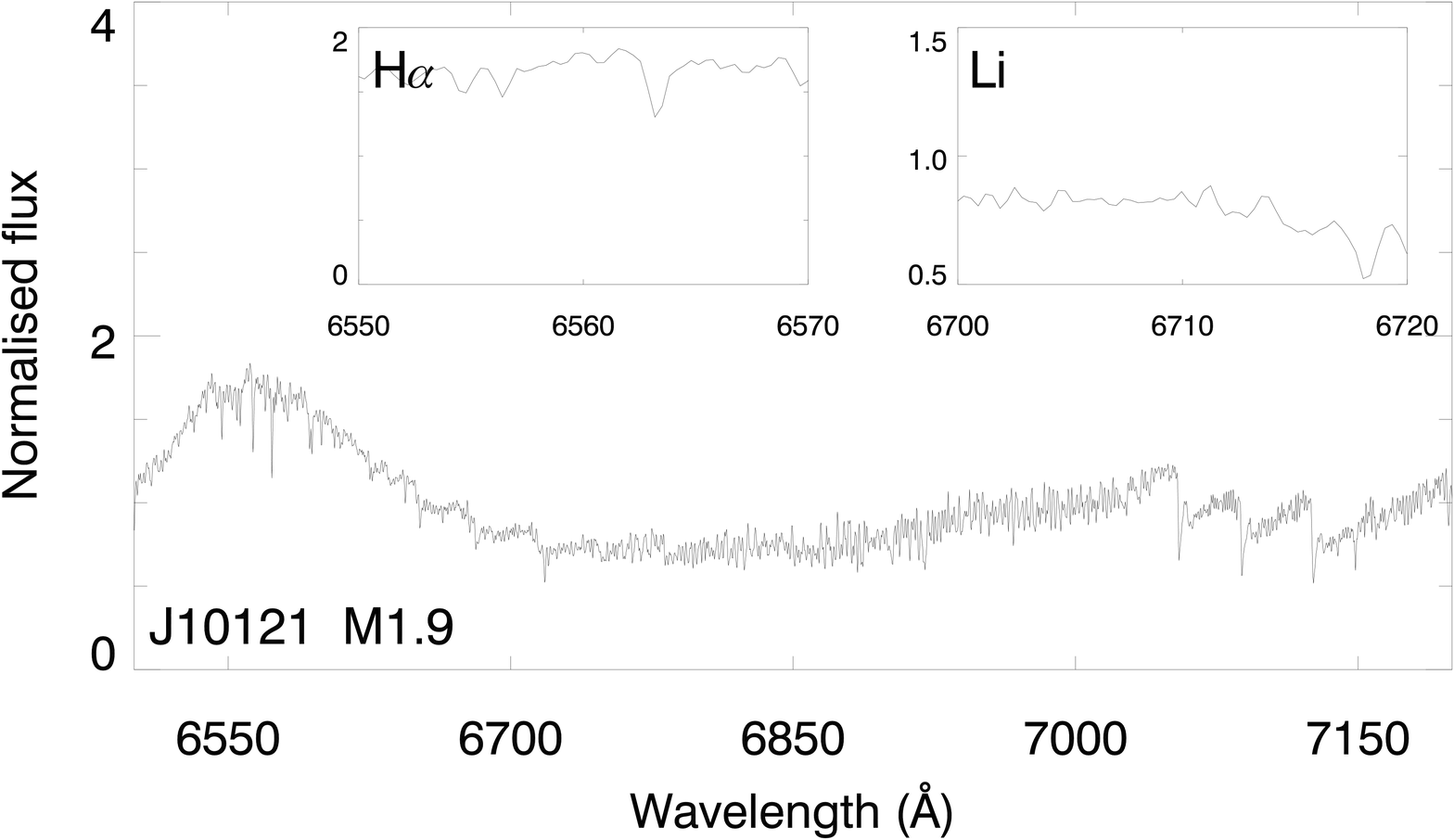} \\  
                        \multicolumn{1}{c}{{\bf Figure 4.} \textit{continued.}} \\
      \end{tabular}
      \end{center}
    \end{minipage}
\end{figure*}

All the observed objects are listed in Tables~\ref{T_BPMG}, \ref{T_ABDMG} and \ref{T_Kinematics}. Here we also compare our RVs and EWs with any previous measurements in the literature, which are all listed in the captions of Tables~\ref{T_BPMG} and~\ref{T_ABDMG}. There are 12 BPMG and 4 ABDMG candidates with previously measured RVs, of which 6 BPMG and 2 ABDMG candidates are in agreement within the combined 2$\sigma$ error bars of both datasets. The five objects in BPMG that have discrepant RVs are J0437, J1643, J1849, J2351 and the unresolved binary pair J0506\,NS. The two objects in ABDMG that have discrepant RVs are J1559 and J1607. A possible reason for RV discrepancies may be that they are in binary systems with varying RVs as they orbit a common center of mass. Longer-term observations are necessary to investigate the binarity of these objects.

\begin{center}
{\tiny
\begin{table*}
\setlength{\tabcolsep}{0.0cm}
\begin{tabular}{p{3.1cm}p{1.9cm}p{2.1cm}p{0.6cm}p{1.0cm}p{1.6cm}p{0.6cm}p{1.5cm}p{1.1cm}p{0.7cm}p{1.4cm}p{1.5cm}p{0.6cm}}
\hline
\hline
Name                   & RV                   &    RV$_{\rm lit}$ & Ref  & $\Delta $RV & $v\sin i$              & Ref & H$\alpha$ EW & H$\alpha_{\rm lit}$ & Ref & Li EW        & Li EW$_{\rm lit}$ & Ref \\
2MASS-                 & (${\rm km\,s}^{-1}$) &                   &      &             & (${\rm km\,s}^{-1}$)   &     & (\AA)        &                     &     & (m\AA)       &                   &     \\
\hline
\multicolumn{13}{c}{BPMG candidates confirmed as members} \\
J01351393$-$0712517    &       $+6.5 \pm 1.8$ &   $+11.7 \pm 5.3$ & a    & $-2.7$      &         $50.7 \pm 5.9$ & g   &      $-4.9$ &              $-5.36$ &   n &        $<23$ & 46.7              &   b \\ 
                       &                      &    $+6.8 \pm 0.8$ & b    &             &                        &     &             &                      &     &              &                   &     \\
J02175601+1225266      &       $+7.0 \pm 1.4$ &                   &      & $-1.4$      &         $22.6 \pm 3.0$ &     &      $-7.2$ &               $-7.0$ &   o &        $<22$ &                   &     \\ 
J10141918+2104297      &       $+3.1 \pm 0.3$ &                   &      & $-0.3$      &                  $<20$ &     &      $-1.1$ &              $-0.91$ &   p &        $<20$ &                   &     \\ 
J05335981$-$0221325    &      $+22.0 \pm 1.3$ & $+21.00 \pm 2.00$ & b    & $+3.2$      &          $5.4 \pm 1.3$ & g   &      $-6.0$ &               $-6.1$ &   o &        $<49$ & $< 49$            &   g \\ 
J16430128$-$1754274    &      $-10.0 \pm 1.5$ &   $+11.3 \pm 3.5$ & c    & $+3.5$      &          $8.7 \pm 2.7$ & g   &      $-2.1$ &              $-1.55$ &   p & $364 \pm 21$ & 300               &   q \\ 
J04373746$-$0229282    &      $+25.1 \pm 1.1$ & $+20.73 \pm 0.46$ & d    & $+4.1$      &                  $6.5$ & l   &      $-2.2$ &               $-2.1$ &   n & $140 \pm 21$ & $120 \pm 30$      &   r \\ 
                       &                      &   $+21.7 \pm 0.3$ & a    &             &                        &     &             &                      &     &              &                   &     \\
                       &                      &   $+21.2 \pm 0.5$ & e    &             &                        &     &             &                      &     &              &                   &     \\
J05015881+0958587      &      $+18.8 \pm 1.5$ &   $+14.9 \pm 3.5$ & f    & $+2.9$      &                  $<20$ &     &      $-6.1$ &              $-6.03$ &   h &        $<53$ & 0                 &   s \\ 
J05241914$-$1601153    &      $+20.6 \pm 4.1$ &   $+17.2 \pm 0.5$ & g    & $+0.0$      &         $50.0 \pm 4.5$ & g   &     $-11.8$ &              $-11.7$ &   o & $217 \pm 29$ & 223               &   t \\ 
J05082729$-$2101444    &      $+22.8 \pm 3.8$ & $+26.60 \pm 0.43$ & g    & $+2.0$      &         $23.5 \pm 1.8$ & g   &     $-20.9$ &              $-24.9$ &   o & $618 \pm 43$ & 481.1             &   b \\ 
J19102820$-$2319486    &       $-7.9 \pm 1.7$ &    $-7.2 \pm 0.2$ & h    & $+3.9$      &         $12.2 \pm 1.8$ & g   &      $-8.7$ &               $-8.2$ &   o &        $<55$ & 23.3              &   b \\ 
\hline
\multicolumn{13}{c}{BPMG candidates rejected as members} \\
J11515681+0731262      &      $-11.1 \pm 2.3$ &                   &      & $-9.8$      &                  $<20$ &     &      $-1.3$ &              $-3.97$ &   p &        $<31$ &                   &     \\ 
J00323480+0729271      &      $+56.3 \pm 1.4$ &                   &      & $+54.6$     &                   14.7 & m   &      $-2.7$ &              $-5.23$ &   h &        $<17$ &                   &     \\ 
J23512227+2344207      &      $+38.6 \pm 1.6$ &    $+2.1 \pm 0.5$ & a    & $+43.1$     &                  $<20$ &     &      $-1.8$ &               $-2.1$ &   o &        $<17$ &                   &     \\ 
J05015665+0108429      &      $+19.7 \pm 1.6$ &                   &      & $+1.9$      &                  $<20$ &     &      $+3.6$ &              $+3.89$ &   p &        $<20$ &                   &     \\ 
J10015995+6651278      &      $-22.1 \pm 1.3$ &                   &      & $-15.6$     &                  $<20$ &     &      $-2.8$ &                      &     &        $<41$ &                   &     \\ 
J08224748+0757171      &       $-3.1 \pm 1.9$ &                   &      & $-16.2$     &                  $<20$ &     &      $-3.4$ &                      &     &        $<38$ &                   &     \\ 
J12115308+1249135      &       $-2.4 \pm 3.8$ &                   &      & $-5.4$      &                  $<20$ &     &      $-1.2$ &                      &     &        $<23$ &                   &     \\ 
J05320450$-$0305291    &      $+26.2 \pm 1.6$ &                   &      & $+7.2$      &                  $<20$ &     &      $-1.3$ &                      &     &  $60 \pm 13$ &                   &     \\ 
J07293108+3556003\,N   &      $+13.1 \pm 1.4$ &                   &      & $+5.7$      &                  $<20$ &     &      $-2.8$ &                      &     &        $<42$ &                   &     \\ 
J07293108+3556003\,S   &      $+14.0 \pm 1.4$ &                   &      & $+6.6$      &                  $<20$ &     &      $-2.9$ &                      &     &        $<30$ &                   &     \\ 
J18495543$-$0134087    &       $+0.9 \pm 2.2$ &  $+119.0 \pm 0.5$ & a    & $+18.6$     &         $37.2 \pm 3.4$ & g   &      $-5.0$ &               $-5.8$ &   o &  $59 \pm 12$ &                   &     \\ 
J07264154+1850346      &      $-22.2 \pm 1.3$ &                   &      & $-34.8$     &                  $<20$ &     &      $-5.2$ &                      &     &        $<53$ &                   &     \\ 
J05064946$-$2135038\,N &      $+36.4 \pm 1.5$ &   $+31.2 \pm 0.9$ & h    & $+15.6$     &                  $<20$ &     &      $-8.0$ &                      &     &        $<41$ &                   &     \\ 
                       &                      &           $+31.7$ & i    &             &                        &     &             &                      &     &              &                   &     \\
                       &                      &   $+21.6 \pm 0.6$ & j    &             &                        &     &             &                      &     &              &                   &     \\
                       &                      &          $+21.43$ & k    &             &                        &     &             &                      &     &              &                   &     \\
J05064946$-$2135038\,S &      $+34.0 \pm 1.4$ &           $+21.2$ & j    & $+13.2$     &                  $<20$ &     &      $-2.0$ &                      &     &        $<41$ &                   &     \\ 
                       &                      &           $+21.4$ & k    &             &                        &     &             &                      &     &              &                   &     \\

\hline
\end{tabular}
\caption{RVs and EWs for observed BPMG candidates. Column 1: objects are named according to their 2MASS identifier. Column 2: our RV measurement. Columns 3 and 4 provide any RV data from available sources in the literature and the reference. Column 5: $\Delta$ RV = RV$-V_{T}\cos\lambda$. Columns 6 and 7: projected rotational velocity ($v\sin i$) and the source reference. If no reference was available, we calculate $v\sin i$ using the calibration described in $\S$\ref{S_Results}. Columns 8 to 10: our H$\alpha$ EW measurement, any alternative literature values and their references. Columns 11 to 13: our Li EW measurement, any alternative literature values and their references. References: (a) \protect\citealt{2012a_Shkolnik}, (b) \protect\citealt{2014b_Malo}, (c) \protect\citealt{2008a_Zwitter}, (d) \protect\citealt{2012a_Bailey}, (e) \protect\citealt{2015a_Macintosh}, (f) \protect\citealt{2007a_Kharchenko}, (g) \protect\citealt{2014a_Malo}, (h) \protect\citealt{2002a_Gizis}, (i) \protect\citealt{1995a_Reid}, (j) \protect\citealt{2014a_Elliott}, (k) \protect\citealt{2015a_Tokovinin}, (l) \protect\citealt{2010a_Houdebine}, (m) \protect\citealt{2012a_Reiners}, (n) \protect\citealt{2009a_Shkolnik}, (o) \protect\citealt{2006a_Riaz}, (p) \protect\citealt{2013a_Lepine}, (q) \protect\citealt{2011a_Kiss}, (r) \protect\citealt{2006a_Feigelson}, (s) \protect\citealt{2009a_Da_Silva}, (t) \protect\citealt{2013a_Malo}.}
\label{T_BPMG}
\end{table*}}
\end{center}
%


\begin{center}
{\tiny
\begin{table*}
\setlength{\tabcolsep}{0.0cm}
\begin{tabular}{p{3.5cm}p{1.7cm}p{2.5cm}p{0.7cm}p{0.9cm}p{1.8cm}p{0.7cm}p{1.9cm}p{1.8cm}p{0.7cm}p{1.4cm}}
\hline
\hline
Name                 & RV                       &   RV$_{\rm lit}$ & Ref & $\Delta $RV & $v\sin i$            & Ref & H$\alpha$ EW & H$\alpha_{\rm lit}$ & Ref   & Li EW        \\
2MASS-               & (${\rm km\,s}^{-1}$)     &                  &     &             & (${\rm km\,s}^{-1}$) &     & (\AA)        &                     &       & (m\AA)       \\
\hline
\multicolumn{11}{c}{ABDMG candidates confirmed as members} \\
J12574030+3513306\,N &          $-14.1 \pm 1.6$ &                  &     & $+2.8$      &                $<20$ &     & $-1.8$       & $-1.36$             & f   & $<51$         \\ 
J12574030+3513306\,S &          $-18.0 \pm 1.4$ &                  &     & $+1.1$      &                  8.0 & d   & $-3.8$       & $-4.27$             & g   & $<47$         \\
J15594729+4403595    &          $-29.5 \pm 3.8$ &  $-15.8 \pm 0.5$ & a   & $-0.4$      &       $54.9 \pm 4.6$ & a   & $-3.2$       & $-3.3$              & h   & $25 \pm 10$   \\
                     &                          &  $-19.6 \pm 0.6$ & b   &             &                      &     &              &                     &     &          \\
J16455062+0343014    &          $-23.3 \pm 1.3$ &                  &     & $-1.5$      &                $<20$ &     & $-1.5$       & $-1.38$             & i   & $184 \pm 22$  \\ 
J12383713$-$2703348  &          $+7.8 \pm 1.2 $ & $+9.60 \pm 0.20$ & a   & $+0.0$      &                  3.6 & a   & $-2.8$       & $-3.1$              & h   & $<63^{\rm j}$ \\ 
J09321267+3358285    &           $+3.5 \pm 1.2$ &                  &     & $+4.8$      &                $<20$ &     & $-2.9$       &                     &     &  $<40$        \\ 
\hline
\multicolumn{11}{c}{ABDMG candidates rejected as members} \\
J09211104+4801538    &          $+10.5 \pm 2.1$ &                  &     & $+17.7$     &                $<20$ &     & $-5.7$       &                     &     & $178 \pm 30$  \\ 
J07445070+0007355    &           $+4.2 \pm 1.5$ &                  &     & $-15.6$     &                $<20$ &     & $-4.6$       &                     &     &  $<20$        \\ 
J09245082+3041373    &          $+13.7 \pm 2.3$ &                  &     & $+13.1$     &                $<20$ &     & $-4.8$       &                     &     & $<30$         \\ 
J10042148+5023135    &           $-1.7 \pm 1.5$ &                  &     & $+8.6$      &                $<20$ &     & $-3.5$       & $-1.07$             & j   & $<20$         \\ 
J13342523+6956273    &          $-14.1 \pm 1.4$ &                  &     & $+9.9$      &                $<20$ &     & $-6.3$       &                     &     & $<38$         \\ 
PYC~J13351+5039\,N   &          $-14.1 \pm 1.3$ &                  &     & $+8.5$      &                $<20$ &     & $-3.4$       &                     &     & $<35$         \\ 
PYC~J13351+5039\,S   &          $-16.3 \pm 1.4$ &                  &     & $+6.4$      &                $<20$ &     & $-3.3$       &                     &     & $<20$         \\ 
J17520294+5637278    &          $-20.7 \pm 1.3$ &                  &     & $+10.2$     &                $<20$ &     & $-5.1$       &                     &     & $<52$         \\ 
J16232165+6149149    &          $-18.3 \pm 3.8$ &                  &     & $+10.9$     &                $<20$ &     & $-4.9$       & $-5.6$              & h   & $<38$         \\ 
J06073185+4712266    &          $+27.1 \pm 1.8$ &                  &     & $+28.9$     &                $<20$ &     & $-6.2$       &                     &     & $40 \pm 16$   \\ 
J09022792+5848142    &           $+0.1 \pm 1.1$ &                  &     & $+11.3$     &                $<20$ &     & $-2.1$       & $-1.99$             & i   & $<38$         \\ 
J09065515+4532299    &           $+9.2 \pm 3.5$ &                  &     & $+14.5$     &       $30.9 \pm 3.0$ &     & $-3.2$       & $-3.60$             & i   & $<51$         \\ 
J14190331+6451463    &          $-12.0 \pm 2.4$ &                  &     & $+13.6$     &       $20.8 \pm 3.0$ &     & $-4.6$       & $-7.1$              & h   & $54 \pm 12$   \\ 
J15471191+4148218    &          $-11.5 \pm 4.1$ &                  &     & $+16.9$     &       $52.9 \pm 3.0$ &     & $-5.5$       &                     &     & $42 \pm 11$   \\ 
J16074132$-$1103073  &          $-19.4 \pm 1.9$ &   $-8.5 \pm 1.2$ & a   & $-5.2$      &                $<20$ &     & $-5.8$       &                     &     & $40 \pm 20$   \\ 
J08304079+0421444    &          $+23.4 \pm 4.9$ &                  &     & $+7.5$      &                $>60$ &     & $-5.2$       &                     &     & $116 \pm 17$  \\ 
J10121768$-$0344404  &           $+7.7 \pm 1.0$ &   $+9.0 \pm 1.4$ & c   & $-4.2$      &                  1.8 & e   & $+0.2$       & $+0.4$              & h   & $<20$         \\ 
\hline
\end{tabular}
\caption{RVs and EWs for observed ABDMG candidates. Column 1: objects are named according to the 2MASS identifier. Column 2: our RV measurement. Columns 3 and 4 provide any RV data from available sources in the literature and the reference. Column 5: $\Delta$ RV = RV$-V_{T}\cos\lambda$. Columns 6 and 7: projected rotational velocity ($v\sin i$) and the source reference. If no reference was available, we calculate $v\sin i$ using the calibration described in $\S$\ref{S_Results}. Columns 8 to 10: our H$\alpha$ EW measurement, any alternative literature values and their references. Column 11: our Li EW measurement. References: (a) \protect\citealt{2014a_Malo}, (b) \protect\citealt{2015a_Bowler}, (c) \protect\citealt{2007a_Kharchenko}, (d) \protect\citealt{2009a_Jenkins}, (e) \protect\citealt{2012a_Reiners}, (f) \protect\citealt{2002a_Gizis}, (g) \protect\citealt{2013a_Lepine}, (h) \protect\citealt{2006a_Riaz}, (i) \protect\citealt{2015a_Ansdell}, (j) \protect\citealt{2009a_Shkolnik}, Li EW = $< 23.55\,$m\AA\ (1$\sigma$ upper limit).}
\label{T_ABDMG}
\end{table*}}
\end{center}

%


\begin{center}
{\tiny
\begin{table*}
\setlength{\tabcolsep}{0.0cm}
\begin{tabular}{p{3.20cm}p{0.65cm}p{1.50cm}p{1.95cm}p{1.95cm}p{1.80cm}p{0.90cm}p{0.70cm}p{1.00cm}p{1.70cm}p{1.2cm}p{1.0cm}}
\hline
\hline
Name                           & Ref  & HJD             & ${\mu_{\alpha}}$ &  ${\mu_{\delta}}$ & $M_{K}$         & $V$   & $K_{\rm s}$  & $f_{\rm X}$ & Distance                & SpT                     & $P_{\rm B2}$ \\
2MASS-                         &      &                 &                  &                   &                 &       &              &                              & (pc)                    & (M-)                    &              \\
\hline
\multicolumn{12}{c}{BPMG candidates confirmed as members} \\
J01351393$-$0712517            & 1    & 291.400         & $+106.5 \pm 5.1$ &   $-60.7 \pm 5.1$ & $5.19 \pm 0.14$ & 13.43 & 8.08         &                      $-3.02$ &  $37.7 \pm 4.3^{\rm a}$ &      $4.1, 4.3^{\rm f}$ & 40.54 \\ 
J02175601+1225266              & 2    & 291.428         &  $+52.3 \pm 1.6$ &   $-53.2 \pm 1.5$ & $4.93 \pm 0.20$ & 14.09 & 9.08         &                      $-2.83$ &          $67.9 \pm 6.1$ &        $3.5, 4^{\rm g}$ & 29.36 \\ 
J10141918+2104297              & 2    & 291.714         & $-139.3 \pm 1.1$ &  $-158.8 \pm 0.8$ & $4.58 \pm 0.09$ & 10.08 & 6.26         &                      $-3.38$ &  $22.0 \pm 1.6^{\rm b}$ & ${\rm K9}, 0.7^{\rm f}$ &  0.30 \\ 
J05335981$-$0221325            & 3    & 372.434         &  $-0.8 \pm 13.8$ &  $-63.8 \pm 13.8$ & $4.60 \pm 0.18$ & 12.42 & 7.70         &                      $-2.81$ &          $41.8 \pm 3.3$ &        $2.9, 3^{\rm h}$ & 51.55 \\ 
J16430128$-$1754274            & 3    & 375.727         &  $-27.4 \pm 3.8$ &   $-51.3 \pm 4.2$ & $4.69 \pm 0.11$ & 12.57 & 8.55         &                      $-3.05$ &          $59.2 \pm 2.8$ &      $1.4, 0.5^{\rm h}$ &  0.00 \\ 
J04373746$-$0229282$^{\rm A}$  & 1    & 376.337         &  $+44.6 \pm 2.1$ &   $-62.9 \pm 2.1$ & $3.96 \pm 0.32$ & 10.53 & 6.41         &                      $-2.62$ &  $31.2 \pm 2.0^{\rm b}$ &              $1.9, 2.2$ &  9.70 \\ 
J05015881+0958587$^{\rm B}$    & 3    & 376.356         & $+36.4 \pm 18.7$ & $-108.0 \pm 18.7$ & $4.39 \pm 0.12$ & 11.51 & 6.37         &                      $-3.15$ &  $33.2 \pm 3.7^{\rm c}$ &              $4.1, 3.8$ & 84.86 \\ 
J05241914$-$1601153$^{\rm B}$  & 3    & 376.386         &  $+20.5 \pm 5.2$ &   $-36.7 \pm 5.2$ & $5.34 \pm 0.35$ & 13.57 & 7.81         &                      $-3.20$ &          $31.7 \pm 4.9$ &      $4.9, 4.5^{\rm h}$ &  6.40 \\ 
J05082729$-$2101444            & 3    & 377.368         &  $+36.5 \pm 4.9$ &   $-15.3 \pm 4.9$ & $6.42 \pm 0.36$ & 14.41 & 8.83         &                      $-3.24$ &          $30.8 \pm 4.9$ &        $5.4, 5^{\rm h}$ &  0.71 \\ 
J19102820$-$2319486            & 3    & 377.772         &  $+19.2 \pm 5.4$ &   $-51.6 \pm 5.4$ & $4.02 \pm 0.11$ & 13.22 & 8.22         &                      $-2.91$ &          $69.2 \pm 3.4$ &        $4.0, 4^{\rm h}$ &  2.21 \\ 
\hline
\multicolumn{12}{c}{BPMG candidates rejected as members} \\
J11515681+0731262              & 2    & 290.745         & $-126.1 \pm 5.6$ &   $110.4 \pm 5.6$ & $5.24 \pm 0.12$ & 12.42 & 7.89         &                      $-2.91$ &          $34.0 \pm 1.8$ &        $1.9, 3^{\rm g}$ &  0.00 \\ 
J00323480+0729271              & 1    & 291.335         & $+105.0 \pm 3.1$ &   $-63.4 \pm 3.0$ & $4.36 \pm 0.13$ & 12.82 & 7.51         &                      $-2.95$ &          $42.7 \pm 2.6$ &        $3.7, 4^{\rm g}$ &  0.00 \\ 
J23512227+2344207              & 1    & 291.361         & $+266.7 \pm 4.8$ &   $-70.5 \pm 4.8$ & $7.79 \pm 0.10$ & 14.18 & 8.82         &                      $-3.17$ &          $16.1 \pm 0.7$ &      $4.0, 4.0^{\rm f}$ &  0.00 \\ 
J05015665+0108429$^{\rm C}$    & 2    & 291.515         &  $+33.8 \pm 5.1$ &   $-95.5 \pm 5.1$ & $4.39 \pm 0.12$ & 11.74 & 6.37         &                      $-4.38$ &          $26.4 \pm 2.1$ &        $4.0, 5^{\rm f}$ & 95.65 \\ 
J10015995+6651278              & 2    & 291.655         &  $-87.4 \pm 4.8$ &   $-87.9 \pm 4.8$ & $5.21 \pm 0.14$ & 12.29 & 8.22         &                      $-3.18$ &          $40.1 \pm 2.6$ &              $0.8, 0.9$ &  0.00 \\ 
J08224748+0757171              & 2    & 291.701         & $-51.4 \pm 11.0$ &  $-75.8 \pm 11.0$ & $5.24 \pm 0.22$ & 14.29 & 9.21         &                      $-3.22$ &          $62.5 \pm 6.2$ &        $3.6, 4^{\rm g}$ &  0.00 \\ 
J12115308+1249135              & 2    & 291.746         &  $-70.1 \pm 1.2$ &   $-60.1 \pm 1.4$ & $5.29 \pm 0.24$ & 12.60 & 8.80         &                      $-2.31$ &          $50.4 \pm 2.1$ &              $0.1, 0.0$ &  0.00 \\ 
J05320450$-$0305291$^{\rm B}$  & 3    & 372.421         &   $+6.7 \pm 2.2$ &   $-50.4 \pm 2.2$ & $3.73 \pm 0.22$ & 11.12 & 7.01         &                      $-3.37$ &          $45.5 \pm 4.5$ &        $3.3, 2^{\rm g}$ &  0.07 \\ 
J07293108+3556003\,N           & 2    & 372.455         &  $-31.2 \pm 2.3$ &  $-101.7 \pm 2.3$ & $4.80 \pm 0.09$ & 11.88 & 7.80         &                      $-3.08$ &          $42.6 \pm 2.6$ &        $1.3, 1^{\rm g}$ &  0.00 \\ 
J07293108+3556003\,S           & 2    & 372.455         &  $-31.2 \pm 2.3$ &  $-101.7 \pm 2.3$ & $3.69 \pm 0.12$ & 11.88 & 7.80         &                      $-3.08$ &          $39.9 \pm 3.1$ &              $1.4, 1.1$ &  0.00 \\ 
J18495543$-$0134087            & 3    & 374.734         & $+40.6 \pm 15.6$ & $-183.8 \pm 16.5$ & $4.57 \pm 0.20$ & 13.38 & 8.84         &                      $-2.86$ &          $71.5 \pm 6.9$ &      $2.3, 2.5^{\rm h}$ &  0.00 \\ 
J07264154+1850346              & 2    & 375.399         &  $-20.0 \pm 4.9$ &   $-61.3 \pm 4.9$ & $4.21 \pm 0.19$ & 13.83 & 9.13         &                      $-3.30$ &          $57.6 \pm 5.2$ &        $2.9, 4^{\rm f}$ &  0.00 \\ 
J05064946$-$2135038\,N         & 3    & 376.375         &  $+37.0 \pm 3.5$ &   $-38.1 \pm 3.6$ & $4.69 \pm 0.06$ & 11.67 & 6.11         &                      $-2.95$ &  $17.3 \pm 2.6^{\rm d}$ &              $4.7, 4.4$ &  0.00 \\ 
J05064946$-$2135038\,S         & 3    & 376.375         &  $+37.0 \pm 3.5$ &   $-38.1 \pm 3.6$ & $4.70 \pm 0.06$ & 10.44 & 6.12         &                      $-2.98$ &  $17.3 \pm 2.6^{\rm d}$ &              $1.9, 2.2$ &  0.00 \\ 
\hline
\hline
\multicolumn{12}{c}{ABDMG candidates confirmed as members} \\
J12574030+3513306\,N$^{\rm B}$ & 1   &  372.602         & $-264.1 \pm 4.7$ &  $-139.9 \pm 3.3$ & $6.07 \pm 0.10$ & 10.54 & 6.55         &                      $-2.98$ &          $17.6 \pm 0.9$ &      $4.3, 4.0^{\rm f}$ &  0.00 \\ 
J12574030+3513306\,S$^{\rm B}$ & 1   &  372.602         & $-264.1 \pm 4.7$ &  $-139.9 \pm 3.3$ & $6.07 \pm 0.10$ & 13.16 & 8.02         &                      $-2.35$ &          $17.6 \pm 0.9$ &      $1.4, 1.0^{\rm f}$ &  0.00 \\
J15594729+4403595              & 2   &  372.671         &  $-72.4 \pm 4.9$ &   $-17.3 \pm 4.9$ & $5.41 \pm 0.21$ & 11.83 & 7.62         &                      $-3.15$ &          $35.0 \pm 2.5$ &        $1.4, 1^{\rm h}$ &  4.66 \\ 
J16455062+0343014              & 1   &  372.685         &  $-42.4 \pm 4.7$ &  $-108.8 \pm 4.7$ & $6.56 \pm 0.18$ & 12.48 & 8.44         &                      $-3.25$ &          $49.1 \pm 2.9$ &              $0.9, 1.3$ & 24.45 \\ 
J12383713$-$2703348            & 2   &  376.599         & $-185.1 \pm 5.1$ &  $-185.2 \pm 5.1$ & $6.60 \pm 0.10$ & 12.44 & 7.84         &                      $-3.27$ &          $25.2 \pm 0.8$ &      $2.6, 2.5^{\rm h}$ & 96.77 \\ 
J09321267+3358285              & 1   &  377.499         &  $-64.6 \pm 4.6$ &   $-99.0 \pm 4.6$ & $5.84 \pm 0.13$ & 14.66 & 9.02         &                      $-3.07$ &          $64.1 \pm 3.5$ &              $3.6, 4.4$ & 20.36 \\ 
\hline
\multicolumn{12}{c}{ABDMG candidates rejected as members} \\
J09211104+4801538              & 1   &  372.464         &  $-67.7 \pm 5.0$ &  $-125.2 \pm 5.0$ & $6.59 \pm 0.11$ & 14.15 & 9.17         &                      $-2.86$ &          $47.0 \pm 2.0$ &              $3.9, 4.4$ &  0.00 \\ 
J07445070+0007355              & 1   &  372.509         &  $-42 2 \pm 5.0$ &  $-129.7 \pm 4.7$ & $7.50 \pm 0.15$ & 14.30 & 9.23         &                      $-3.13$ &          $39.2 \pm 2.1$ &              $4.1, 4.6$ &  0.00 \\ 
J09245082+3041373              & 1   &  372.542         & $-104.2 \pm 4.9$ &  $-165.6 \pm 4.9$ & $6.93 \pm 0.10$ & 13.52 & 8.67         &                      $-3.18$ &          $33.3 \pm 1.2$ &              $3.4, 4.1$ &  0.00 \\ 
J10042148+5023135              & 1   &  372.576         & $-140.0 \pm 2.3$ &  $-194.8 \pm 2.3$ & $4.10 \pm 0.13$ & 11.67 & 7.20         &                      $-2.79$ &          $58.4 \pm 3.0$ &      $3.0, 3.0^{\rm f}$ &  0.00 \\ 
J13342523+6956273              & 1   &  372.625         & $-103.6 \pm 4.9$ &   $-14.6 \pm 4.9$ & $5.73 \pm 0.17$ & 13.60 & 8.73         &                      $-3.14$ &          $47.0 \pm 3.1$ &              $3.1, 4.0$ & 71.01 \\ 
PYC~J13351+5039\,N$^{\rm D}$   & 1   &  372.653         &  $-95.8 \pm 4.6$ &   $-45.9 \pm 4.6$ & $5.31 \pm 0.35$ & 13.34 & 8.37         &                      $-2.87$ &          $53.3 \pm 3.6$ &        $3.8, 4^{\rm g}$ &  0.00 \\ 
PYC~J13351+5039\,S$^{\rm D}$   & 1   &  372.653         &  $-95.8 \pm 4.6$ &   $-45.9 \pm 4.6$ & $5.31 \pm 0.35$ & 12.72 & 9.43         &                      $-2.55$ &          $53.4 \pm 3.6$ &        $2.7, 3^{\rm g}$ &  0.00 \\ 
J17520294+5637278              & 2   &  372.713         &  $+39.5 \pm 8.1$ &   $-29.8 \pm 8.1$ & $6.24 \pm 0.24$ & 13.32 & 8.38         &                      $-3.24$ &          $16.7 \pm 2.1$ &      $4.1, 3.5^{\rm h}$ &  0.02 \\ 
J16232165+6149149$^{\rm B}$    & 2   &  372.730         &  $-59.6 \pm 5.7$ &   $+49.4 \pm 5.7$ & $6.32 \pm 0.19$ & 13.88 & 9.21         &                      $-2.99$ &          $35.9 \pm 2.5$ &      $3.1, 2.5^{\rm h}$ &  0.00 \\ 
J06073185+4712266              & 1   &  374.422         &   $-6.2 \pm 4.7$ &    $-7.8 \pm 5.1$ & $7.18 \pm 0.10$ & 14.35 & 8.89         &                      $-2.77$ &          $33.1 \pm 1.2$ &              $4.8, 5.2$ &  0.00 \\ 
J09022792+5848142              & 1   &  374.480         &  $-45.7 \pm 4.6$ &   $-84.1 \pm 4.6$ & $5.42 \pm 0.15$ & 13.30 & 8.95         &                      $-2.84$ &          $70.4 \pm 4.4$ &              $2.4, 2.4$ &  0.00 \\ 
J09065515+4532299              & 1   &  374.514         &  $-63.8 \pm 4.9$ &  $-103.8 \pm 4.9$ & $6.02 \pm 0.13$ & 13.27 & 9.04         &                      $-2.80$ &          $58.3 \pm 2.9$ &              $1.8, 2.0$ &  0.00 \\ 
J14190331+6451463              & 2   &  374.596         & $-104.4 \pm 9.7$ &    $+9.7 \pm 4.9$ & $7.04 \pm 0.14$ & 14.15 & 9.56         &                      $-2.95$ &          $37.0 \pm 1.6$ &        $3.3, 3^{\rm f}$ &  0.00 \\ 
J15471191+4148218              & 1   &  374.651         &  $-58.2 \pm 5.5$ &   $-15.3 \pm 5.5$ & $6.62 \pm 0.29$ & 14.77 & 9.65         &                      $-2.99$ &          $54.7 \pm 6.3$ &              $3.9, 3.8$ &  0.00 \\ 
J16074132$-$1103073            & 2   &  374.694         &  $-69.7 \pm 6.0$ &  $-145.6 \pm 6.0$ & $7.38 \pm 0.11$ & 14.19 & 8.99         &                      $-3.19$ &          $36.8 \pm 1.4$ &        $4.1, 4^{\rm f}$ &  3.98 \\ 
J08304079+0421444              & 1   &  375.366         & $-66.2 \pm 11.7$ & $-122.5 \pm 12.2$ & $6.46 \pm 0.14$ & 14.13 & 9.01         &                      $-2.95$ &          $52.9 \pm 3.1$ &              $4.0, 4.8$ &  2.80 \\ 
J10121768$-$0344404$^{\rm C}$  & 2   &  375.467         & $-151.8 \pm 1.0$ &  $-243.8 \pm 0.9$ & $5.52 \pm 0.03$ &  9.59 & 5.01         &                      $-4.85$ &  $15.7 \pm 3.8^{\rm e}$ &      $1.9, 2.0^{\rm f}$ & 27.87 \\ 
\hline
\end{tabular}
\caption{Photometric and kinematic data for our observed candidates. Column 1: objects are named according to their 2MASS identifier; (A) qualifies as a member because it is a companion to 51 Eri, which is known to be a member of BPMG (\citealt{2006a_Feigelson}), (B) unresolved binary (as quoted in source paper), (C) H$\alpha$ in absorption, (D) no entry in 2MASS: $K_{\rm s}$ magnitude from \protect\cite{2012a_Schlieder} and no $V$ magnitude, $V-K_{\rm s}$ interpolated from table 5 in \protect\cite{2013a_Pecaut}. Column 2 indicates the literature source from which the candidate was taken: $1=$ \protect\cite{2012a_Shkolnik}, $2=$ \protect\cite{2012a_Schlieder}, $3=$ \protect\cite{2013a_Malo}. Column 3 gives the Heliocentric Julian Date of our observation from 2456000 days. Columns 4 and 5 are the proper-motions in right ascension and declination from the PPMXL catalog (\protect\citealt{2010a_Roeser}). Columns 6 to 8 are the $M_{K}, V$ and $K_{\rm s}$ magnitudes, where $V$ magnitudes are from APASS (\citealt{2012a_Henden}) and $K_{\rm s}$ magnitudes are from 2MASS (\citealt{2003a_Cutri}). Column 9 provides the calculated X-ray to bolometric luminosity ratios as described in $\S$\ref{S_Confirming}. Column 10: distances are calculated from kinematic parallaxes described in $\S$\ref{S_Confirming} - previously published trigonometric distances (in pc) of: (a) $37.9 \pm 2.4$ (\protect\citealt{2012a_Shkolnik}), (b) $23.1 \pm 1.0$, $29.4 \pm 0.3$ (\protect\citealt{2007a_van_Leeuwen}), (c) $24.9 \pm 1.3$ and $30.3 \pm 11.2$ (\protect\citealt{2007a_van_Leeuwen} and \protect\citealt{2014a_Riedel}, respectively), (d) $19.2 \pm 0.5, 19.2 \pm 0.5$ (\protect\citealt{2014a_Riedel}), (e) $7.9 \pm 0.1$ (\protect\citealt{2007a_van_Leeuwen}). Column 11: Spectral-types calculated from TiO5 molecular band indices and compared to any TiO5-based spectral-types from (f) \protect\cite{2009a_Shkolnik}, (g) \protect\cite{2012a_Schlieder}, (h) \protect\cite{2006a_Riaz}, or are otherwise estimated from $V-K_{\rm s}$ and interpolation of table 4 in \protect\cite{2013a_Pecaut}. Column 12 provides the MG membership probabilities from BANYAN II (see $\S$\ref{S_Confirming}).}
\label{T_Kinematics}
\end{table*}}
\end{center}

All BPMG and all but one ABDMG candidates confirmed as members in $\S$\ref{S_Confirming} have published H$\alpha$ EWs that are within 0.5\,\AA\ of our measurements, with the exception of J0508, which has a very broad ($\sim\,20$\,\AA) H$\alpha$ emission line. There is less agreement in H$\alpha$ EWs amongst candidates that we rejected (for example J1151, J0032, J1004 and J1419). Discrepancies in H$\alpha$ measurements may be due to the choice in locating the continuum level around H$\alpha$, or that we are observing H$\alpha$ variability of a few \AA\ on timescales of several years (\citealt{2012a_Bell}). 

For objects confirmed as BPMG or ABDMG members in this work there is broad consistency in Li EWs, with the exception of J0508- where \cite{2014b_Malo} report a Li EW = $484.1 \pm 10.1\,$m\AA, compared to our measurement of $618 \pm 43$\,m\AA. Either measurement would still be consistent with an undepleted Li abundance (\citealt{2007a_Palla}). Spectral types calculated from TiO5 molecular band indices are generally within 0.5 sub-classes of published values and we note that, in cases where there are difference of more than half a sub-class, some published spectral types are rounded to the nearest integer sub-class and the difference in spectral-class may be less pronounced. Where there were no published TiO5-based spectral types we used $V-K_{\rm s}$ and tables 4 (for BPMG, calibrated for $5-30\,$Myr stars) and 5 (for ABDMG, calibrated for MS stars) in \cite{2013a_Pecaut} to linearly interpolate the predicted spectral type. These agreed to within half a sub-class for all BPMG candidates and to within one sub-class for all ABDMG candidates (see Table~\ref{T_Kinematics}).

\section{Confirming membership for BPMG and ABDMG candidates}\label{S_Confirming}

The criteria for MG membership that we adopt in this paper are: (i) the candidate should have kinematics consistent with those defined by previously defined members of the BPMG or ABDMG; (ii) the ``kinematic parallax'' implied by the MG velocity and the candidate's position and proper motion should agree with any trigonometric parallax; (iii) the candidate should be young enough to exhibit H$\alpha$ in emission and have $L_{\rm x}/L_{\rm bol} \sim 10^{-3}$; (iv) the position of the candidate in the absolute magnitude versus colour diagram (using either its trigonometric or kinematic parallax) should be consistent with other members of the MG.

A moving group with a common 3D velocity will appear to have a convergent point on the sky. The line of sight velocity (i.e. the RV) of a MG member is $V_{\rm T}\cos\lambda$, where $V_{\rm T}$ is the magnitude of the velocity of the MG and $\lambda$ is angle between the sky position of the target star and the convergent point of the MG. The convergent points (05h 19m 48s, $-$60d 13m 12s and 06h 11m 34s, $-$47d 43m 39s) and values of $V_{T}$ (21.4 and $31.2\,{\rm km\,s}^{-1}$) are derived for the BPMG and ABDMG respectively, using the lists of confirmed MG members in \cite{2014a_Gagne}. We require that our candidates have $|\Delta {\rm RV}| = |{\rm RV} - V_{\rm T}\cos\lambda| < 5\,{\rm km\,s}^{-1}$ to be considered genuine members. However, we also include J0437 (= GJ~3305~AB) as a member of BPMG (see Table~\ref{T_BPMG}), with $\Delta {\rm RV} = +7.1\,{\rm km\,s}^{-1}$ because it is a low-mass common proper-motion companion to 51 Eri, which is a known BPMG member. Although this is a visual binary separated by 6\,AU (\citealt{2012b_Delorme}) we were unable to resolve the two components. Previously published, higher precision RV measurements for GJ~3305~AB (e.g., \citealt{2012a_Bailey}; \citealt{2012a_Shkolnik}; \citealt{2015a_Macintosh}, see Table~\ref{T_BPMG}) would satisfy our $|\Delta {\rm RV}| < 5\,{\rm km\,s}^{-1}$ criterion.

The proper motion of a candidate (taken from the PPMXL catalog, \citealt{2010a_Roeser}) should correspond to the tangential velocity predicted by MG membership. This defines a kinematic parallax that should be within $2\sigma$ of any measured trigonometric parallax: four of the confirmed BPMG members and one confirmed ABDMG member have trigonometric parallaxes and all satisfy this criterion. There were no cases in which kinematic criteria for a BPMG candidate matched ABDMG, or vice versa.

It is universally observed that M-dwarfs in young ($<150$\,Myr) clusters exhibit chromospheric H$\alpha$ in emission, and H$\alpha$ emission is observed in all stars in the Hyades and Praesepe ($\sim\,600\,$Myr) with spectral types later than K7 (\citealt{2014a_Douglas}). Therefore we demand that our young MG candidates must have H$\alpha$ in emission. One object each from the BPMG and ABDMG candidates (J0501 and J1012, respectively) were rejected on the basis of having H$\alpha$ in absorption, although they passed the kinematic tests.

\begin{figure}
\begin{center}
\includegraphics[scale=0.23]{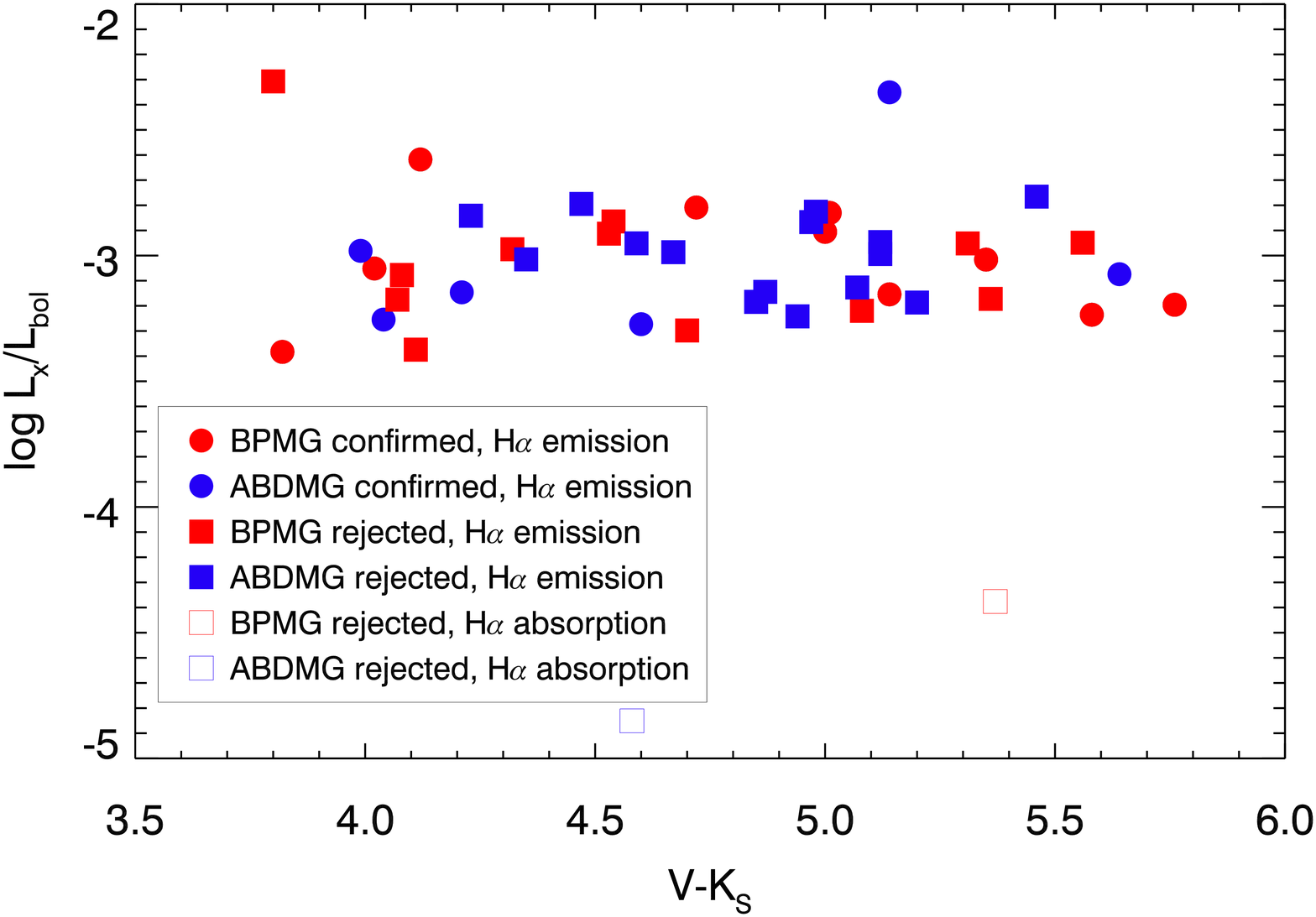}
\end{center}
\caption{Xray-to-bolometric luminosity ratios versus $V-K_{\rm s}$ colour for the entire observed sample of BPMG and ABDMG candidates. Closed symbols denote objects with H$\alpha$ in emission and open symbols have H$\alpha$ in absorption. Objects judged as new MG members are represented by circles and square symbols are objects that fail the membership tests (see $\S$\ref{S_Confirming}). All objects with H$\alpha$ in emission were found to have $L_{\rm x}/L_{\rm bol}$ consistent with young, magnetically-active stars. One candidate in BPMG and one in ABDMG were rejected as members because they have H$\alpha$ in absorption. Both of these have $\log L_{\rm x}/L_{\rm bol} < -4$, consistent with low magnetic activity.}
\label{F_Xray}
\end{figure}

As well as displaying H$\alpha$ in emission, young, magnetically-active, low-mass stars should have large X-ray-to-bolometric luminosity ratios. Using the formulation in \cite{1995a_Fleming} and \cite{2001a_Stelzer} we calculated $L_{\rm x}/L_{\rm bol}$ ratios using HR1 and count rates from the ROSAT BSC and FSC catalogs. The ROSAT positional error circle is $\sim 15$'', therefore we searched a 30'' circle around each candidate, and it may be possible that there is some X-ray contamination in some cases. All candidates displaying H$\alpha$ in emission have $L_{\rm x}/L_{\rm bol}$ values within a factor of 3 of the saturation limit for X-ray luminosity ($\sim 10^{-3}$, \citealt{2000a_James}). The objects we finally classify as likely BPMG and ABDMG members all have $L_x/L_{\rm bol} \geq 10^{-3.38}$. The two candidates with H$\alpha$ in absorption also have $L_{\rm x}/L_{\rm bol} < 10^{-4}$ (see Figure~\ref{F_Xray} and Table~\ref{T_Kinematics}).

The $v \sin i$ data are unable to further constrain MG membership. Whilst rapidly rotating stars could signify youth, M-dwarfs spin down less rapidly than their solar-mass counterparts. \cite{1999a_Terndrup} observe that M-dwarfs in the Pleiades (age $\sim 125\,$Myr) have $v\sin i$ ranging between $< 10\,{\rm km\,s}^{-1}$ and $100\,{\rm km\,s}^{-1}$ and $< 10\,{\rm km\,s}^{-1}$ and $50\,{\rm km\,s}^{-1}$ in Hyades M-dwarfs (age $\simeq 625\,$Myr).

Objects that belong to either the BPMG or ABDMG should broadly map a sequence that is consistent with previously confirmed members in a colour versus absolute-magnitude diagram (CMD), given their calculated kinematic parallax. Figures~\ref{F_BPMG_CMD} and~\ref{F_ABDMG_CMD} are $M_{K}$ vs $V-K_{\rm s}$ CMDs for the BPMG and ABDMG (respectively) for candidates that satisfied all criteria previously discussed in this section, compared with BPMG and ABDMG members from \cite{2004a_Zuckerman}, \cite{2008a_Torres} and objects with $> 90$ per cent membership probability in Malo et al. (2013; 2014a; 2014b); all of which satisfy $|\Delta {\rm RV}| < 5\,{\rm km\,s}^{-1}$. Isochronal models from Bell et al. (2014, herein B14, using the interior models of \citealt{2011a_Tognelli}) are overplotted at 10 and 20\,Myr for the BPMG sequence in Figure~\ref{F_BPMG_CMD} and at 100 and 150\,Myr for the ABDMG sequence in Figure~\ref{F_ABDMG_CMD}. We find the B14 isochrones best match the BPMG and ABDMG sequence and agree with recent LDB ages for both the BPMG (\citealt{2014a_Binks}) and Tucana-Horologium (\citealt{2014a_Kraus}; \citealt{2015a_Bell}). Although there is scatter of as much as $\sim\,1$ magnitude amongst objects in both CMDs we cannot rule out membership for any of our candidates because the scatter could be a combination of i) unresolved binarity (up to $\sim\,0.75\,$mag brighter than single stars); ii) variability in the sources ($\sim 0.3\,$mag, \citealt{2014a_Soderblom}); iii) photometric uncertainties (generally $\leq 0.1\,$mag) and/or iv) a possible age spread in the MGs (unknown).

\nocite{2014a_Bell}
\nocite{2015a_Baraffe}

Of the 10 confirmed BPMG members, 6 (out of 10) were from \cite{2013a_Malo} and two each were from Schlieder et al. (2012, out of 10) and Shkolnik et al. (2012, out of 4). Of the 6 ABDMG objects confirmed as M-dwarf members, 4 (out of 16) were from \cite{2012a_Schlieder} and 2 (out of 7) were from \cite{2013a_Malo}. A number of the candidates that are confirmed by their RV also show strong Li absorption. The presence of Li is a strong indicator that the age of an M0 (M5) dwarf is less than 150 (50)\,Myr (\citealt{2014a_Jeffries}).

We also tested the membership of these objects using other commonly used and available methodology. The BANYAN II web-tool\footnote{\url{http://www.astro.umontreal.ca/~gagne/banyanII.php}} predicts MG membership probabilities based on both kinematic and positional data (\citealt{2014a_Gagne}). We used right ascension, declination, proper-motions and RVs for inputs (and only parallaxes if a trigonometric parallax was available). The BANYAN II probabilities are presented in the last column of Table~\ref{T_Kinematics}. Only 2 of our confirmed BPMG members, J0501 and J0533, have membership probabilities of $>\,50$ per cent based on BANYAN II; 6 objects were more likely to be field stars than members of the BPMG and J0508 had a much higher probability of being a Columba MG member ($>\,50$ per cent). Membership probabilities derived in BANYAN II were $<\,25$ per cent for all but one of our confirmed ABDMG members (see Table~\ref{T_Kinematics}). We caution that some of these low membership probabilities may be because BANYAN II uses the proximity of candidates to the $X,Y,Z$  coordinates of the MG, where the MG centroid and dispersion are defined by a list of ``bona fide members'', as a membership criterion.  As \cite{2014a_Gagne} concede, it is possible that these lists are incomplete (we note there are  comparatively few BPMG and ABDMG members in the northern hemisphere) and are probably biased towards closer stars (particularly M-dwarfs), with the result that the spatial dispersion of MGs are probably underestimated. If so, this could lead to artificially lowered membership probabilities for more distant objects, such as those included in our study. We also note that two of our likely members in BPMG (J1643 and J0508) with very low BANYAN II probabilities have strong Li lines that indicate they must be very young objects. In what follows we adopt our membership criteria, but report the BANYAN II probabilities in Table~\ref{T_Kinematics}.

\section{Discussion}\label{S_LDBs}

\subsection{Revisiting the LDB of BPMG}\label{S_LDB_BPMG}

\begin{figure}
\begin{center}
\includegraphics[scale=0.18]{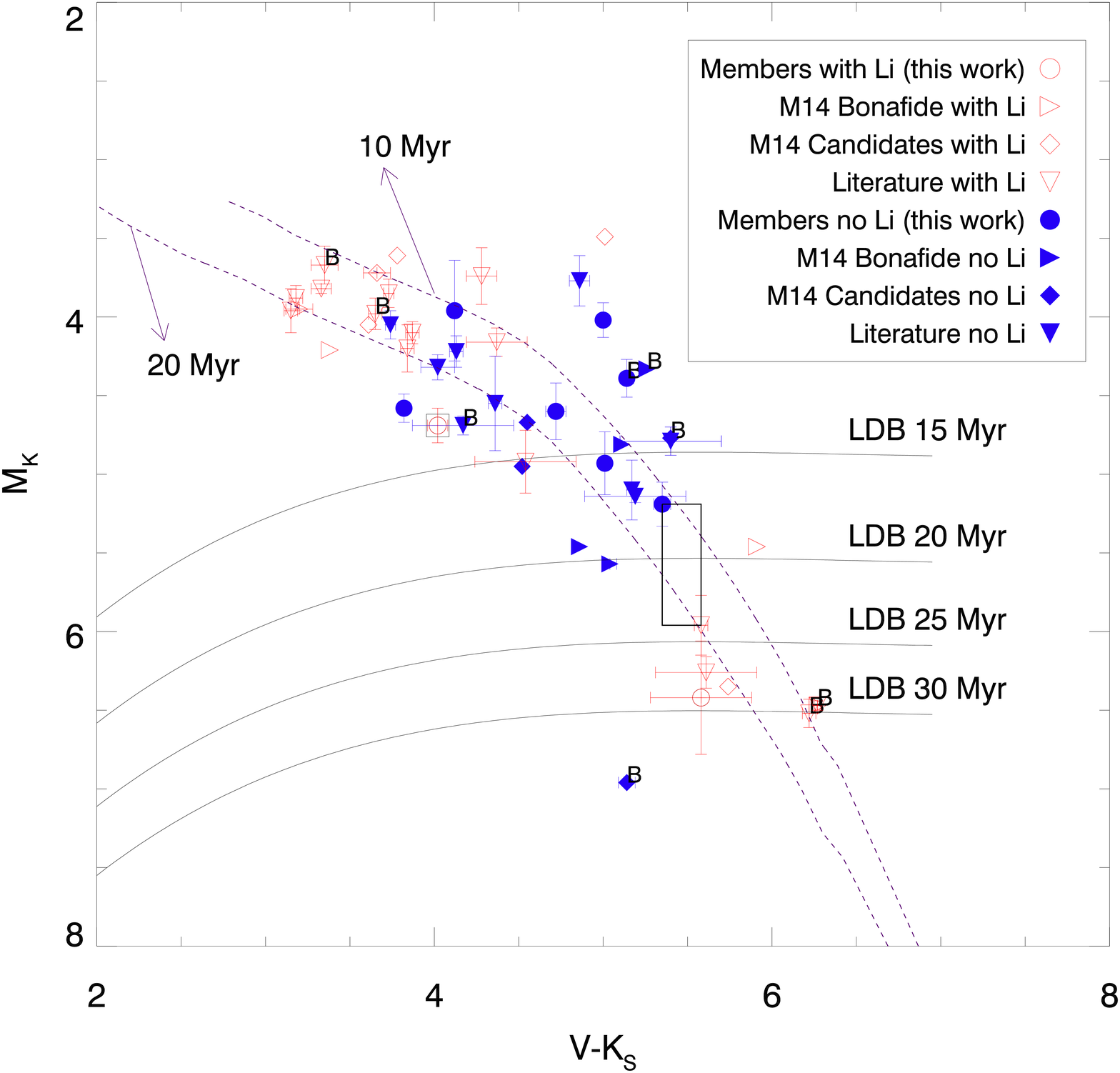}
\vskip+3ex
\includegraphics[scale=0.24]{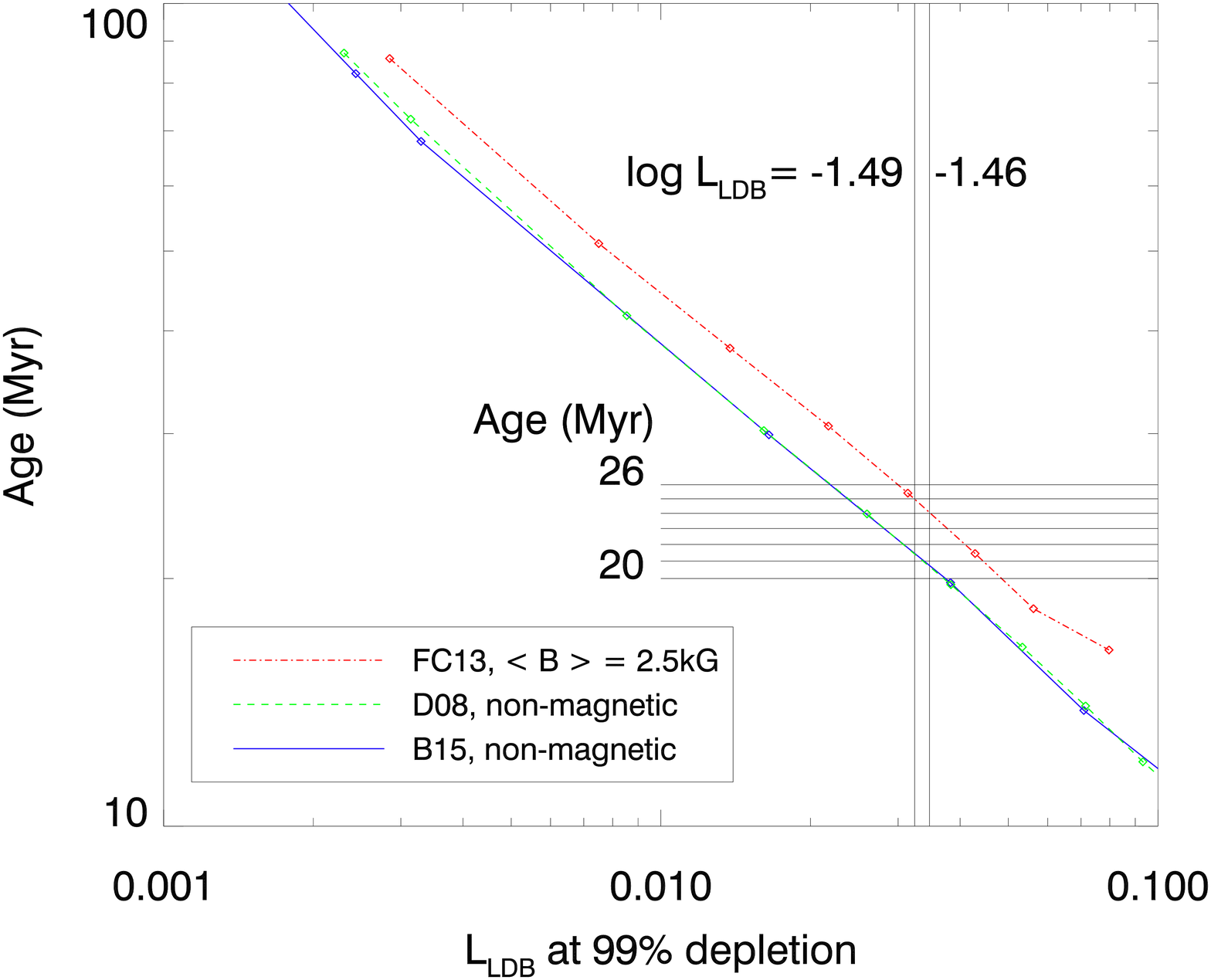}
\end{center}
\caption{Top: The revised CMD for BPMG using the newly identified members in \protect\cite{2014b_Malo} and objects we qualify as members. Any of the observed candidates that we class as members in this work that have zero probability of membership based on the BANYAN II analysis are represented by black, open squares. The dot-dashed lines correspond to the isochrones from the \protect\cite{2014a_Bell} models at 10 and 20\,Myr. M14 Bonafide = objects referred to as bonafide BPMG members in \protect\cite{2014b_Malo}, M14 Candidates = objects referred to as candidate BPMG members in \protect\cite{2014b_Malo}. Bottom: Comparison of the LDB location using the Dartmouth stellar evolutionary models. The red dot-dash line represents a surface magnetic field of 2.5\,kG and the green dotted line (D08) is the non-magnetic Dartmouth evolutionary model (\protect\citealt{2008a_Dotter}). The \protect\cite{2015a_Baraffe} models (blue solid line, B15) demonstrate the model consistency amongst non-magnetic models. The difference between the magnetic models and the non-magnetic models, for a given $L_{\rm LDB}$, is $\sim 3$\,Myr but $\leq 0.5\,$Myr between the non-magnetic models.}
  \label{F_BPMG_CMD}
\end{figure}

Binks \& Jeffries (2014, herein BJ14) used the LDB method to estimate an age of $21 \pm 4\,$Myr for the BPMG. The LDB method works by measuring the age-dependent lowest luminosity where almost-complete Li depletion is observed in confirmed low-mass members (or alternatively the highest luminosity at which Li remains undepleted). The principal advantage of the LDB technique is that different evolutionary models predict very similar relationships between the luminosity at the LDB and age -- i.e. the technique is model-insensitive. However, recent work on the BPMG by \cite{2014b_Malo} using a partly different sample of low-mass members and models which incorporate magnetic inhibition of convection (due to \citealt{2013a_Feiden}) has arrived at an older LDB age of $26 \pm 3\,$Myr. This age discrepancy lies beyond any model-dependence identified in BJ14. Here we re-examine the LDB of the BPMG using the new members identified in this paper plus newly identified low-mass members from \cite{2014b_Malo}. Our aim is to establish whether the older is age due to a difference in methodology, a difference in the sample of low-mass stars used or a difference due to the adoption of magnetic models.

\nocite{2014a_Binks}

Figure~\ref{F_BPMG_CMD} shows the $M_{K}$ vs $V-K_{\rm s}$ diagram with all the new members from this paper and from \cite{2014b_Malo}. Stars with (and without) Li are identified as having an Li EW $>$ ($<$) 200\,m\AA, corresponding to $<$ ($>$) a factor of 100 in Li depletion from the initial value at birth (\citealt{2007a_Palla}). The \cite{2014b_Malo} sample does not change our estimation of the LDB location and the slightly lower LDB luminosity quoted by Malo et al. (2014b, $\log L_{\rm LDB}/L_{\odot} = -1.49$ vs $-1.46$ in B14) results in only a 1\,Myr difference. The primary difference appears to be that the magnetic models yield an older age. Using the magnetic models we obtain an age of $24 \pm 4\,$Myr (see the bottom panel of Figure~\ref{F_BPMG_CMD}).

This model dependence is caused by the magnetic field which inflates the radii and reduce core temperatures for pre-main sequence stars at a given age, delaying the onset of Li. Both \cite{2014a_Jackson} and \cite{2015a_Somers} have also consistently predicted that coverage by dark starspots would have a similar effect on LDB ages. It would require $\sim\,25$ per cent coverage by dark spots to obtain the same LDB age inferred by the magnetic models. At present whilst it is clear that these stars are magnetically active, it is unclear how effective magnetic inhbition of convection is in fully convective stars. Similarly, whilst we know (from rotational modulation) that young M-dwarfs have significant starspot coverage, it is still uncertain just how much flux is blocked at the surface (e.g. \citealt{2013a_Jackson}). Any process that acts to inflate radii during pre-main sequence evolution will make LDB ages older.

\subsection{An LDB age for ABDMG?}\label{S_LDB_ABDMG}

\begin{figure*}
\begin{center}
\includegraphics[scale=0.3]{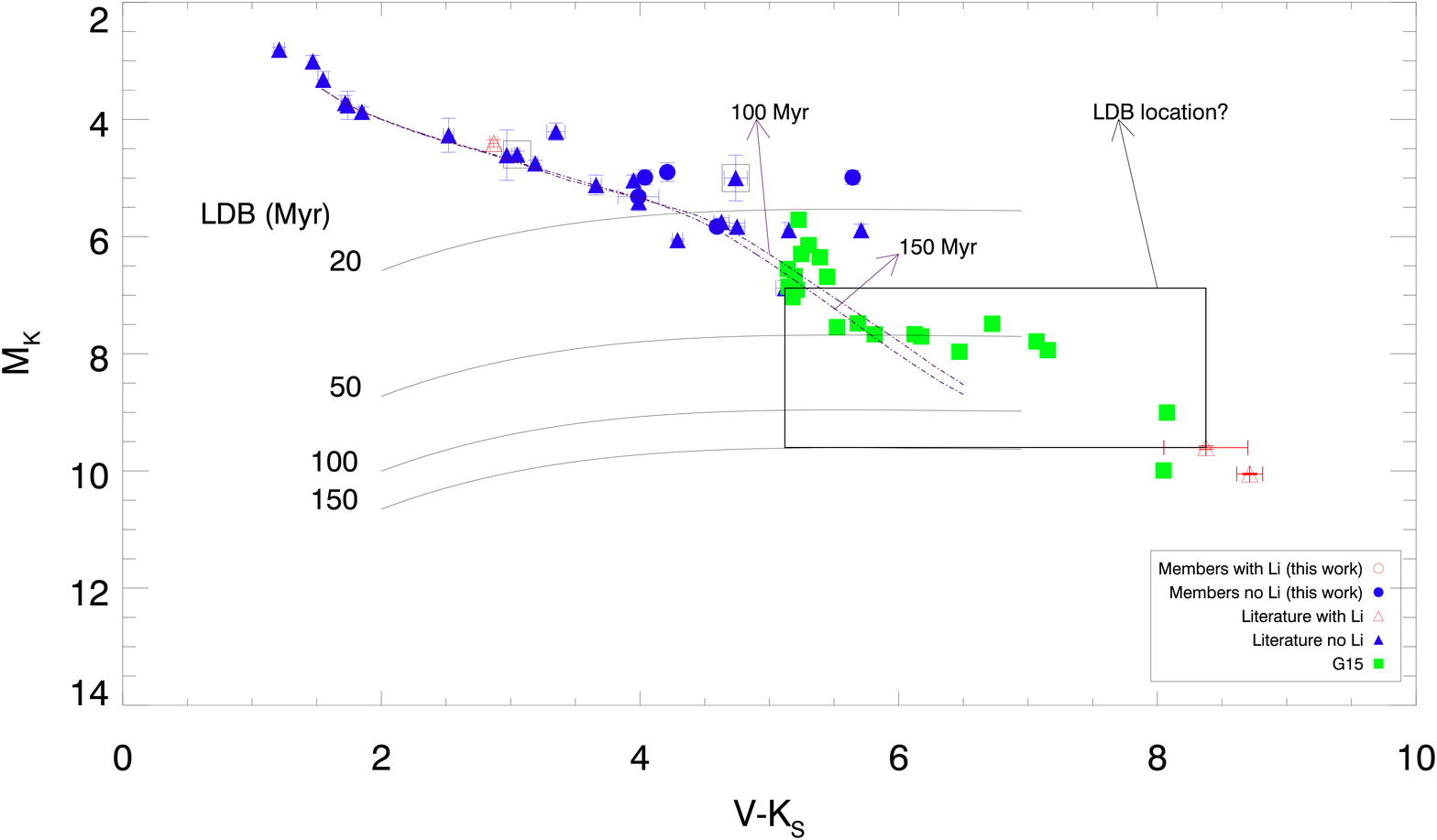}
\end{center}
\caption{Current constraints on the location of the LDB in ABDMG. Any of the observed candidates that we class as members in this work that have zero probability of membership based on the BANYAN II analysis are represented by black, open squares. Green squares are the proposed candidates of ABDMG with spectral types between M4 and M9 in Gagn{\'e} et al. (2015, G15). The LDB luminosity tracks are for 99 per cent depletion and the isochrones are the same as in Figure~\ref{F_BPMG_CMD} but correspond to 100 and 150\,Myr.}
\label{F_ABDMG_CMD}
\end{figure*}

The addition of the 6 confirmed ABDMG objects observed at the INT in this paper represents a $\sim 50$ per cent increase in the number of M-dwarf ABDMG members with an Li measurement, although most are upper limits. \cite{2014a_Gagne} report an M8 ABDMG candidate (2MASS~J0019262+4614078, herein J0019+4614) with an RV of $-19.5 \pm 2.0\,{\rm km\,s}^{-1}$, which gives a value of $\Delta {\rm RV} = 3.2\,{\rm km\,s}^{-1}$. \cite{2009a_Reiners} detected a strong Li feature around the 6708\AA\ line in J0019+4614. An Li EW value is not reported, however an estimation of the EW by eye from the top-left panel in their figure 3 suggests it is $\gtrsim 625$\,m\AA, consistent with an undepleted Li abundance (\citealt{2007a_Palla}). \cite{2015b_Bowler} have identified a M7.5$\pm$0.5 secondary component to J15594729+4403595 (hereafter J1559~B), which is confirmed as an ABDMG member in this work and has an Li EW of $710\,$m\AA. Although our RV measurement of $-29.5 \pm 3.8\,{\rm km\,s}^{-1}$ satisfies ABDMG membership, previously published RVs of $-15.8 \pm 0.5$ and $-19.6 \pm 0.6$ in \protect\citealt{2014a_Malo} and \protect\citealt{2015a_Bowler} (respectively) are inconsistent. It is possible that the object is a spectroscopic binary and we do not rule out membership based on these differences in RV. \cite{2015a_Bowler} measure a kinematic distance of 27\,pc to J1559~B, and a $K_{\rm s}$ apparent magnitude of $11.76 \pm 0.03$, which gives $M_{K} = 9.60 \pm 0.03$ and a $V-K_{\rm s}$ colour of $8.38 \pm 0.33$ by linearly interpolating the spectral type using table 5 in \cite{2013a_Pecaut}.

Both the confirmed objects observed at the INT and confirmed members in the literature are placed onto a $M_{K}$ versus $V-K_{\rm s}$ CMD in Figure~\ref{F_ABDMG_CMD}. There is a gap of several magnitudes in both colour and magnitude between an Li-poor object (2MASS~J04141730$-$0906544, $V-K_{\rm s} = 5.12$, $K = 8.76 \pm 0.02$, trigonometric distance of $23.8 \pm 1.4$\,pc and $M_{K} = 6.88$) and an Li-rich object (J1559~B, see above). The corresponding LDB age is 35--150\,Myr using the non-magnetic models of \citealt{2015a_Baraffe}, or an age range of 40-165\,Myr using the same magnetic models that \cite{2014b_Malo} used to interpret the LDB of the BPMG. Whilst this value is poorly constrained, it does provide at least some indication of a lower limit to the age, and is consistent with the $149^{+51}_{-19}$\,Myr age recently provided by \cite{2015a_Bell}, which is based on fitting empirical isochrones and is independent of the LDB technique. Should J1559~AB turn out not to be an ABDMG member, then J0019+4614 ($V-K_{\rm s} = 8.71$, $K_{\rm s} = 11.50 \pm 0.01$, kinematic distance of 19.5\,pc and $M_{K} = 10.05 \pm 0.03$) would provide an age upper limit of 196\,Myr using the \cite{2015a_Baraffe} models, or 218\,Myr using the magnetic models from \cite{2014b_Malo}. The present situation for an LDB age for ABDMG is far from satisfactory. There is a striking void of RV-confirmed ABDMG objects between M4 and M8. Should such stars exist in ABDMG, an assessment of their Li content would almost certainly improve the location of the LDB and provide a more precise age. 

\section{Conclusions}\label{S_Conclusions}

In this paper we have used optical spectroscopy to test the membership status of previously reported M-dwarf candidates of the BPMG and ABDMG. Ten BPMG and six ABDMG candidates are confirmed as members based on i) measured RVs which are  within $5\,{\rm km\,s}^{-1}$ of the expected RV required for MG membership; ii) high levels of magnetic activity by virtue of observing H$\alpha$ in emission and $L_{\rm x}/L_{\rm bol}$ values that are consistent with very youthful M-dwarfs; iii) the kinematic parallaxes implied by cluster membership place the candidates close to the sequence of known members in an absolute magnitude versus colour diagram. We measure RVs for the first time for 12 BPMG and 19 ABDMG candidates, 2 and 4 of which we confirm as members, respectively. Lithium measurements are obtained for the first time for 16 BPMG and 22 ABDMG candidates, of which 2 and 5 qualify as members, respectively. Although the majority of our proposed new MG members returned low membership probabilities (\citealt{2014a_Gagne}) this may be because BANYAN II uses spatial location as part of its membership assessment and many of our objects lie beyond the previously considered spatial extents of these MGs.

Whilst we do not observe any new BPMG members that improve the location of the lithium depletion boundary (LDB), several new members bolster its position on a CMD (see Figure~\ref{F_BPMG_CMD}) and we find that magnetic inhibition of convection manifested either as dark spot coverage or as an interal magnetic field increase the age of the BPMG by $\sim 15$ per cent. A tightly constrained LDB for the ABDMG remains elusive, although large scale surveys such as the BANYAN All-Sky Survey (BASS, \citealt{2015a_Gagne}) and the Planets Around Low-Mass Stars (PALMS, \citealt{2015a_Bowler}) survey are uncovering dozens of M4-M9 ABDMG candidates, which with membership confirmation and an Li measurement would significantly improve the LDB age of the ABDMG.

A strong test for MG membership is that the parallaxes of candidates should be consistent with the kinematic parallax implied by MG membership. Four of our confirmed BPMG members have trigonometric parallaxes that support evidence for membership. The first data release from the {\it Gaia} mission is expected in 2017, and parallaxes for all M-dwarfs ($V\,<\,16$) in BPMG and ABDMG should be available to within $10\,\mu$as precision (\citealt{2008a_Lindegren}). We anticipate that, combined with suitable spectroscopic measurements, {\it Gaia} will be able to provide conclusive membership status for many M-dwarf MG candidates that await confirmation.

\section{Acknowledgements}

ASB and RDJ would like to thank Cameron Bell for providing useful suggestions which have improved this manuscript. Based on observations made with the Nordic Optical Telescope (46-102), operated by the Nordic Optical Telescope Scientific Association and with the Isaac Newton Telescope (I/2013A/1) operated by the Isaac Newton Group, at the Spanish Observatorio del Roque de los Muchachos of the Instituto de Astrofisica de Canarias.  The research leading to these results has received funding from the European Union Seventh Framework Programme (FP7/2007-2013) under grant agreement No. 312430 (OPTICON).  This research has made use of the SIMBAD database, operated at CDS, Strasbourg, France. ASB acknowledges the support of the STFC.

\bibliography{Paper}
\end{document}